\documentclass[11pt]{report}

  \usepackage[T1]{fontenc}
  \usepackage{graphicx}
  \usepackage[skins]{tcolorbox}
  \usepackage[Lenny]{fncychap}
  \usepackage{ydrop}
  \usepackage{amssymb}
  \usepackage{hyperref}
  \usepackage{xcolor}
  \usepackage[draft,nomargin,inline,index]{fixme} 
  \usepackage{graphicx}
  \usepackage{aas_macros}
  \usepackage[inline]{enumitem}
  \usepackage[title]{appendix}
  \usepackage{caption}
  \usepackage{array}
  \usepackage{xspace}
  \usepackage{amsmath,amsfonts,amssymb}
\fxsetup{theme=color,mode=multiuser}

  \FXRegisterAuthor{ggw}{aggw}{GGW}

  \FXRegisterAuthor{jcw}{ajcw}{JCW}

  \FXRegisterAuthor{elr}{aelr}{ELR}

  \FXRegisterAuthor{jlh}{ajlh}{JLH}

  \FXRegisterAuthor{pss}{apss}{PSS}

  \FXRegisterAuthor{mmb}{ammb}{MMB}

  \FXRegisterAuthor{cp}{acp}{CP}

  \FXRegisterAuthor{ck}{ack}{CK}

  \FXRegisterAuthor{ed}{aed}{ED}

  \FXRegisterAuthor{az}{aaz}{AZ}

  \FXRegisterAuthor{ms}{ams}{MS}

  \FXRegisterAuthor{dl}{adl}{DL}

  \FXRegisterAuthor{bga}{abga}{BGA}

  \FXRegisterAuthor{lw}{alw}{LW}

  \FXRegisterAuthor{yy}{ayy}{YY}

  \FXRegisterAuthor{hwl}{ahwl}{HWL}

  \FXRegisterAuthor{sjc}{asjc}{SJC}

  \FXRegisterAuthor{ra}{ara}{RA}

  \FXRegisterAuthor{gca}{agca}{GCA}

  \FXRegisterAuthor{kt}{akt}{KT}

  \FXRegisterAuthor{lk}{alk}{LK}

  \FXRegisterAuthor{pyf}{apyf}{PYF}

  \FXRegisterAuthor{sb}{asb}{SB}

  \FXRegisterAuthor{kw}{akw}{KW}

\FXRegisterAuthor{hej}{hj}{HJ}

  \usepackage{natbib}
  \bibpunct{(}{)}{;}{a}{}{,}

  \newcommand{\lya}{Ly$\alpha$\xspace}
  \newcommand{\unitcgssb}  {erg\,s$^{-1}$\,cm$^{-2}$\,arcsec$^{-2}$\xspace}
  
  \newcommand{\unitcgslum} {erg\,s$^{-1}$\xspace}
  
  \date{}
  
  \newtcolorbox{hibox1}{
  fonttitle=\bfseries\Large,
  grow to left by=-1cm,
  grow to right by=-1cm,
  colback=blue!5!white,
  title=Key Point}

  \newtcolorbox{hibox2}{
  fonttitle=\sffamily\bfseries\large,
  title=Key Point,
  leftrule=3mm,
  arc=0mm,
  grow to left by=-1cm,
  grow to right by=-1cm,
  colback=red!5!white,
  colframe=red!75!black} 
  
  \newtcolorbox{hibox3}{
  arc=0mm,
  leftrule=3mm,
  grow to left by=-1cm,
  grow to right by=-1cm,
  colback=blue!5!white,
  colframe=black!75!white}
  
  \newcommand*{\titleGP}{\begingroup 
    \centering 
    \vspace*{\baselineskip} 

    \rule{\textwidth}{1.6pt}\vspace*{-\baselineskip}\vspace*{2pt} 
    \rule{\textwidth}{0.4pt}\\[\baselineskip] 

    {\huge POLARIMETRY WITH \\[0.5\baselineskip] THE GIANT MAGELLAN TELESCOPE}\\[0.3\baselineskip] 

    \rule{\textwidth}{0.4pt}\vspace*{-\baselineskip}\vspace{3.2pt} 
    \rule{\textwidth}{1.6pt}\\[\baselineskip] 

    \scshape 
    A Scientific Justification \\ \& \\ 
    Technical Requirements Document \\[\baselineskip] 
    October, 2026 \par 

    \vspace*{1\baselineskip} 
    {\Large G. Grant Williams$^{1}$, Ramya M Anche$^{1}$, B-G Anderson$^{2}$, G. C. Anupama$^{3}$, Stefano Bagnulo$^{4}$, Alex Carciofi$^{5}$, Seok Jun Chang$^{6}$, Ewan Douglas$^{1}$, Jennifer L. Hoffman$^{7}$, Helen Jermak$^{8}$, Christoph Keller$^{9}$, Ludmilla Kolokolova$^{10}$,
    Hee-Won Lee$^{11}$, Douglas C. Leonard$^{12}$, Nikos Mandarakas$^{13}$, Maxwell Millar-Blanchaer$^{14}$, Claudia Oliveira$^{5}$, Chris Packham$^{15}$, Rafael Ribeiro$^{5}$, Enrique Lopez-Rodriguez$^{16,17}$, Manisha Shrestha$^{18}$, Paul Smith$^{1}$, Konstantinos Tassis$^{13}$,  Jake D. Turner$^{19,20}$, Lifan Wang$^{21}$, J. Craig Wheeler$^{22}$, Klaas Wiersema$^{23}$, Padma Yanamandra-Fisher$^{24}$, Yujin Yang$^{25}$, \& Ann Zabludoff$^{1}$ \\}
    
    \vspace*{2\baselineskip}
    
    {\small $^{1}$University of Arizona, USA, $^{2}$McDonald Observatory, USA, $^{3}$Indian Institute of Astrophysics, India, $^{4}$Armagh University, UK, $^{5}$University of Sao Paulo, Brazil, $^{6}$Max-Planck Institute of Astrophysics, Germany, $^{7}$University of Denver, USA, $^{8}$Liverpool John Moores University, UK, $^{9}$National Solar Observatory, USA, $^{10}$University of Maryland, USA, $^{11}$Sejong University, Korea, $^{12}$San Diego State University, USA $^{13}$Department of Physics, University of Crete and IA-FORTH, Greece, $^{14}$University of California, Santa Barbara, USA, $^{15}$University of Texas, San Antonio, USA $^{16}$KIPAC, Stanford University, USA, $^{17}$University of South Carolina, SC, $^{18}$Monash University, Aus, $^{19}$Department of Astronomy and Carl Sagan Institute, Cornell University, Ithaca, NY, USA, $^{20}$NHFP Sagan Fellow}, $^{21}$Texas A\& M University, USA, 
    $^{22}$University of Texas, USA,  $^{23}$University of Hertfordshire, UK,  $^{24}$Space Science Institute, USA,  $^{25}$Korea Astronomy and Space science Institute, Korea.

    \vfill 

  \endgroup}

\begin{document}

\titleGP
\tableofcontents

\chapter{Introduction}
\label{sec:intro}

\drop{T}he photons we collect from astronomical sources are encoded with information about the physics behind their origin. To fully probe that physics, we must maximize the information we extract from those photons. At optical and near-infrared wavelengths, this is typically done with the use of filters or a disperser to extract flux as a function of color. An often disregarded piece of information that is also carried by those photons is the polarization state. Astronomical polarimetry is capable of revealing details about the source geometry and magnetic fields that are often unresolvable and, therefore, inaccessible by any other means. Polarimetry is typically a photon-starved science because crucial model discriminators are often only revealed when precisions of $0.1\%$ in the degree of polarization are achieved. For spectropolarimetry, this necessitates the collection of over one million photons at each spectral bin, which demands the use of large aperture telescopes for typical objects of interest. The future completion of Giant Segmented Mirror Telescopes (GSMTs) promises the long-awaited light grasp to initiate an unprecedented era of discovery with astronomical polarimetry.


%
%
%
%
%
%
%
%
%
%

\section{The Power of Polarimetry}
\label{sec:polarimetry}
    
Polarimetry provides unique information about celestial objects and exposes astrophysical processes that cannot be revealed with imaging or spectroscopy alone. Spectropolarimetry extracts the maximum information from the source photons. Rather than measuring just flux (e.g.,\ imaging) or flux as a function of wavelength (e.g.,\ filtered imaging or spectroscopy), spectropolarimetry measures the flux as a function of wavelength \textit{and} as a function of the polarization state.

For sources that are intrinsically non-spherical (in geometry or composition) or have significant magnetic fields, polarimetry is often the only direct measure of those physical characteristics. Polarization studies have, for example, established the unified model of active galactic nuclei \citep[AGN;][]{1985ApJ...297..621A, 1993ARAA..31..473A}, revealed the elusive magnetic field in the Milky Way and external galaxies \citep{hiltner1949a, 1949Sci...109..166H, 1951ApJ...114..206D}, established that asymmetries are common in supernovae \citep[see][]{2008ARA&A..46..433W}, and provided details of magnetic structure in unresolved chromospherically active stars \citep{1992A&A...265..669D}. Even when the source being studied is unpolarized, the multiple exposures required for polarimetry can yield high
signal-to-noise images or spectra.

The 2020 Decadal Survey ``Pathways to Discovery in Astronomy and Astrophysics for the 2020s'' \citeyearpar{2021pdaa.book.....N} recognized the importance of polarimetry in enabling future discovery and an
NOAO-sponsored report \citep{2016arXiv161001661N} ranked spectropolarimetry as one of the priorities for characterizing explosive transients in the era of the Vera C. Rubin Observatory and beyond.

\vspace{5mm}
\begin{hibox1}
Polarimetry provides unique information about celestial objects and exposes astrophysical processes that cannot be revealed with imaging or spectroscopy alone. Polarimetry will support several goals of the 2020 Decadal Survey.
\end{hibox1}


\vspace{5mm}
\section{The Benefits of GMT Over TMT or ELT}
\label{sec:gmt_vs_tmt}

\begin{figure}
    \centering
    \includegraphics[width=1\textwidth]{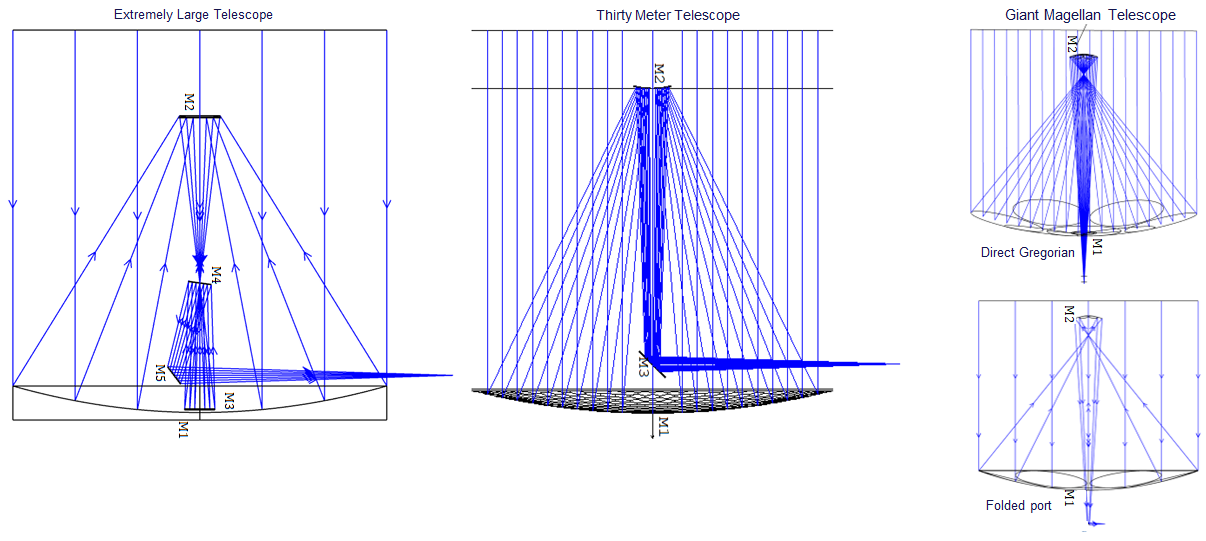}
    \caption{Optical design of all the three GSMTs from Ansys\textsuperscript{\textregistered} Zemax. The Nasmyth configuration with the fold mirror is shown for TMT and ELT, and both the direct Gregorian (DG) and folded port (FP) are shown for GMT.}
    \label{fig:optical-design}
\end{figure}

Of the three GSMT designs shown in Figure \ref{fig:optical-design}, i.e.,\ the Thirty Meter Telescope (TMT), the Extremely Large Telescope (ELT), and the Giant Magellan Telescope (GMT), the GMT will likely be the most capable of low-amplitude, high-precision polarimetry because of its axisymmetric design. The other telescopes use fold mirrors to redirect the source light to their science instruments. Fold mirrors are not axisymmetric and therefore, they introduce significant instrumental polarization, which is often higher than the polarization signal from the astrophysical target itself. In principle, instrumental polarization can be calibrated out but, in practice, this calibration becomes very problematic for large alt-az telescopes since flexure and derotation of the instrument result in an instrumental polarization that changes with time \emph{during an observation}.  Despite this challenge, there are groups within both the TMT and ELT consortia who are advocating for a polarimetric capability for those telescopes (see \citealt{2014SPIE.9150E..13A,anche2018analysis} \& \url{https://www.eso.org/sci/facilities/eelt/science/doc/SC_EELT_polarimetry_072009.pdf}). 
 The optical design of the GMT direct Gregorian (DG) focus is completely axisymmetric, and therefore, any instrumental polarization it introduces will be extremely small, giving GMT a significant advantage over the other GSMTs for \emph{all} polarimetric science.
 If the GMT is not capable of collecting polarization information, valuable insight into the nature of many astrophysical targets will be inaccessible.   
Failing to incorporate a polarimetric capability for GMT would be an enormous missed opportunity.

\vspace{5mm}
\begin{hibox1}
The axisymmetric design of the Giant Magellan Telescope with the direct Gregorian focus gives it a significant advantage over the Thirty-Meter Telescope or the European Extremely Large Telescope for all polarimetric science.
\end{hibox1}
\vspace{5mm}

\vspace{5mm}
\section{Scientific Justification and Technical Requirements}
\label{sec:intro_justification}

In this document, we provide a scientific justification and the technical requirements for polarimetry on the GMT. Many of the science cases given in the \href{http://www.gmto.org/resources}{GMT Science Book 2018}\footnote{Available at \url{https://giantmagellan.org/for-scientists/}} would benefit from a polarimetric capability. Additional science cases for polarimetry with GSMTs are provided by groups associated with the Thirty Meter Telescope \citep[TMT,][]{2014SPIE.9150E..13A} and the Extremely Large Telescope (ELT, \href{https://www.eso.org/sci/facilities/eelt/science/doc/SC_EELT_polarimetry_072009.pdf}{``E-ELT Spectropolarimetry: The Science Case''}) \footnote{Available at \url{https://www.eso.org/sci/facilities/eelt/science/doc/SC_EELT_polarimetry_072009.pdf}}.

    The larger aperture of the GMT over the telescopes that currently have a polarimetric capability significantly increases the potential scientific return through:
    
    \begin{itemize}
     \item Improved S/N on current 8 \& 10-m class targets for the same exposure time.
     \item Shorter exposure times to achieve the same S/N on current 8 \& 10-m class targets.
     \item Similar S/N on considerably fainter targets, thereby significantly increasing the number of targets and the ability to study the statistics of specific targets.
      \item The potential for precision low-level polarimetry (well below 0.1\%).
   \end{itemize}

This document is intended to provide a broad variety of science cases for polarimetry on the GMT.  We did not limit the science cases to just those that have the same technical requirements, in fact, the technical requirements dif\/fer considerably.  Some of the science cases benefit from imaging polarimetry, others from spectropolarimetry.  Some require observations at optical wavelengths and some in the near and far infrared.  Some require low or moderate resolution spectropolarimetry while others require high resolution spectropolarimetry. This demonstrates that providing a polarimetric capability on the GMT is not limited to a specific instrument type but would be broadly beneficial, perhaps even suggesting that all future instruments incorporate a polarimetric capability.  Ultimately, we aim to initially provide a general-purpose capability to the community.  


\chapter{Transients}
\label{sec:transients}

    






\drop{A}n introduction to some elements of spectropolarimetry of transients is given in Chapter 8 of the GMT Science Book (2017). Relevant transients include all types of supernovae, long and short gamma-ray bursts, neutron star/neutron star and neutron star/black hole mergers (revealed through gravitational wave detections), tidal disruption events, classical and recurrent novae, and fast blue optical transients. Recent dedicated surveys have identified a large number of potentially interesting transients. To fully reveal the physics that produces these transients, it's critical to maximize the information content of the data that is collected. Therefore, it's important to obtain polarimetric and spectropolarimetric observations to complement the more common photometry and spectroscopy. Follow-up observations at very early and late times, when the transients may be quite faint, will be needed to disentangle the intrinsic polarization from the interstellar polarization.  The rarest and most interesting sources will likely be distant and faint.  Large aperture telescopes will be needed to provide quality follow-up observations. Because they are photon-starved techniques, polarimetry and spectropolarimetry of the most scientifically interesting sources will require 25-m class telescopes. All the transients discussed here would benefit from multi-epoch spectropolarimetry with cadences ranging from hours (e.g.\ GRBs) to days or weeks (e.g. SNe). 
   \section{Supernovae}
    
In a comprehensive \emph{Annual Reviews of Astronomy and Astrophysics} article entitled ``Spectropolarimetry of Supernovae,'' \citet{2008ARA&A..46..433W} noted:

\begin{quote}``Virtually all supernovae are significantly aspherical near maximum light \ldots''\end{quote}  

\noindent Despite the far-reaching implications of this statement, polarimetry of supernovae is still relatively rare. Such observations increased significantly following the successful polarization studies of SN~1987A \citep{1991ApJ...375..264J,1991ApJS...77..405J}, and
the field expanded throughout the 1990s and 2000s. However, even today, studying SNe with polarimetry, especially spectropolarimetry, is a relatively young field with significant discovery space lying ahead.  Rubin/LSST will greatly expand the potential for a better understanding of the asymmetries in supernovae that polarimetry reveals.  This is the case not because of the sheer number of SNe that LSST will discover (far too many to follow up), but more because of the number of very early discoveries and unusual events.
    
As first suggested by \citet{shapiro:1982p902}, linear polarization represents a key diagnostic to assess the level of asymmetry in a supernova explosion and characterize its nature. The opacity in young supernova atmospheres is dominated by electron scattering, which is highly polarizing.  Any asymmetry in the distribution of the scattering electrons results in a net polarization, with higher polarization generally indicating a greater departure from spherical symmetry (although this can vary with the viewing angle and properties of the electron-scattering atmosphere, e.g., \citealt{leonard:2005p330,Dessart:2011p3784}).

While observations of SN~1987A, the closest SN in the modern era, revealed asymmetries at late times through direct imaging \citep{wang:2002p671}, all other supernovae have been too distant to directly resolve even with the highest resolution imaging. Therefore, polarimetry is the most direct tool to assess the geometry of young SNe; it is capable of probing down to size scales that, if they {\it could}  be resolved, would correspond to an angular resolution of $10 \mu$-arcsec for a typical SN photosphere of $10^{15}$~cm at 10 Mpc \citep{2008ARA&A..46..433W}.

Nature is three-dimensional (3D) and the ability to discriminate this structure gives a special window into the physics of stellar explosions. With standard techniques of photometry and spectroscopy, we can study the 3D structure of explosions only in the case of nearby young supernova remnants. For more distant events, the most direct way to glean information about geometrical structure is through spectropolarimetry. Combined with line-of-sight Doppler shifts, spectropolarimetry yields information on the 3D structure of an explosion that is otherwise only a point of light. Only spectropolarimetry can resolve different chemical structures that are ejected at different angles, a critical clue to the nature of the underlying explosion or perhaps the interaction with a neighboring star. Polarimetric observations also provide key information on the orientation of asymmetric explosions, which constrains modeling and provides us with different viewpoints on explosions with complex geometries. 
 
 Most importantly, multi-epoch observations significantly enhance the power of spectropolarimetry, providing a transformative leap in our ability to interpret the physical phenomena underlying these complex time-dependent explosions. The expansion of the ejecta over time, combined with the receding photosphere, creates a scenario in which neither the light source nor the scattering region is constant. Multi-epoch spectropolarimetric observations thus probe changes in supernova polarization with \textit{both wavelength and time}, allowing us to reconstruct detailed characteristics of the explosion that are difficult or impossible to extract by other means.
    
    \subsection{Type Ia (Thermonuclear) Supernovae}
    
    There are a variety of reasons why Type Ia supernovae (SNe~Ia) could be asymmetric. The progenitor, especially in the case of a white dwarf merger, could be rotating and aspherical, the explosion itself could be asymmetric, or the ejecta could strike and immerse a companion. Careful monitoring of any asymmetry with spectropolarimetry could yield information on any or all of these aspects. Even small effects of asymmetry could impact the use of SNe~Ia for precision cosmology.
    
    Spectropolarimetric observations of normal SNe~Ia show that the level of intrinsic continuum polarization is typically quite low ($\sim$ 0.1\%), and that the polarization in the lines of Si~\textsc{ii} and Ca~\textsc{ii} is significant at early times ($\sim$ 1.0\%), but quickly diminishes after the peak in the light curve \citep[e.g.,][]{2010ApJ...722.1162M,2016ApJ...828...24P,2017ApJ...835..100M,2007Sci...315..212W}. Many SNe~Ia also show variable polarization in the Fe~\textsc{ii} lines \citep[e.g.,][]{2009A&A...508..229P}. Polarization is especially prominent in the high-velocity components of these lines \citep[e.g.,][]{2013MNRAS.433L..20M}. The high-velocity features are strongest before maximum light and have characteristic velocities that are $\sim 8000$ km s$^{-1}$ higher than the photospheric velocity. Although nearly ubiquitous in typical SNe~Ia (and absent in SN~1991bg-like events), these remarkable features are unexplained. Suggestions include some aspect of the explosion physics or interaction with a very nearby circumstellar medium.
    
        The current sparse data on SNe~Ia show that the continuum polarization is typically small, suggesting that the ejecta are not strongly distorted near maximum light. This statement already rules out certain models of violently merging white dwarfs but is consistent with standard thermonuclear combustion models \citep{bulla_phd}. An important exception are the relatively faint SN~1991bg-like events that show appreciable continuum polarization rising into the red. This is one clue among several that this sub-class of SNe~Ia may result from different physics than typical SNe~Ia. In normal SNe Ia, the line polarization points to compositional inhomogeneities of intermediate mass-elements with clumpy spatial distributions that partially obscure the photosphere and are not necessarily aligned with any global asymmetry of the ejecta \citep{2008AJ....136.2227C}.
    
    Asymmetries have been identified in both subluminous and overluminous SNe Ia \citep[e.g.,][]{2001ApJ...556..302H,2012A&A...545A...7P,2010ApJ...714.1209T}, and there have been some attempts to correlate polarization results and intrinsic luminosity \citep{2005ApJ...632..450L}.  \citet{2010Natur.466...82M} even argued that the spectral diversity in Type Ia supernovae is a consequence of the random viewing angles from asymmetric explosions. 
    
    Additional studies will be critical to statistically understand the predominance and impacts of the asymmetries and whether there are any correlations with other observable parameters.  A particular issue is whether there are connections between the geometries of the progenitor systems and the combustion physics. Since SNe~Ia are treated as standardizable candles, any correlation of polarization measurements and light curves may help reduce the scatter or remove outliers. Despite early time asymmetries revealed by spectropolarimetry, there has not yet been any substantial impact on calibrating Type Ia for cosmological studies \citep{2017ApJ...835..100M}. This may change as more subtle systematic effects are revealed \citep{riess2018a} and tensions rise between supernova studies and studies of the cosmic background radiation.
    
    To understand the origin and impact of the observed polarimetric asymmetries in SNe~Ia, it is important to get frequent observations of the rapidly-changing conditions on the rise to maximum. The sensitivity of the GMT will allow observations during the early faint phases for relatively nearby events and to include more distant events to increase the current woeful statistical sampling. The large aperture of the GMT is crucial to advancing this field.
    
    \subsection{Core Collapse Supernovae}
    
    Polarimetric observations have been obtained for all types of core collapse SNe: IIP, IIL, IIn, IIb, Ib, \& Ic. Virtually all core-collapse supernovae reveal evidence of strong asymmetry \citep{2008ARA&A..46..433W}. The general trend is that the line and continuum polarization tend to get stronger with time as inner layers are revealed. Stripped-envelope events also tend to have stronger polarization than do SNe Type IIP, which have large blanketing hydrogen envelopes. These data suggest that large asymmetries may be inherent to the explosion process itself, perhaps from asymmetries in the neutrino-deposition process or due to magnetic jets. Jets are especially suggested in the supernovae associated with GRBs.

	Time-resolved spectropolarimetry of core-collapse supernovae has been shown to be remarkably powerful in disentangling the asymmetries in individual components of the ejecta; for example, different elements may be ejected at different angles and with different velocities. In SN~2011dh, a Type IIb explosion, \citet{Mauerhan:2015p4467} found that fast-rising plumes of radioactive $^{56}$Ni explain both the observed polarization profiles and the agreement between the early-time continuum polarization and the subsequent H$\alpha$ and \ion{He}{1} line polarization. In SNe that are interacting with dense circumstellar material (particularly those of Type IIn), line polarization probes the geometry, composition, and density of the CSM and illuminates the late stages of the progenitor's evolution  (e.g., \citealt{2000ApJ...536..239L};  \citealt{2008ApJ...688.1186H};  \citealt{2019MNRAS.488.3089K}; \citealt{2020MNRAS.498.3835B}). \citet{2024MNRAS.529.1104B} compiled multi-epoch spectropolarimetric observations of a sample of 14 SNe IIn and found that many show signatures of dusty CSM and/or multiple shell ejections, and that their asymmetric geometries strongly favor binary progenitors.

\begin{figure}[ht!]
\rotatebox{90}{
 \scalebox{0.7}{
\includegraphics[width=\textwidth]{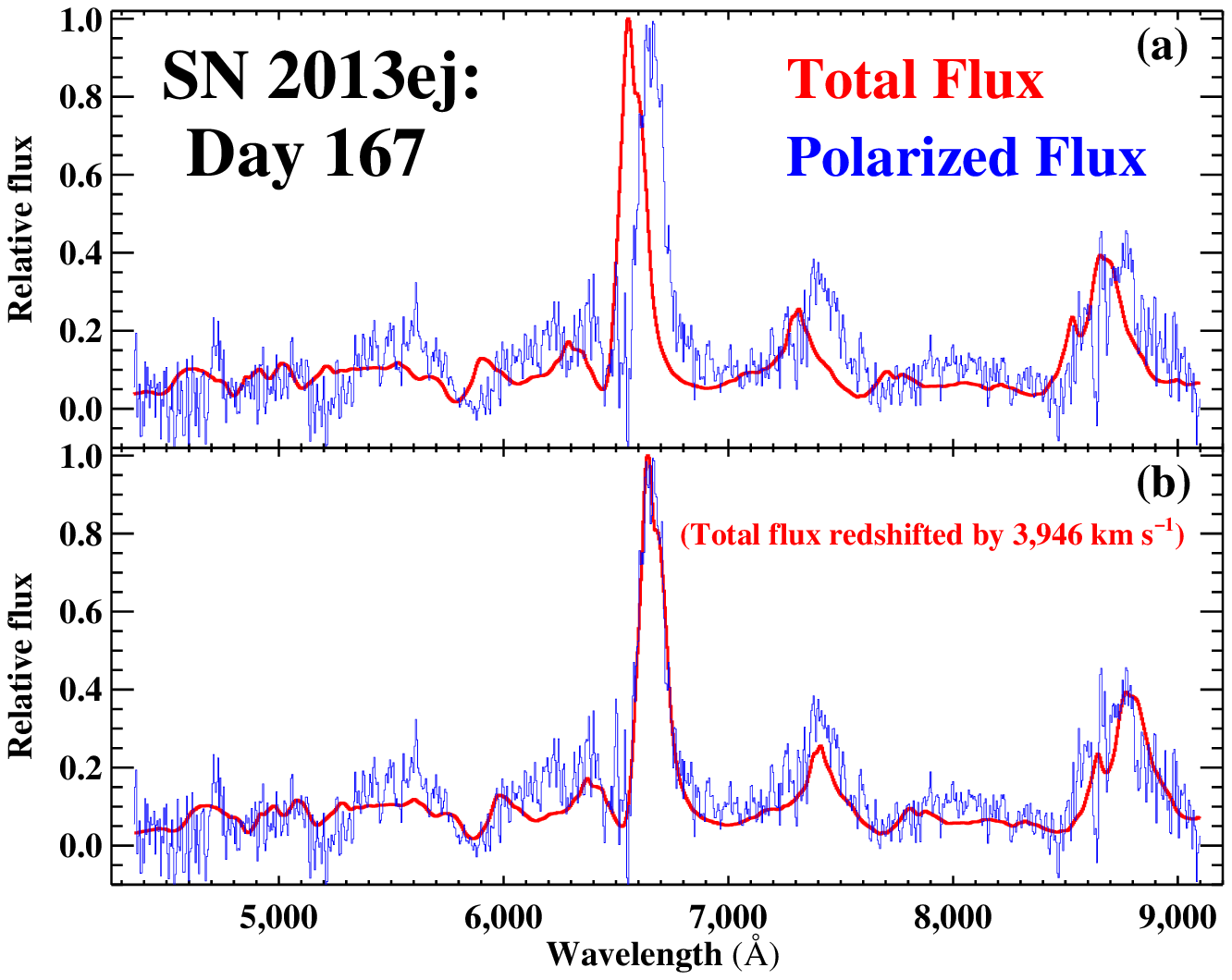}}}
\caption{Polarized flux (blue, noisier data) and total flux (red, thick lines)
spectra of SN 2013ej, taken 167 days after explosion, from
\citealt{2021ApJ...921L..35L}.  (a) Observed data. (b) Same as in (a) but with
the total flux spectrum now artificially redshifted by 3946 km s$^{-1}$.  The
polarized flux is a remarkable replica of the total flux but redshifted by
$\sim 4000$ km/s in SN 2013ej at 167d after explosion, implicating a blob of
$^{56}$Ni synthesized by, and ejected from, the explosion at greater than $\sim 4500{\rm\ km\ s^{-1}}$.}
\label{fig1_1}
\end{figure}

A prime example of the power of spectropolarimetry -- and, in particular, of
data obtained during the faint, later-stage ``nebular phase'', comes from VLT
spectropolarimetric observations of the Type~II-P SN~2013ej taken more than 160 days
after explosion.  For this late epoch, the SN's polarized flux (i.e., its observed
polarization percentage multiplied by its total flux spectrum) served as a
direct probe of {\it reflected} light at nebular times.  Remarkably, as shown
in Figure~\ref{fig1_1}, the polarized flux of SN~2013ej was found to be a replica of its
total flux, but systematically redshifted by $\sim 4000\ {\rm km\ s^{-1}}$. Such
a striking correspondence had never been observed before, although data capable
of revealing it have heretofore been only rarely obtained. This redshift
demands recession, at high velocity, of the scattering source from the region
responsible for the total flux spectrum. By following the analysis of
\citealt{2021ApJ...921L..35L}, the observations implicate a blob of $^{56}{\rm
Ni}$, traveling away from the explosion site at greater than $\sim 4500{\rm\
km\ s}^{-1}$, a value that strains the maximum possible from standard
neutrino-driven CC SN simulations \citep{2021A&A...645A..66O}. This
demonstrates the particular power of nebular-phase spectropolarimetry, since
extracting such information is not possible from an examination of either the
light curve (due to the corrupting effect of wind interaction) or the spectra
(because of time-dependent ionization \citep{2017MNRAS.472.5004U} or the
emission bias towards denser material).  While such observations strain the
capability of even 8-10 meter class telescopes (i.e., the observations of
SN~2013ej required three hours on the VLT and one of the brightest SNe of the
decade), they would be routinely available to the GMT.

Despite these recent advancements, spectropolarimetric studies of SNe are still in their infancy, remaining mostly limited to single, relatively bright objects with little information on time evolution. These studies will become much more sophisticated with the increased sensitivity of GMT, which will enable the examination of SNe when they are fainter, either at early and late stages or when they are more distant. GMT can serve to change the study of spectropolarimetry from one of sparse data on a few events to one with a substantial statistical sample. In this way, spectropolarimetry of core-collapse supernovae can lead to deep insights into the progenitor evolution (e.g.\ differences between single star and binary star evolution and among the many varieties of supernovae) and the explosion mechanism. 

    \subsection{Superluminous Supernovae}
    
    Superluminous supernovae (SLSNe) are rare but fantastically bright ($M < -21.0$) supernovae that give us a new window into the evolution and explosion of massive stars. Many SLSNe are an order of magnitude brighter than SNe~Ia and nearly a hundred times brighter than typical core-collapse supernovae. These events produce as much energy in radiation alone as more normal supernovae do in kinetic energy, about $10^{51}$ ergs. Hydrogen-rich SLSNe~II clearly derive their large luminosity from the collision of the exploding ejecta with circumstellar material (CSM) under conditions where kinetic energy is efficiently converted to radiation (the diffusion time is comparable to the dynamical expansion time). Some discussions discriminate between SLSNe~II that show distinct narrow emission lines (hence called SLSNe~IIn) and others that show broad hydrogen lines. The nature of the underlying explosion is less clear since these objects are shrouded by the substantial CSM. Hydrogen-deficient SLSNe~I lack any obvious evidence for CSM interaction near maximum light (but may show evidence for later collision with CSM), and their explosion mechanism and the origin of their substantial luminosity are both unknown.

	Many mysteries still surround the nature of SLSNe, including the origin of the differences between varieties of hydrogen-rich events (SLSNe~II), and the fundamental cause of the great luminosity of the hydrogen-deficient events (SLSNe~I). As is true in other classes of supernovae discussed above, spectropolarimetry promises to reveal critical information. Most SLSNe are very distant and therefore relatively faint when discovered.  This has limited the number of polarimetric observations of SLSNe to only a few to date \citep[e.g.][]{2016ApJ...831...79I, 2017ApJ...837L..14L, 2018ApJ...853...57B, 2019MNRAS.482.4057M, 2020ApJ...894..154S, 2022A&A...666A..30P,2023A&A...674A..81P}. Once again, the GMT can substantially add to the sample of SLSNe and thus reveal the nature of these spectacular events.
    
    As an example, a currently popular scenario to account for the luminosity of SLSNe~I is that a magnetar acts as a central engine depositing energy into the ejecta over a significant timescale \citep{kasen2010a, woosley2010a,dessart2012a}. If the magnetar central engine is responsible, the explosion should appear elongated resulting in a polarimetric signal. Alternative scenarios that may also produce a polarimetric signal include a pair-instability explosion, accretion onto a central black hole \citep{dexter2013} and CSM interaction \citep{chatzopoulos2012a}.
    
    \citet{2015ApJ...815L..10L} observed the SLSNe~Ic LSQ14mo during five epochs starting before maximum light and covering 26 days using V-band imaging polarimetry with the VLT/FORS2.  Assuming a favorable viewing angle, they found no evidence of asymmetries in the photosphere. \citet{2016ApJ...831...79I} provide the first spectropolarimetric observations of a hydrogen-free superluminous supernova, SN~2015bn.  Observations were obtained at two epochs (-23.7 \& 27.5 days) and revealed departures from spherical symmetry along a single persistent dominant axis.
    
    Advancing our knowledge of the central engine for SLSNe will require a large sample to allow us to study the predominance of asphericity and to statistically account for viewing angle.  Given the apparent magnitude of these SLSNe, this can only be done on very large aperture telescopes like the GMT.

\section{Other Transient Sources}

\subsection{Gamma-ray Bursts}
Gamma-ray bursts (GRBs), which occur at cosmological distances, are one of the most energetic explosions in the universe. GRBs are broadly divided into two groups based on how long the gamma-ray duration lasts, short ($<2$ s) and long ($>2$ s). Long gamma-ray bursts (LGRBs) are well-established to be produced by the explosion of very massive stars whereas short-duration GRBs are thought to be the result of the merger of binary neutron stars or a neutron star and a black hole. In both cases, the explosion launches a relativistic jet which produces short-lasting high-energy prompt emission.  The interaction of this jet with the ambient medium produces an afterglow which is detected in various wavelengths ranging from gamma rays to radio \citep{Piran_1999, Zhang_2004}. One of the major open questions in LGRB science is the driving mechanism of these explosions. Magnetic field properties of the jets can shed light on this open question \citep{Lazzati_2006, Toma_2013, Covino_2016, Kobayashi_2019}. The emission mechanism for the GRB jets is synchrotron radiation which is expected to produce a high level of linear polarization. In most cases, the expected optical circular polarization is negligible, though, in at least one case, significant optical circular polarisation was unexpectedly detected (\citealt{Wiersema_2014}). Linear polarimetry at an early time can provide important clues to improve our understanding of the magnetic field strength and the structure of the jets. At the same time, a polarization signal is also expected at a later time near the jet break (a break in the lightcurve decay, associated with the combination of the jet Lorentz factor and its opening angle) which helps researchers to gain more knowledge about the jet opening angle. \cite{Rossi_2004}, showed that polarization can play a crucial role in differentiating among various afterglow models for the magnetic field order within and normal to the shock, the internal jet energy structure, the viewing angle towards the jet, and the jet opening angle.

In the past, due to the lack of robotic telescopes, most of the optical linear polarimetry observations of the GRBs focused on late-time observations to study jet breaks. These observations resulted in the detection of polarization at the few percent level ($0.5 - 3 \%$) e.g., GRB 990510 \citep{Covino_1999, Wijers_1999}, GRB 010222 \citep{Björnsson_2002}, GRB 011211 \citep{Covino_2002}, GRB 071010A \citep{Covino_2008}, GRB 091018 \citep{Wiersema_2012}, GRB 121024A \citep{Wiersema_2014}. For most cases, along with polarization detection, a rotation in the position angle is also seen. Since the optical afterglows of GRBs are faint and the level of polarization low, a polarimeter on GMT can play a crucial role in observing this phase of GRB afterglows. This would significantly increase the number of polarimetric observations thereby revealing the details of the jet structure for many systems.

The relatively recent increase in the availability of robotic telescopes and instruments designed to observe the polarization of transients has made early (seconds to minutes after the GRB) polarimetric observations of GRB afterglows possible. There is an expectation of a high level of polarization of the early afterglow when it is dominated by reverse shock emission from the interaction of the jet and the ambient medium \citep{2003ApJ...594L..83G}. This reverse shock emission contains the imprints of the magnetic field structure of the jet and, if the magnetic field is well ordered, then a high level of polarization is expected. This has been seen for GRB~120308A with a measured peak polarization value of $28 \%$ \citep{Mundell_2013}. A high level of polarization closer to $10\%$ was observed for GRB~090102 at early times \citep{Steele_2009}. Interestingly, \citet{Shrestha_2021} observed a polarization value of $14.6 \%$ a few hours after the burst which is unusual as the expectation is that the forward shock dominates at this time. They found that later injection of ejecta could have created a reverse shock emission causing an increase in polarization. There are a limited number of polarization detections for GRBs as shown by \citet{Shrestha_2022}. Robotic telescopes, even those able to respond to GRB alerts in under a minute \citep[e.g.][]{2020SPIE11445E..3DJ}, are still limited to the brighter sources; thus a polarimeter on GMT will help to increase this number which will enable a detailed population study.  

\subsection{Gravitational Wave Sources / Kilonovae}

It has long been thought that short-duration GRBs are the result of the merger of binary neutron stars. A smoking gun for that scenario came with the discovery of GW170817/GRB~170817A, detected as a binary neutron star merger at a distance of just $\sim$40 Mpc by the LIGO and Virgo gravitational wave observatories.   \citet{2017NatAs...1..791C} used VLT/FORS2 to obtain imaging polarimetry of GW170817 during five epochs spanning 1.5 to ten days after the merger.  They found that the optical transient had a degree of linear polarization and a position angle that were consistent with field stars implying that the measurement was the result of Galactic interstellar polarization (ISP). Therefore the kilonova emission (AT2017gfo) is intrinsically unpolarized down to a significantly low level in the optical ($\lesssim0.5\%$; \citealt{2017NatAs...1..791C}). This has also been seen for various simulations of kilonova emissions from binary neutron star mergers such as performed by \citet{Bulla_2019} and \citet{Shrestha_2023}. However, some simulations of binary neutron star mergers predict a detectable polarization signal at near-infrared wavelengths \citep{Bulla_2019, Shrestha_2023}. Although an off-axis afterglow is predicted to be polarized \citep{2002ApJ...570L..61G}, there was no evidence for the detection of polarized optical afterglow emission from this transient. At radio wavelengths, an afterglow for GW 170817 was detected, and a strict limit on the radio polarisation could be derived, strongly constraining the order and strength of magnetic fields parallel and normal to the shock (\citealt{Corsi}) - again demonstrating the power of polarimetry to establish key parameters that can not be measured in other ways. 

\citet{2017NatAs...1..791C} argue that the lack of late-time optical polarization of the kilonova component is consistent with the properties of the lanthanide-rich material, in which the electron scattering opacity (which is responsible for the polarization in SNe) is lower than the total opacity by three orders of magnitude.  At early times, when the temperature is higher and the emission spectrum is blue, the electron scattering optical depth is close to one, and the polarization should be a function of the degree of asymmetry of the photosphere.
Since this is just the beginning of gravitational-wave 
multi-messenger astronomy (SN~1987A ushered in particle multi-messenger astronomy), the future promises to have many very interesting events.  Because the sensitivity of GW detectors has improved and will continue to improve, far more distant sources will be found, for which the associated kilonovae will be $>21$ mag \citep{Mochkovitch} and thus require 25-m class telescopes for detailed follow-up. Future polarization measurements will help clarify the picture of the progenitor system, particularly for neutron star/black hole mergers for which the ejecta are expected to be more anisotropic and thus have a higher degree of polarization \citep{Bulla_2021, Li_2019}.


\subsection{Tidal Disruption Events}

A tidal disruption event (TDE) occurs when a star wanders too close to the central supermassive black hole (SMBH) in a galaxy and is ripped apart by the SMBH's tidal forces. This interaction produces a flare that can span X-ray to radio wavelengths, as stellar material forms an accretion disk around the SMBH. After being predicted nearly half a century ago \citep[e.g.,][]{1975Natur.254..295H, 1976MNRAS.176..633F}, tens of these events have now been observed \citep[e.g.,][]{2021ARA&A..59...21G}, with 100s or 1000s expected in the LSST era.

A TDE produces more light than most supernovae, outshines the host galaxy for a period of months to years, and encodes information on black hole mass and spin. TDEs are thus key to discriminating among models for massive black hole formation and growth as well as to testing relativistic effects in regimes of strong gravity.
TDEs light up otherwise dormant galactic nuclei, 
addressing certain biases in black hole demographics derived solely from AGN. TDEs may also reveal intermediate-mass BHs (IMBHs), binary BHs, and recoiling BHs.


TDEs are intrinsically rare, occurring roughly every $10^4$ to $10^5$ years in a given galaxy. They typically peak at $M_g \sim -19$ and 
emit optical/UV light over months \citep{2023ApJ...955L...6Y}. This  timescale makes TDEs ideal LSST targets. LSST will likely discover
more than 1000 TDEs brighter than $m = 24$ per year, well within the reach of GMT's
capabilities. 

GMT optical/NIR spectroscopic follow-up of TDEs will enable the following new science: (1) an accurate census of dormant SMBHs in the local Universe, with the potential to identify correlations between SMBH properties and host-galaxy mass, bulge, type, and more; (2) demographic constraints on dormant SMBHs out to $z \sim 0.7$, a currently inaccessible redshift range; and (3) identification of IMBHs in dwarf galaxies.  Through GMT imaging- and spectro-polarimetry, we will gain (4) insights into the geometry and evolution of the emitting accretion disk, tidal streams, and reprocessing environment.

With a polarimetric capability, the GMT would be a powerful resource for understanding these intrinsically faint and asymmetric sources. 
To date, only a handful of TDEs have been studied using polarimetry. Below we summarize those findings, which indicate why future observations with large-aperture telescopes are needed to differentiate among physical models.

Previous work shows that TDE  polarization measurements provide clues about the system geometry. 
Optical imaging polarimetry of the TDE AT~2020mot reveals a peak linear polarization of 25 ± 4\% \citep{2023Sci...380..656L}, consistent with highly polarized synchrotron radiation from a relativistic jet. Yet the radio emission measured at later times disfavors this scenario. \citet{2023Sci...380..656L} therefore suggest that the polarized flux arises from shocks 
formed when the stellar material collides with itself during accretion disk formation.

Modeling shows that multi-epoch polarimetric observations can constrain the viewing angle of TDEs \citep{2023A&A...670A.150C}. 
Optical linear and circular polarization measurements of TDE AT 2022cmc, taken with the Very Large Telescope/FORS2
within a $\sim$ week of discovery, are consistent with zero \citep{2023ApJ...943L..18C}.
These nondetections
suggest that the thermal component of this TDE---the disk plus outflows---is being 
viewed pole-on, assuming an axially symmetric geometry. 

The evolution of TDE polarization in time varies among sources. For example, the polarization strength of TDE AT~2019dsg is 
9.2\% ± 2.7\% near
lightcurve peak and 
decreases to less than 2.7\% one month later \citep{2020ApJ...892L...1L}. While the high polarization at early times is consistent with that of TDE Swift~J164449.3+573451 \citep{2012MNRAS.421.1942W} and TDE Swift~J2058+0516 \citep{2020MNRAS.491.1771W}, the low polarization at later times resembles TDE OGLE16aaa \citep{2019MNRAS.482.5023H}. \citet{2020ApJ...892L...1L} argue that AT~2019dsg's polarization evolution arises 
from relativistic jet emission or a non-isotropic accreting disk, instead of from dust effects.

Spectropolarimetry provides a complementary probe of the origin of the polarization and the evolving geometry of the system. For example, a VLT program focusing on 
spectropolarimetry of optical (thermal) TDEs \citep{2022NatAs...6.1193L, leloudas_2022_7099738} finds that polarization strength can peak in the wings of the broad emission lines, while the line cores are depolarized.
These authors conclude that electron scattering, not synchrotron emission or dust polarization, is responsible.
Polarization decreases quickly with time, suggesting that the system becomes more axisymmetric and that the disk forms rapidly. Spectropolarimetry of TDEs combined with adequate modeling can thus shed light on the nature of TDEs.

\subsection{Novae}

Novae are interacting close binary systems with an accreting white dwarf primary and a mass-losing companion. The companion is generally a main sequence star, although in some cases the secondary is an evolved star. Nova outbursts are a result of a thermonuclear runaway reaction in the accreted layer on the WD surface accompanied by ejection of the accreted matter with velocities $>300$~km s$^{-1}$ \citep[e.g.][]{2012BASI...40..161A}. Based on the rate of decline from the peak magnitude, novae are classified into several speed classes ranging from the very fast ($t_2 < 10$ days) to the very slow ($t_2 >100$ days; with $t_2$ being the time taken to decline by 2 magnitudes from maximum). Fast novae are believed to occur on massive white dwarfs that are either CO WDs or ONe WDs. The WD could be magnetic in nature, and depending on the magnetic strength the accretion disc may be fully disrupted (polars), or partially disrupted (intermediate polars). Theoretical models \citep[e.g.][and references therein]{2016PASP..128e1001S} indicate nova outbursts occur for accretion rates of $\dot{\rm{M}}\le 10^{-9}\, \rm{M}_\odot\,\rm{yr}^{-1}$ for classical novae (events with one recorded nova outburst) to $\sim 10^{-7}\, \rm{M}_\odot\,\rm{yr}^{-1}$ for recurrent novae (events with more than one recorded nova outburst). 

The temporal evolution of a nova outburst spectrum is a reflection of the physical conditions of the ejecta such as expansion velocity, temperature, density, optical depth, clumpiness, etc. The Tololo scheme by \citet{1991ApJ...376..721W} spectroscopically classified novae into the FeII or He/N class based on the strongest non-Balmer features present in the spectrum around maximum. Recent observations over the past decade have shown there is no clear distinction between the types with cases of several systems transitioning from the FeII to He/N, He/N-FeII or He/N-FeII-He/N. Most fast novae are dominated by the He/N type of spectrum, although the very early phases of a few of these events (e.g. U Sco) have shown the presence of Fe II lines, albeit for a very short duration. The FeII types are generally slow novae, with several events transitioning to the He/N type at the later phases. This clearly indicates the FeII and He/N lines originate in different regions. Although not clearly established, it is quite likely that the He/N spectrum is from the nova ejecta and the FeII spectrum is from a circumbinary material ionised by the nova outburst \citep{2013A&A...559L...7S}. 

Direct imaging of nova remnants several years after outburst and ejecta shapes inferred by modeling the line profiles indicate non-spherical structures such as prolate shells with rings and blobs, bipolar structures, etc. Optical interferometric observations \citep[e.g.][]{2012BASI...40..267C, 2014ASPC..490..243C} and radio VLBI observations \citep[e.g.][]{2006Natur.442..279O, 2007ApJ...665L..63B, 2022A&A...666L...6M} indicate the structures are formed fairly early on during an outburst. 

Several novae are seen to form dust in the ejecta during decline (typically below 3 magnitudes from maximum). An optically thick dust formation is accompanied by a sharp drop in the optical magnitudes with a corresponding increase in the infrared magnitudes, while the formation of optically thin dust does not cause any significant dip in the optical magnitudes \citep[e.g.][]{1998PASP..110....3G}. 

Gamma-ray and radio emission studies of novae have shown the presence of internal shocks, most likely due to multiple ejections at varying velocities, although in some cases, like the recurrent novae with giant secondaries they are due to shock interaction of the nova ejecta with the circumstellar material \citep[see][and references therein]{2014Sci...345..554A, 2021ApJS..257...49C}.

Linear polarization is seen in novae during outburst due to several factors or combinations of factors such as the ejection mechanism, geometry, clumpiness, scattering by dust grains, electron scattering, and polarization in resonance lines. Temporal polarimetric studies of several novae have shown variability in polarization levels as well as position angles, with some objects evolving from just interstellar polarization to developing intrinsic polarization as the outburst progresses. Examples include classical novae such as V339 Del \citep{2017Ap.....60...19S, 2019ApJ...872..120K}, V705 Cas, V4362 Sgr, V2313 Oph and BY Cir \citep{2002A&A...384..504E} and V5668 Sgr \citep{2018A&A...611A...3H}; and recurrent novae during outburst: RS Oph \citep{1990MNRAS.243..144C}, U Sco \citep{2000A&A...355..256I, 2013A&A...559A.121A}, V745 Sco \citep{2006AJ....132..433K} and T Pyx \citep{2019A&A...622A.126P}.

It is observed that slow novae have the largest observed change in polarization measurements due to the shaping of the nova remnant. While modeling the emission line profiles provides details of the geometry of the shell, polarization can provide the tilt of the shell in the plane of the sky \citep[e.g. V5668 Sgr][]{2018A&A...611A...3H}. Spectropolarimetric observations provide information about the continuum and line polarizations. High-velocity components in the emission lines have been detected in some novae. These components are, in some cases, interpreted as being due to the formation of a jet-like structure. Such a structure could be due to wind ejection, or increased accretion during the re-formation of the accretion disc \citep{2006AJ....132..433K, 2013A&A...559A.121A}. 

Polarization levels seen in novae during outburst are generally low, and therefore polarimetric observations are by far quite few, and also limited to the brighter objects and/or the brighter phase of the nova outburst. The future large telescopes, such as the GMT, will enable observations of the fainter systems, as well as during the transition and nebular phases, when the shaping of the shell is clear. Re-formation of the accretion process is also likely to begin during these phases. Polarimetric studies during these phases would be very useful to understand the nova explosion geometry as well as the accretion process. Moderate to high resolution spectropolarimetric observations hold the key to distinguish the components giving rise to the FeII and He/N spectra, and detection of asymmetry in the ejected material.

Though several novae have shown linear polarization during outburst, detection of intrinsic polarization at quiescence is not common. For example, although RS Oph showed intrinsic linear polarization during outburst, no intrinsic polarization was detected at quiescence \citep{2019AcA....69..361N}. Likewise, spectropolarimetric observations of the recurrent nova T CrB at quiescence indicated no intrinsic polarization \citep{2022NewA...9701859N}.  Detection of polarization at quiescence has, by and large, been confined to systems with a magnetic white dwarf. A strong magnetic field prevents the formation of an accretion disc, leading to accretion along the magnetic poles of the WD (polars). In some magnetic WDs, the magnetic field is strong enough to just disrupt and truncate the inner regions of the accretion disc, leading to accretion via accretion curtains onto the magnetic poles regions of the WD (intermediate polars). In both scenarios, the gas is shocked to high temperatures producing circularly polarized cyclotron radiation that is the source of the observed continuum polarisation in these systems. Time resolved polarimetric studies of polars and intermediate polars indicate spin modulated circular polarization \citep[e.g.][]{1994A&A...283..163P, 2017PASA...34...17A, 2018MNRAS.473.4692P}. Such studies can be used to study the spin as well as the magnetic field strengths of the WD. 

\section{Instrument Requirements}

Transient science can benefit from either imaging polarimetry or spectropolarimetry.  In general, linear polarimetry is best but circular polarimetry can be useful for some specific science cases, particularly those concerning large-scale ordered magnetic fields.  For imaging polarimetry, it's best to observe in standard bandpass filters from the optical to the near-infrared.  The imaging field-of-view (FOV) can, in principle, be relatively small, $\sim$ 30 arcsec.  However, it can be very beneficial to simultaneously observe local standards which would require a larger FOV of order a few arcminutes.  The larger FOV would also require calibration for instrumental polarization as a function of position on the detector (see e.g. \citealt{Gonzalez} for the case of FORS2 on VLT). Spectropolarimetry requires a long-slit of one to a few arcminutes in length for good sky subtraction.  The spectral coverage that includes the most well-studied features ranges from 3800\AA--1.0$\mu$m with a spectral resolution of $\textrm{R} \sim 500 - 1000$.  Higher spectral resolutions of $\textrm{R} \sim 3500$ are useful in resolving CSM lines in interacting SNe and in reducing the effects of sky emission line residuals.  The polarimetric sensitivity should be $P \leq 0.1\%$ or better per resolution element in a one-hour exposure on a $V < 20.0$ point source.  For the most part, a MOS mode is not required. \\ \\
\noindent \textbf{Science Program:} \newline
Transients

\vspace{2ex}
\noindent \textbf{Linear and/or Circular Polarimetry:} \newline
Linear with some specific cases requiring circular.

\vspace{2ex}
\noindent \textbf{Wavelength Coverage:} \newline
From the Balmer break at approximately 3500~\AA~to beyond the Ca II triplet or approximately 9000~\AA

\vspace{2ex}
\noindent \textbf{Field-of-View:} \newline
Most transient sources are point sources.  For imaging polarimetry it would be helpful to see several field stars, perhaps no smaller than an arcminute.

\vspace{2ex}
\noindent \textbf{Observation Type (e.g.\ imaging polarimetry, spectropolarimetry, high-contrast):} \newline
Moderate-resolution spectropolarimetry is best. Imaging polarimetry is acceptable. 

\vspace{2ex}
\noindent \textbf{Target Brightness (magnitudes):} \newline
Signal-to-noise of 10 down to 19th magnitude.

\vspace{2ex}
\noindent \textbf{Spectral Resolution:} \newline
$R \approx 500 - 1000$

\vspace{2ex}
\noindent \textbf{Spatial Resolution:} \newline
Seeing limited 

\vspace{2ex}
\noindent \textbf{Required S/N or S/N per Wavelength Bin:} \newline
$\mathrm{S/N} \approx 10$ at 19th magnitude

\vspace{2ex}
\noindent \textbf{Level of Polarization per Spatial and/or Spectral Element:} \newline
$P \approx 0.1\%$

\vspace{2ex}
\noindent \textbf{Acceptable Instrumental Polarization and Absolute Error (IP(\%)  $\pm$ error(\%)):} \newline
$P < 0.05\%$

\vspace{2ex}
\noindent \textbf{Acceptable Linear to Circular Crosstalk:} \newline
Unknown

\vspace{2ex}
\noindent \textbf{Additional Requirements:} \newline
None

\chapter{Stars and their environments}
\label{sec:stars}
    
%
%

\section{Stellar Wind Accretion in Symbiotic Stars}

\begin{figure}[ht!]
\includegraphics[width=\textwidth]{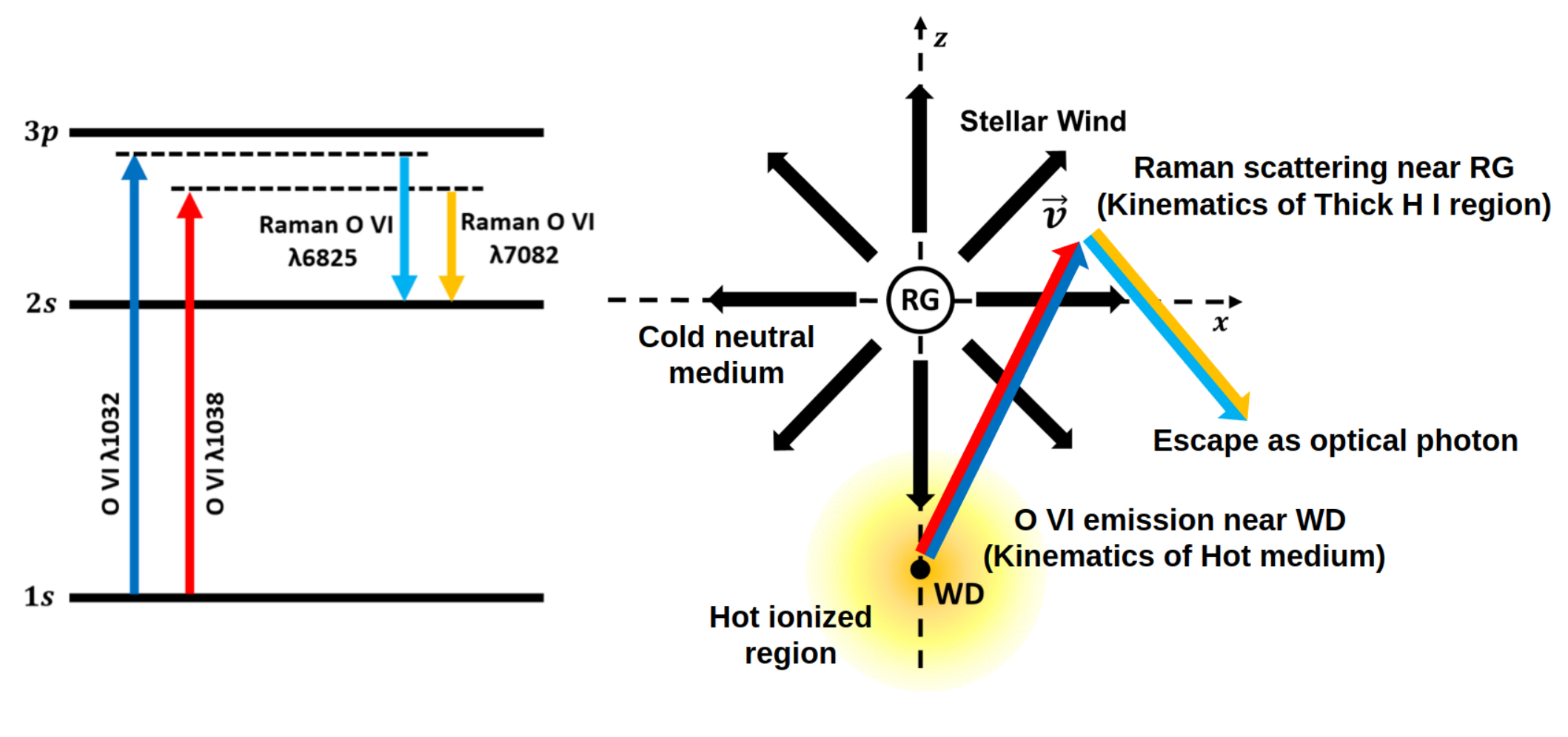}
\caption{Schematic illustrations of Raman scattering with atomic hydrogen of \ion{O}{6} $\lambda\lambda$ 1032, 1038 (left) and Raman scattered \ion{O}{6} in symbiotic stars (right).
The red, blue, light blue, and orange colors represent \ion{O}{6} $\lambda\lambda$ 1032, 1038  and Raman \ion{O}{6} $\lambda\lambda$ 6825, 7082, respectively.
\ion{O}{6} emission is generated in the hot ionized medium near a white dwarf (WD, filled circle) and scattered by atomic hydrogen in the cold neutral medium near a red giant (RG, open circle).
Raman \ion{O}{6} photons include the physical properties of the hot medium intrinsically and the kinematics of the cold medium via scattering.
}
\label{fig:raman_schematic_illustration}
\end{figure}

\begin{figure}[ht!]
\includegraphics[width=\textwidth]{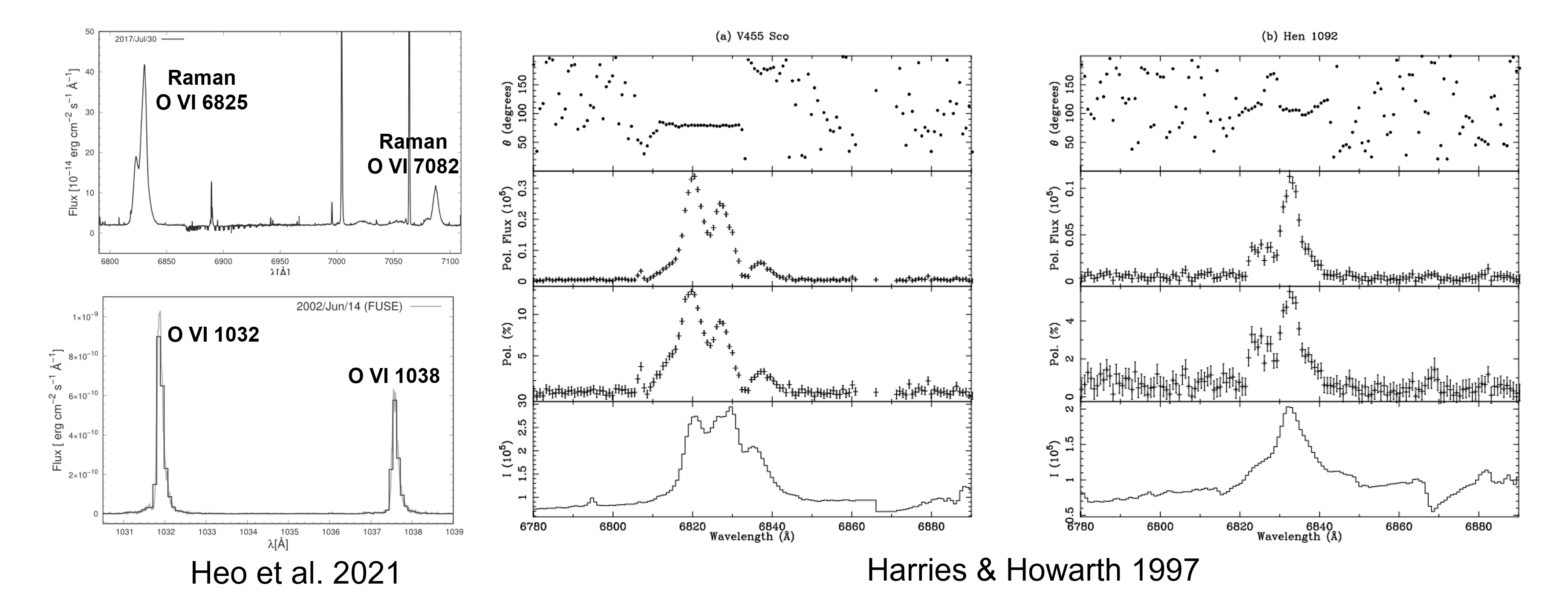}
\caption{
Spectroscopic and spectropolarimetric observation data of Raman \ion{O}{6} $\lambda\lambda$ 6825, 7072
in the literature.
On the left side, the top and bottom panels show the spectra of Raman \ion{O}{6} and \ion{O}{6} emission lines of RR~Tel, respectively, in \cite{raman_heo21}.
The center and right panels display the spectropolarimetric data of V455~Sco and Hen~1092 in \cite{raman_harries97}.
}
\label{fig:raman_observation}
\end{figure}

\begin{figure}[ht!]
\includegraphics[width=\textwidth]{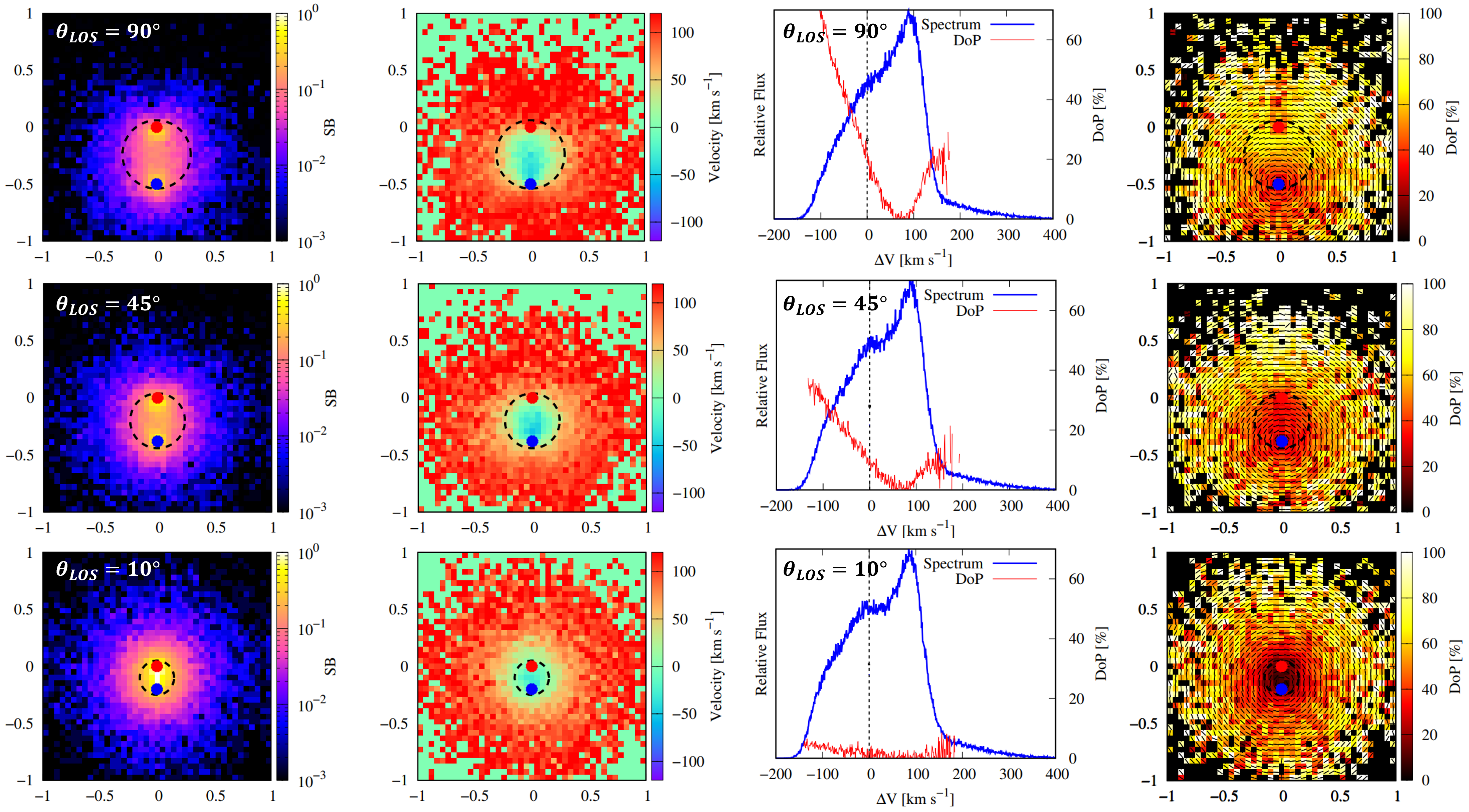}
\caption{3D Monte Carlo Simulation results of Raman \ion{O}{6} $\lambda$ 6825 using the code 'STaRS' \citep{raman_chang20} adopting the stellar wind model in \cite{raman_lee97}.
Here, a simple wind model is taken assuming a point \ion{O}{6} emission source is placed at the location of the white dwarf (WD, blue dot) surrounded by a radial wind from the red giant (RG, red dot) with the velocity proportional to $(1-R_*/r)^{-1}$ as shown in the schematic illustration of Figure~\ref{fig:raman_schematic_illustration}.
The columns from left to right are surface brightness, velocity map, the degree of polarization, and polarization map.
The red and blue dots in the maps represent the projected positions of a red giant and white dwarf, respectively.
The black arrows in the fourth column indicate the direction of the polarization vector.
Each row represents three lines of sight, $\theta_{LOS}$ = 90$^\circ$ (first), 45$^\circ$ (second), and 10$^\circ$ (third), where $\theta_{LOS}$ is the angle from the axis aligned the white dwarf and red giant.
}
\label{fig:raman_images}
\end{figure}

\begin{figure}[ht!]
\includegraphics[width=\textwidth]{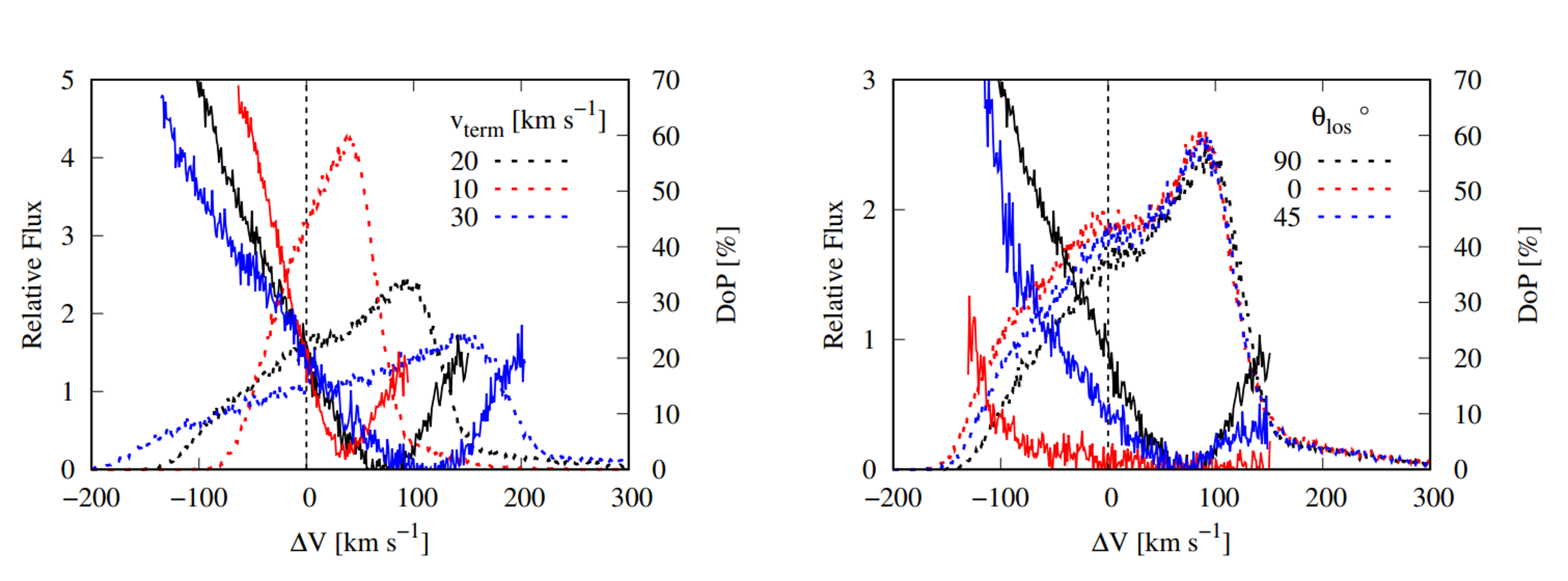}
\caption{The simulated spectra of Raman \ion{O}{6} $\lambda$ 6825 for three $\rm v_{term}$ (left) and $\theta_{LOS}$ (right). The solid and dashed black lines are identical to the spectrum and the degree of polarization, respectively, in the first row of Figure~\ref{fig:raman_images}
}
\label{fig:raman_spectra}
\end{figure}

Symbiotic stars, wide binary systems consisting of a white dwarf and a mass-losing giant, are unique objects that provide the opportunity to study the accretion processes taking place through gravitational capture of the slow stellar wind from the giant component \citep{raman_keynon09}. Symbiotic stars are classified into `S(tellar)'
type and `D(usty)' type, where `D' type symbiotic stars exhibit an IR excess indicative of a warm dust
component. The mass donors of `S' type symbiotic stars are usually normal giants, whereas 
`D' type symbiotics have a Mira or an OH/IR source as the mass donor. The orbital parameters are known for
`S' type symbiotic stars with an orbital period in the range of several years. The binary separations
of `D' type symbiotics are poorly known, probably exceeding several decades.

Many symbiotic activities, including various outbursts and X-ray emission, are attributed to the stellar wind
accretion. Since symbiotic stars are candidates for Type~Ia progenitors, measuring the accretion rate from the giant to the white dwarf is crucial in the prediction of when the binary system will explode. The mass loss rate of the giant component in symbiotic stars may be in the range of $10^{-6}-10^{-8}
{\rm\ M_\odot\ yr^{-1}}$. \cite{raman_lee22} presented their hydrodynamical study for a typical `S' type symbiotic star
and proposed that about 10\% of the giant stellar wind material is captured by the white dwarf component.

The spectrum of symbiotic stars includes the strong UV and optical emission lines from the hot component near the white dwarf, the stellar continuum of a red star, and the molecular absorption features from the cold component near the giant. Two interesting broad emission features at 6525 and 7082 \AA\ are identified as resulting from an inelastic scattering process with atomic hydrogen (i.e., Raman scattering) of \ion{O}{6} resonance doublets at 1032 and 1038 \AA, respectively \citep{raman_schmid89}.
The left panel of Figure~\ref{fig:raman_schematic_illustration} illustrates schematically the Raman scattering process of \ion{O}{6} photons and hydrogen atoms in the ground state, and the right panel shows how Raman scattered lines 
are associated with the physical properties of both hot and cold mediums in symbiotic stars. 
An \ion{O}{6} photon excites an electron from the ground state 1s of a hydrogen atom to a virtual state near 3p.
When the electron is de-excited to the 2s state, an optical photon redward of H$\alpha$ is created due to Raman scattering.
\ion{O}{6} $\lambda\lambda$ 1032, 1038 are important coolants of the plasma expected to be found around the hot component with $T > 10^5 \rm K$, while the information from the cold component with $T < 10^4 \rm K$ is imprinted on the Raman scattered photons.
The cross sections for these Raman processes are quite small at $\sim 10^{-23}{\rm\ cm^2}$, requiring a very thick neutral region illuminated by an intense \ion{O}{6} emission source. 
This condition is ideally met in symbiotic stars, where the cool neutral region is found near the base of the slow stellar wind from the giant, and copious \ion{O}{6} emission is expected near the accreting white dwarf.

Two important characteristics of Raman features are their broad profiles and strong linear polarization, both of which are 
attributed to the inelastic and electric dipole nature of the scattering. For an incident \ion{O}{6} photon with frequency $\nu_i$, the frequency $\nu_o$ of a 
Raman \ion{O}{6} photon is related by 
\begin{equation}
\nu_o = \nu_i - \nu_{\rm{Ly}\alpha}
\end{equation}
where $\nu_{\rm{Ly}\alpha}$ is the frequency of Ly$\alpha$. Therefore, a range $\Delta\nu_i$ of \ion{O}{6} incident frequencies corresponds to
the same frequency range $\Delta\nu_o$. However, the line width of the Raman line,
\begin{equation}
{\frac{\Delta\nu_o}{\nu_o}}= \left({\frac{\nu_i}{\nu_o}} \right){\frac{\Delta\nu_i}{\nu_i}},
\end{equation} 
becomes broader by a factor $(\lambda_o/\lambda_i)\sim 6$ times that of \ion{O}{6}. Because of the line broadening
by a factor of six, the resultant Raman line profiles are determined mainly by the relative kinematics of the \ion{O}{6} emission region with respect to the neutral region and are almost independent of the observer's line of sight.
Figure~\ref{fig:raman_observation} shows the spectroscopic and spectropolarimetric data of Raman \ion{O}{6} in \cite{raman_heo21, raman_harries97}.
In the top left panel, Raman \ion{O}{6} lines are clearly seen to be broader than other emission lines.

Raman \ion{O}{6} features consist of purely scattered radiation, with the mean scattering number being only a few, implying strong polarization retained in the emergent Raman radiation. 
\cite{raman_harries96,raman_harries97} reported a broad range of the degree of polarization of Raman \ion{O}{6} $\lambda$ 6825 from $\sim 0\%$ to $\sim$ 30\% through spectropolarimetry of 23 symbiotic binary systems.
The center and right panels of Figure~\ref{fig:raman_observation} show
Raman \ion{O}{6} $\lambda$ 6825 linearly polarized with a degree in the range  2-10 \%.
The Raman photons in the blue part are more strongly polarized than those in the red part, 
as the red and blue photons are formed in different parts of the neutral medium.
Furthermore, they noted various spectral shapes of Raman \ion{O}{6}, single-, double-, and triple-peak profiles.

A proper interpretation of the profiles and polarization of Raman \ion{O}{6} features requires 
refined radiative transfer studies.
\cite{raman_lee97,raman_harries97} adopted a Monte Carlo technique to investigate the polarization and spectroscopic behaviors of Raman \ion{O}{6} using a simplified stellar wind model. 
Figure~\ref{fig:raman_images} shows the simulation results of a 3D Monte Carlo radiative transfer study using the code 'STaRS' \citep{raman_chang20} and taking the simple stellar wind model in \cite{raman_lee97}. 
The sightline angle $\theta_{\rm LOS}$ is defined as the angle between the line of sight and the axis connecting the white dwarf and red giant.
In the first column, Raman \ion{O}{6} is bright near the WD and the RG because the \ion{O}{6} emission is formed mainly near the WD, and the optically thick neutral medium is present near the RG.
The blue Raman photons are formed dominantly in the region between the WD and the RG, where the stellar wind appears as an inflow in the rest frame of the WD. This is illustrated in the second column of Figure~\ref{fig:raman_images}.
Therefore, the polarization of blue Raman photons tends to depend strongly on the line of sight.
In the third column, the spectral profile does not depend on $\theta_{\rm LOS}$, whereas the degree of polarization in the blue part shows a strong dependence.
For $\theta_{\rm LOS} = 90^\circ$ corresponding to the axis connecting the WD and RG components perpendicular to the line of sight, the blue part of Raman \ion{O}{6} is maximally polarized due to
the dipole nature of Raman scattering.
The fourth column of Figure~\ref{fig:raman_images} shows the weakening of polarization as the Raman scattering process becomes closer to backward or forward scattering processes when $\theta_{\rm LOS}$ decreases.

In a theoretical model consisting of an accretion disk in the binary orbit plane and 
bipolar outflows, the blue and central components are expected to be polarized in the direction perpendicular to the orbital plane, whereas the red part will be polarized in the parallel
direction. Raman \ion{O}{6} 6825 and 7082 are strongly polarized with polarization flip in the red part, shown in the previous studies by \cite{raman_harries96}.
\cite{raman_heo16} considered the \ion{O}{6} emission from the collimated outflows to analyze Raman \ion{O}{6} profiles in D-type symbiotic stars, V1016-Cyg.

The position angle flip observed in the red part of Raman \ion{O}{6} is also explained by an alternative theoretical model where the outflowing \ion{O}{6} emission region is replaced by an additional neutral clumpy region. In this case, \ion{O}{6} line photons originate in the accretion flow around the white dwarf and enter the neutral region in front of the giant or the neutral components moving out of the binary system along the polar directions.
\cite{raman_lee19} suggested the existence of the neutral outflow through Raman \ion{0}{6} profiles in S-type symbiotic stars, AG~Dra.
In addition, \cite{raman_lee07,raman_heo21} proposed that Raman \ion{O}{6} profiles are mainly attributed to the kinematics of the accretion disk.
In this case, when the velocity width of intrinsic \ion{O}{6} emission is broader than the stellar wind velocity, the overall polarization of Raman \ion{O}{6} will be very weak.

 A recent study by \cite{raman_skopal23} proposes the formation of a neutral region in the orbital plane
during outbursts in symbiotic stars. According to this model, Raman \ion{O}{6} features are also
formed in the orbital plane as well as in the neutral region surrounding the giant component.
The neutral region in the orbital plane is present only during outbursts; the polarization
structures of Raman \ion{O}{6} will be very different from those observed during the quiescent phase.

One notable point is that the position angle reflects the binary orbital phase. Because many symbiotic stars are a wide binary system, it is quite difficult to determine their binary 
orbital phase. Spectropolarimetric monitoring observations may yield important information regarding the binary orbital phase \citep[e.g.][]{raman_schild96}.
\begin{hibox1}
Spectropolarimetry of Raman \ion{O}{6} features will provide precious information regarding the mass transfer process, the bipolar structure associated with the collimated outflows, and the binary orbital phase. Eventually, carrying out a spectropolarimetric monitoring program will allow us to obtain a good deal of information regarding the variation in the mass transfer process and the binary orbital phase.
\end{hibox1}

\section{Instrument Requirements}
\noindent \textbf{Science Program:} \newline
The investigation of stellar wind accretion in symbiotic stars using spectropolarimetry of Raman \ion{O}{6} features.

\vspace{2ex}
\noindent \textbf{Linear and/or Circular Polarimetry:} \newline
Linear.  Raman-scattered \ion{O}{6} features are known to show strong linear polarization, which is closely
related to the stellar wind accretion processes in these systems. In particular, the position
angles will indicate the relative orientation of the accretion flows and stellar winds
with respect to the binary axis connecting the white dwarf and the giant component.

\vspace{2ex}
\noindent \textbf{Wavelength Coverage:} \newline
3700 \AA\ to 7200 \AA.  Raman \ion{O}{6} 6825 and 7088 are spectral features of primary interest, however, Raman \ion{He}{2} features formed blueward of hydrogen Balmer lines will also be interesting. Therefore, the desired
 wavelength range is the entire optical region from 3700 \AA\ to 7200 \AA.
 
\vspace{2ex}
\noindent \textbf{Field-of-View:} \newline
N/A

\vspace{2ex}
\noindent \textbf{Observation Type (e.g.\ imaging polarimetry, spectropolarimetry, high-contrast):} \newline
Spectropolarimetry

\vspace{2ex}
\noindent \textbf{Target Brightness (magnitudes):} \newline
 N/A

\vspace{2ex}
\noindent \textbf{Spectral Resolution:} \newline
$\lambda/\Delta\lambda = 10,000 - 30,000$.  In order to investigate the change of the position angle that is expected in the red parts of Raman \ion{O}{6} features, a spectral resolution exceeding $\lambda/\Delta\lambda = 10,000$ is required, and additional information would be obtained when the spectral resolution is $\lambda/\Delta\lambda = 30,000$ or higher.  Previous spectropolarimetric work has been done with ESPADONS at CFHT and BOES at BOAO.

\vspace{2ex}
\noindent \textbf{Spatial Resolution:} \newline
Seeing limited 

\vspace{2ex}
\noindent \textbf{Required S/N or S/N per Wavelength Bin:} \newline
$\mathrm{S/N} \approx 5$ at 18th magnitude

\vspace{2ex}
\noindent \textbf{Level of Polarization per Spatial and/or Spectral Element:} \newline
$P \approx 0.1\%$

\vspace{2ex}
\noindent \textbf{Acceptable Instrumental Polarization and Absolute Error (IP(\%)  $\pm$ error(\%)):} \newline
$P < 0.05\%$

\vspace{2ex}
\noindent \textbf{Acceptable Linear to Circular Crosstalk:} \newline
N/A

\vspace{2ex}
\noindent \textbf{Additional Requirements:} \newline
N/A

\chapter{Interstellar and Circumstellar Media}
\label{sec:ism}
Interstellar polarization has seen a remarkable renaissance in the last few decades, primarily driven by the maturation and extensive testing of the radiative grain alignment paradigm: "Radiative Alignment Torque" (RAT) theory \citep{dolginov1976,draine1997,lazarian2007a,bga2015b}.  With this quantitative and predictive theoretical paradigm, polarization observations of the interstellar medium (ISM) can now be analyzed with confidence, allowing reliable measurements of the ISM magnetic fields, as well as a number of grain and environmental parameters. The recent selection by NASA of the \textit{"PUFFINS"} Pioneer-class mission \citep{anche2025} reinforces this.  Several \textit{ab initio} codes for modeling the polarization have been developed (primarily by the group of T. Hoang at KASI, Korea) and are being implemented in general user codes, such as POLARIS \citep{reissl2016}.  

Interstellar polarization was first discovered in 1949 \citep{hiltner1949a,hall1949} and was identified already by W. Albert Hiltner, in his second publication on the subject \citep{hiltner1949b}, as being due to elongated dust grains aligned with the Galactic magnetic field. While a detailed theory of grain alignment was developed soon after \citep[]["DG"]{davis1951a}, it turned out to be difficult to quantitatively verify and was eventually shown to not be able to explain the observed polarization \citep[e.g.][]{jones1984,hough2008}. Starting from a suggestion by \citet{dolginov1976}, a new theory of grain alignment, based on the transfer of angular momentum from the radiation field to the dust, was developed. This RAT paradigm has now been expanded and extensively tested.   As we argue herein, it provides the theoretical basis for detailed studies of several aspects of interstellar medium physics.  

ISM polarization has, to date, commonly been performed with medium-sized telescopes and broad- or single-band instruments, generally targeting bright stars (V$\lesssim$ 10-12 mag.). The advent of a detailed understanding of the grain alignment physics now allows and demands a more specific target selection (location, extinction, sampling selection) as well as the ability to achieve higher signal-to-noise (S/N) over the entire optical spectrum, preferably at medium (R$\sim$1000) resolution. ISM spectro-polarimetry is still a relatively poorly explored field, providing a significant discovery potential of new effects.

Traditionally, most ISM polarimetry has focussed on single-band measurements. However, even from the start\citep[e.g.][]{serkowski1973} the wavelength dependence has been known and utilized by astronomers \citep[e.g.][]{whittet1978,whittet1992, wilking1980,whittet2001,bga2007,siebenmorgen2018}, to probe the relationship between the dust distribution and characteristics along the line of sight and the observed polarization. Because of the complications introduced by line-of-sight variations in magnetic field geometry and dispersion (turbulence), single-band photo-polarimetry is prone to degeneracies between grain and environmental parameters.
Because different wavelengths - to first order - respond the same to the magnetic field variations, spectro-polarimetry provides an important probe of the ISM and dust physics.

The convex spectral dependence of the broadband extinction polarization, from the space-ultraviolet (UV) to the near-infrared (NIR), can, generally, be described by the so-called Serkowski function:
\begin{equation}
p(\lambda)=p_{\rm max}\cdot \exp \left[ -K \cdot \ln^2 \left(\frac{\lambda}{\lambda_{\rm max}}\right) \right]
\end{equation}
\noindent
where the maximum polarization, p$_{\rm max}$, occurs at wavelength $\lambda_{\rm max}$ - usually located in the 0.4-0.6 $\mu$m range and $K$ controls the width of the curve.  As discussed by e.g. \citet{codina1976, wilking1982,whittet1992} the $K$ parameter is likely correlated with $\lambda_{\rm max}$.

As has been shown by, e.g., \citet{kim1995, clayton2003} the Serkowski curve can be reproduced by scattering models based on regular grains size distributions \citep[e.g.][]{mathis1977} with modifications for the fractional alignment efficiency as a function of grain size. Full optical/NIR spectro-polarimetry can, therefore, probe the grain size distribution, the radiation field characteristics, and the dust parameters as functions of grain size and environment.   
As shown by \citet{whittet2001,bga2007}, the location of the peak polarization, $\lambda_{\rm max}$, is generally directly related to the extinction. This is consistent with the RAT condition that grains are aligned if exposed to radiation with wavelengths less than the grain diameter because the reddening of the light into a cloud makes the remaining aligning radiation successively redder. The location of $\lambda_{\rm max}$ also depends on the local value of the total-to-selective extinction, R$_V$ as a tracer of the average grain size, whether primarily \citep{whittet1978,siebenmorgen2018}, or secondarily \citep{bga2007}. It is, however, important to recognize, when discussing the future of high S/N ISM spectro-polarimetry, that the Serkowski form is a phenomenological parametrization derived from observations.

As is well known from far infrared (FIR) and mm-wave mapping \citep[e.g.][]{wright2010, dame2001}, the interstellar medium is extensive but can vary dramatically over small distances.   Modern MHD simulations \citep[e.g.][]{walch2015} indicate that the true scale of the ISM may be even smaller than current surveys can resolve. Also, as indicated by, e.g.,\ the results of \cite{siebenmorgen2018}, multiple clouds along the line of sight can significantly change the polarization spectrum.  

To utilize the powerful tool of spectro-polarimetry to probe these structures, we, therefore, need to be able to select our background stars with high-spatial accuracy in 3D or at higher extinction, implying fainter targets than is possible today.

\citet{vaillancourt2020} showed that the $\lambda_{\rm max}(A_V)$ relation bifurcates at around A$_V$=4-5 mag.  Using \textit{ab inito} RAT-based modeling, they showed that the "upper branch" in the relation corresponds to dense gas where grain growth has taken place. This technique directly probes grain growth, independent of uncertainties in the dust emissivity \citep{ysard2013}.  Because of the limited sensitivity of their observations (performed with the Lick/Kast instrument), only a handful of stars could be observed at $A_V>$5 mag., precluding any mapping of the effects.   With larger telescopes utilizing, e.g., the 2MASS catalog as target selectors, the power of full-spectrum optical/NIR polarimetry for cloud structure and evolution could be fully explored. Such observations could, in turn, provide valuable target pools and synergistic data to MIR dust spectroscopy with JWST.

For many interstellar phenomena, precise spatial localization and resolution are critical; for instance, theoretical studies \citep{hoang2020} indicate that the grain alignment in shock fronts may be dominated by mechanical ("MET") alignment \citep{hoang2018} rather than the radiative mechanism. As shown by, e.g., \citet{reach2005}, the fronts in supernova remnants have dimensions of several arc seconds (as traced by the fluorescent emission from shocked H$_2$).

A related issue is the 2175\AA\ extinction feature's origin and behavior. The carrier of this prominent but highly variable feature in the ISM extinction curve has not yet been conclusively identified, limiting our understanding of the extinction in the Milky Way and external galaxies. In the [very] limited sample of ISM UV polarization to date (a total of 30 lines of sight observed between the WUPPE mission and HST/FOS), two lines of sight show polarization in the feature \citep{martin1999}.   While the small sample makes it nigh impossible to form firm conclusions, \citet{blasberger2017} have argued for an association with PAH emission and reflection nebulae. The detection of red diffuse emission around ``star 46'' \citep{bga2013} in the IC 63 region \citep{lai2020} may support this claim. However, to connect this to the polarization - and thus dust - characteristics and variability will require the ability to collect high S/N data from precisely located - and often faint - stars.

Because the polarization spectrum relates to the grain size distribution and the spectral energy distribution of the aligning radiation field, complementary information is available (for extinction polarization) from the ultraviolet to the near infrared. As noted above, the [space] ultraviolet polarization has been only poorly explored to date.  This is about to change with the selection of the \textit{PUFFINS} mission by NASA (expected launch in 2030).  In addition, a number of polarimetry missions, primarily focused on stellar astrophysics are under development/consideration, e.g. the SMEX/MIDEX \textit{Polstar} mission proposal \citep{scowen2022} and the \textit{Pollux} instrument for the Habitable Worlds Observatory \citep{neiner2026}.  These advances in UV spectro-polarimetry will undoubtedly open up new needs and avenues for ground-based optical/NIR polarimetry.

Finally, the structure and evolution of circumstellar envelopes, e.g., of asymptotic giant branch stars, can be powerfully probed by spectroscopy \citep[e.g.][]{mauron2000} and spectro-polarimetry \citep{2024ApJ...963...76A}. Extinction polarization towards background stars can directly address grain alignment, grain size distribution, and other characteristics without the uncertainties of the line of sight variations in dust temperature and emissivity inherent in FIR/mm-wave data. As these shells are typically only of the order of a few arc minutes, faint background stars are needed, challenging for even medium-sized telescopes (or even 10m class ones).


As the above examples show, extinction polarization provides an important complement to FIR emission polarization mapping by being insensitive to dust temperature and emissivities and probing pencil beams, which can be directly compared to line spectroscopy of the gas. 

Taken together, and as explored below, these considerations argue for the need for full optical-NIR spectro-polarimetry capabilities on large telescopes.

\begin{hibox1}
Because polarization is a ``photon-hungry'' technique - especially for ISM applications with typical polarization values of $\lesssim$1\%, large telescopes and sensitive instruments are crucial to take full advantage of recent theoretical and observational progress.
\end{hibox1}

\section{RAT Alignment in a Nutshell}

As noted above, radiative grain alignment was first proposed by \citet{dolginov1976} and developed by a number of researchers led by Bruce Draine, Alex Lazarian, and their collaborators. Good surveys of the mechanism can be found in \citep[e.g.][]{lazarian2007a, hoang2009a,bga2015b}. To first order, the Radiative Alignment Torque (RAT) paradigm predicts that 
\begin{description}
\item{} a \textbf{paramagnetic} grain will be aligned if
\item{$\Rightarrow$} exposed to an anisotropic radiation field with
\item{$\Rightarrow$} wavelengths less than the grain diameter
\end{description}

For a grain with net helicity (an irregular grain will by necessity be helical to a degree), the right- and left-hand circular polarization of a radiation field will have different scattering efficiencies, and therefore the radiation transfers angular momentum to the grain. This effect has been confirmed in the laboratory by \citet{abbas2006}. If the grain is paramagnetic, the distribution of free spins in the solid will rearrange to minimize the total energy of the system by trading mechanical angular momentum for spin-flips, magnetizing the grain, which is known as the Barnett effect \citep{purcell1979} and is the inverse of the Einstein-de Haas effect \citep{einstein1915} known from solid state physics.  The interaction between the external magnetic field and the magnetized grain, under continued radiative torques, then aligns the grain's angular momentum vector with the magnetic field.   This is referred to as ``external alignment.''
Because a grain, given a torque in a random direction, will spin in that direction unless internal energy dissipation takes place, an effect referred to as ``internal alignment'' must also occur to ensure that the grain is presenting a fixed projection along the line of sight.   Under angular momentum conservation, it is easy to show that the lowest energy state for a rotating solid is when the angular momentum is aligned with the axis of maximum inertia (smallest physical axis). As \citet{purcell1979} showed, the most efficient process to achieve the required energy dissipation is the so-called Barnett relaxation, which occurs when the spin distribution in the grain cannot react fast enough to changes in the rotation orientation of the grain (for an asymmetric grain not rotating around a symmetry axis, the angular momentum will nutate around the symmetry axes \citep{purcell1979} causing a time-dependent magnetization. If the nutation is fast enough, the Barnett effect becomes dissipative). Nuclear relaxation \citep{lazarian1999} will also play a role.

The resulting alignment of the grains is a balance between the driving and damping of the alignment. The strength of the magnetic susceptibility of the grain material is a critical parameter in the time-scale for alignment \citep[e.g.][]{hoang2008}, as is the strength and color of the radiation field \citep[e.g.][]{medan2019}, the amount of H$_2$ formation on the grains, and several other factors.  

The inclusion of ``super-paramagnetic grains'' \citep[cf.][]{goodman1995b}, their sizes and fraction as a function of overall all grain size, will affect the detailed shape and height of the polarization spectrum \citep{mathis1986}.   Large and diamagnetic (carbonaceous) grains have small or no internal alignment, similarly affecting the polarization spectrum.

The dominant disalignment is gas-grain collisions \citep[e.g.][]{draine1998,vaillancourt2020,soam2021a}, allowing polarization, and especially spectro-polarimetry to constrain gas density and temperature on pencil beams.

As discussed by \citet{hoang2019} and \citet{tram2021}, large dust grains exposed to a strong radiation field can be spun up to rotation velocities where the centrifugal forces exceed the grain's tensile strength. Complementing the FIR emission polarimetry of this effect with extinction polarization data would allow more direct estimates of the grain sizes involved.
It is interesting to note that the grains seen in the envelope of IRC+10216 are found from NIR scattering observations \citep{groenewegen1997}, FIR emission \citep{maercker2022}, and optical spectro-polarimetry \citep{2024ApJ...963...76A} to be large. It's still unclear whether that is because these grains are compact and have high tensile strength or because they are not spun up efficiently enough (given their low susceptibility).

\section{Magnetic Fields in ISM and CSM Environments}

As shown by \citet{hiltner1949b}, dust-induced polarization from the interstellar medium traces the magnetic field. Dust-induced polarization is not directly sensitive to the magnetic field strength, but the variations in field direction, measured by polarization, can be used to estimate the field strength \citep[e.g.][]{davis1951a,chandrasekhar1953,houde2009}.   For this, general large-scale mapping is more important than very high spatial resolution.   Also, typically, single-band measurements are sufficient.   However, extra-dense sampling can be beneficial for specific regions with small-scale structures. Since these techniques rely on statistical samples, regions would benefit from denser sampling, which is only possible with large telescopes.

\section{Dust and ISM/CSM Environment Parameters}

The detailed grain alignment efficiency depends on the mineralogy and structure of the dust, as well as the radiation and other environmental parameters. Several specific questions of dust physics and more general ISM characteristics can be addressed by probing polarization in regions with specific characteristics. Here, we highlight a few examples.

Evolved stars provide unique dust and radiation field characteristics. Asymptotic giant branch (AGB) stars can have evolved oxygen-dominated ([O]/[C]$>$1) atmospheres with silicate and metal oxide dust, S-type stars ([O]/[C]$\approx$1) have mixed dust mineralogy \citep[including MgS dust,][]{hony2009}, and carbon-rich ([O]/[C]$<$1) envelopes are dominated by amorphous carbon and SiC dust \citep{ivezic1996a}.  These provide unique laboratories for studying mineralogically ``pure'' dust populations.

Other classes of evolved stars have extended circumstellar envelopes, e.g., R CrB stars \citep{bieging2006}. While the polarimetry of the central star of these extended objects is primarily sensitive to the dust geometry in the envelopes (via scattered light), true background stars provide the opportunity to explore the magnetic fields, grain characteristics, and alignment processes. While many such circumstellar envelopes (CSEs) span arc minutes in the sky, to probe more than individual background stars typically requires a limiting magnitude beyond V$\sim$15-16 mag.

Recent observations of the transition region from the stellar wind to the ISM - the astrosheeth - in IRC+10216 (Andersson et al., 2025, in prep.) revealed a significant \textit{enhancement} of the polarization efficiency ($p/A_V$) in this narrow region, even though the elevated density and temperatures expected in these shocked regions \citet{sahai2023} would nominally imply enhanced grain disalignments.  This enhanced alignment efficiency might represent the predicted, but not yet observationally confirmed, "Mechanical Torque" alignment process \citep[eg.][]{hoang2018}.  Such transition regions are observable around many evolved stars, but since they are even smaller than the circumstellar envelope, even fainter stars are needed to characterize them.  Shocks are also important sites of grain processing and physics, but also have very small extents in the sky, even for relatively low-velocity dissociative shocks \citep[e.g.][]{reach2005}.   

The polarization efficiency in the ISM is generally found to be limited to $p/A_V \lesssim 3\%\; \mathrm{mag}^{-1}$. However, as noted by \citet{panopoulou2019} for select regions in the Galaxy, the polarization efficiency can be significantly higher. As shown by \citet{medan2019}, the polarization efficiency in the Local Bubble wall is correlated with the integrated intensity from the surrounding O and B stars \citep[see also][]{gontcharov2019}. In their sample - selecting only lines of sight with good S/N - 10 stars show $p/A_V > 10\%\; \mathrm{mag}^{-1}$ (45 stars show $p/A_V > 5\%\; \mathrm{mag}^{-1}$). To explore these high polarization efficiencies, the stars and sight lines around the existing stars should be explored with sensitive spectro-polarimetry to probe the dust and environmental effects. To probe sight lines within an arc minute of these high-polarization efficiency stars will typically require observations of V~$\lesssim$~17~mag.

\section{Polarization of Spectral Features}

There are few broadband features in the optical spectrum associated with dust \citep[the Diffuse Interstellar Bands have not been conclusively identified and have not been found to show polarization,][]{adamson1995,cox2011,clayton2014}. However, extending the consideration to the near- and mid-infrared, a number of features are of significant interest.  Because of the wavelengths and associated observational constraints (e.g., the need to chop-nod in the MIR), these will only be very briefly discussed here. 

The 9.7~$\mu$m silicate feature is strongly polarized, while the 3.4~$\mu$m feature, due to aliphatic CH, is not \citep{chiar2006}.
One line of sight may show polarization in the 11.3~$\mu$m PAH feature \citep{zhang2017}, although a second, unidentified, feature at 10.3 $\mu$m is also found in the spectrum. These data do show the possibility of probing PAH features with MIR polarimetry. 
Several ice lines (e.g., water 3.1 $\mu$m and CO 4.7 $\mu$m) also show polarization; the relative shape of the intensity and polarization profiles may be used to constrain mantle volume (Hough et al. 2008)

\section{Line Polarization by Ground-State Alignment}


As shown by e.g.\ \citet{yan2006} and \citet{yan2013}, some spectral lines are expected to show polarization when the affected atoms are embedded in a magnetic field and exposed to anisotropic radiation.   Such ``ground state alignment'' polarization is predicted to be able to yield very high levels of polarization.   While the phenomenon is well known and established in laboratory laser physics, it has so far only been found in one astronomical source: the post-AGB binary system 89 Her \citep{zhang2017}. Several well-known optical lines are predicted to be observable, including the \ion{Na}{1}, \ion{K}{1} and \ion{Al}{2}.
If this technique can be observationally more broadly confirmed, it would provide a very potent tool for magnetic field studies in diffuse gas. This research would be uniquely enhanced by large telescopes since it will require sensitive polarimetry at high spectral resolution to not only resolve the spectral line but also - ideally - component structure from separate line-of-sight velocity components. This requires a spectral resolution of at least R~$\sim$~60,000 and better R~$\gtrsim$~100,000.
A more extensive discussion of the use of ground state alignment for ISM studies can be found in \citet{bga2022b}

\subsection{Instrument Requirements}
\label{requirements_lyNeb2}

\noindent \textbf{Science Program:} \newline
Interstellar and Circumstellar Media

\vspace{2ex}
\noindent \textbf{Linear and/or Circular Polarimetry:} \newline
Linear polarimetry

\vspace{2ex}
\noindent \textbf{Wavelength Coverage:} \newline
3600\AA -- 9000\AA\ and NIR (3.4 $\mu$m\ ) -- MIR\ (6 -- 12 $\mu$m\ )

\vspace{2ex}
\noindent \textbf{Field-of-View:} \newline
$\sim$few\arcmin\

\vspace{2ex}
\noindent \textbf{Observation Type (e.g.\ imaging polarimetry, spectropolarimetry, high-contrast):} \newline
Imaging polarimetry and spectropolarimetry

\vspace{2ex}
\noindent \textbf{Target Brightness (magnitudes):} \newline
Mostly V$\sim$ 11-18
\vspace{2ex}

\noindent 
\textbf{Spectral Resolution:} \newline
$R \sim 1000$ (for continuum polarimetry) to $R \sim 100,000$ (for GSA polarization)

\vspace{2ex}
\noindent \textbf{Spatial Resolution:} \newline
A few x the PSF width.  Mostly point-source observations.

\vspace{2ex}
\noindent \textbf{Required S/N or S/N per Wavelength Bin:} \newline
Imaging polarimetry: 5$\sigma$ for 0.1\% polarization \newline
Spectropolarimetry: 5$\sigma$ per measurement aperture per integrated line flux

\vspace{2ex}
\noindent \textbf{Level of Polarization per Spatial and/or Spectral Element:} \newline
$\sim$ 1\%

\vspace{2ex}
\noindent \textbf{Acceptable Instrumental Polarization and Absolute Error (IP(\%)  $\pm$ error(\%)):} \newline
$<$ 0.3$\pm$0.05\%

\vspace{2ex}
\noindent \textbf{Acceptable Linear to Circular Crosstalk:} \newline
N/A

\vspace{2ex}
\noindent \textbf{Additional Requirements:} \newline
N/A

\chapter{Galaxies and Large Scale Structure}
\label{sec:galaxy_assembly}
\section{The Galaxy-Black Hole Connection: Active Galactic Nuclei}
	\label{Ch5_AGN}

\begin{hibox1}
    Observations and modeling of the total and polarized SED of AGN in a wide IR (1-13 $\mu$m) wavelength range will provide crucial probes to 1) distinguish dust properties in and around the AGN; 2) investigate the jet-host galaxy interaction; and 3) study the evolution, dynamics, and morphology of the dusty torus from a MHD framework.
\end{hibox1}

The interface between active galactic nuclei (AGN) and their host galaxies is a region of a few parsecs in size with a molecular gas and dust flow cycle, a so-called `torus' \citep[e.g.,][]{1988ApJ...329..702K,2011ApJ...741...29D,2011MNRAS.415..741S,2012ApJ...758...66W}. The torus and accretion disk fuel the accretion onto the supermassive black hole (SMBH) and launch outflows, which fundamentally connect the black holes to their host galaxies. This region is particularly interesting for several reasons: (1) from infrared (IR) to X-ray, aside from synchrotron emission, the total flux spectral energy distributions (SEDs) of radio-loud and radio-quiet AGN are virtually identical \citep[e.g.,][]{2011MNRAS.414.1082M,2018ARAA..56..625H}, and (2) advected magnetic fields (B-fields) are strongly implicated in almost all models for the launching and collimations of jets, accretion onto the SMBH, and outflows \citep[e.g.,][]{1982MNRAS.199..883B,2014Natur.515..376G}. The B-fields allow for both the transport of angular momentum in the disk from sub-pc scales to the black hole and the formation of jets and outflows. Although the dusty torus represents part of the AGN accretion flow at pc-scales, the role of B-fields in this pc-scale accretion flow is poorly understood.

It has been suggested that the torus originated in a magneto-hydrodynamic (MHD) wind confined and accelerated by the accretion disk's B-field \citep[e.g.,][]{1992ApJ...385..460E,1994ApJ...434..446K,2006ApJ...648L.101E,2012ApJ...749...32K,2020ApJ...904....9K,2022ApJ...926...50T}. The MHD wind at sub-pc scales can lift the plasma from the midplane of the accretion disk to form a geometrically thick distribution of dusty clouds surrounding the central engine (accretion disk and SMBH). At pc-scales, the torus is the particular region of an outflow (at high latitudes) and inflow (at low latitudes), where the clouds are dusty and optically thick. Within the MHD model, we can define the modified $\beta$-plasma parameter, which is the ratio between turbulent kinetic and magnetic energy: for $\beta<1$, B-fields dominate, and for $\beta>1$, kinetic energy dominates. Thus, the observable signatures are: a) the disk's midplane is expected to be magnetically dominated ($\beta<1$) with a toroidal B-field, and b) the pc-scale wind is expected to be dominated by radiation pressure ($\beta>1$) with a B-field structure parallel to the wind. Although large efforts have been made in the development of MHD wind models in AGN \citep{2018A&A...615A.164V,2022ApJ...926...50T}, the B-field strength and geometry of the central~pc are barely constrained. A quantitative study of B-fields in the central pc-scale of AGN is required to put the MHD models on firmer footing.

\begin{figure}[ht!]
\includegraphics[width=\textwidth]{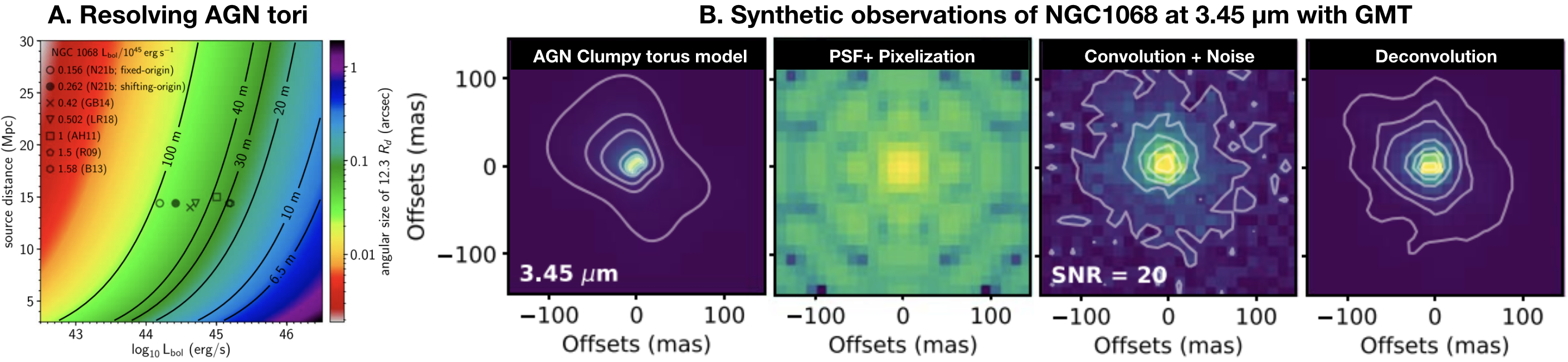}
\caption{GMT can resolve the dusty torus of nearby AGN in the $1-13$ $\mu$m wavelength range. 
\textbf{A.} Expected angular size of the AGN tori as a function of the bolometric luminosity and the distance \citep{2021ApJ...919..136N}. Solid lines trace the diffraction limit at $10\,\mu$m of telescopes between $6.5$ and $100$m diameters. \textit{AGN tori with luminosity and distance below a contour will produce a resolved dusty torus by the corresponding telescope}. The symbols show the various bolometric luminosities for NGC\,1068 at a distance of D$=14$ Mpc. 
\textbf{B.} Synthetic observations of NGC\,1068 at $3.45\,\mu$m with the GMT \citep{2021ApJ...923..127N}. Simulations were obtained using \textsc{CLUMPY} models and the torus parameters by \citet{2018ApJ...859...99L}. The AGN Clumpy torus model and the PSF of GMT at $3.45\,\mu$m were convolved, and a noise level was added to show the observations with s signal-to-noise of 20. A deconvolved image is also displayed to compare with the model. \textit{The extended emission of the AGN tori is resolved with the GMT}.}
\label{fig4_1}
\end{figure}

To study the dusty torus, infrared (IR) wavelengths are key to offer information about the dust grains that produce the obscuration to the active nuclei from some lines-of-sight (LOS). The dust grains in the torus absorb radiation from the active nuclei and re-radiate it as thermal emission in the IR, peaking at $\sim20-30\,\mu$m~\citep{2016MNRAS.462.2618F}. Except for some objects observed using IR interferometric observations with the Very Large Telescope Interferometer (VLTI) \citep[e.g.][]{2004Natur.429...47J,2009MNRAS.394.1325R,2013AA...558A.149B,2016A&A...591A..47L,2022Natur.602..403G,2022A&A...663A..35I}, evidence for such dusty tori has been mostly indirect, as the angular size of the AGN tori is much smaller than the angular resolution of 8-m class telescopes ($<0.1" $ at a typical distance of $\sim10$\, Mpc). The combination of the adaptive secondaries and potential instrumental capabilities in the IR window at the GMT will allow us to obtain a direct image of the dusty torus in nearby AGN. Figure \ref{fig4_1}-A shows the range of bolometric luminosities and AGN distances at which the 30-m class telescopes will resolve the AGN tori, which can be used as selection criteria to elaborate AGN surveys with the GMT. For example, Figure \ref{fig4_1}-B shows the synthetic observations of the dust emission arising from the AGN torus of NGC\,1068 at $3.45\,\mu$m with the GMT. The extended emission along the walls of the torus directly radiated by the central engine is clearly resolved with the GMT. These synthetic observations were computed using radiative transfer through a 3D distribution of dusty matter around the AGN with the HyperCubes of AGN Tori \citep[HyperCAT;][]{2021ApJ...919..136N,2021ApJ...923..127N}. Thus, \textit{30-m class telescopes are uniquely situated to resolve the hot dust of the AGN tori.}

The dust in the torus is magnetically aligned by the presence of pc-scale ordered B-fields surrounding the AGN. The magnetic field can induce a preferential orientation of dust grains within the dusty clouds that can give rise to a measurable degree of polarization. NIR (1-5 $\mu$m) observations provide magnetic information about the hot dust from the directly radiated clumps within the torus. Warm (T $\sim$ 300 K) dust from the shielded areas of the clumps and non-directly illuminated clumps at farther distances from the active nuclei are characterized by Mid-IR (MIR; 7.5-13 $\mu$m) observations. IR (1-13 $\mu$m) polarimetric observations will offer the unique opportunity to perform a tomographic study of the magnetic field and to test the MHD outflow wind model of AGN. This will allow us to study the evolution, dynamics, and morphology of the dusty torus and its interaction with the active nuclei and host galaxy within an MHD framework. Specifically, polarimetric observations in the $2-5$ $\mu$m wavelength range allow us to characterize the MHD flow in the inner edge of the torus. \textit{NIR polarimetry has the advantage of providing information about the magnetic field at the inner edge of the torus, where the accretion disk is being fed by the matter in the AGN tori.}

\begin{figure}[ht]
\includegraphics[angle=0,trim=0cm 1cm 0cm 0cm,scale=.28]{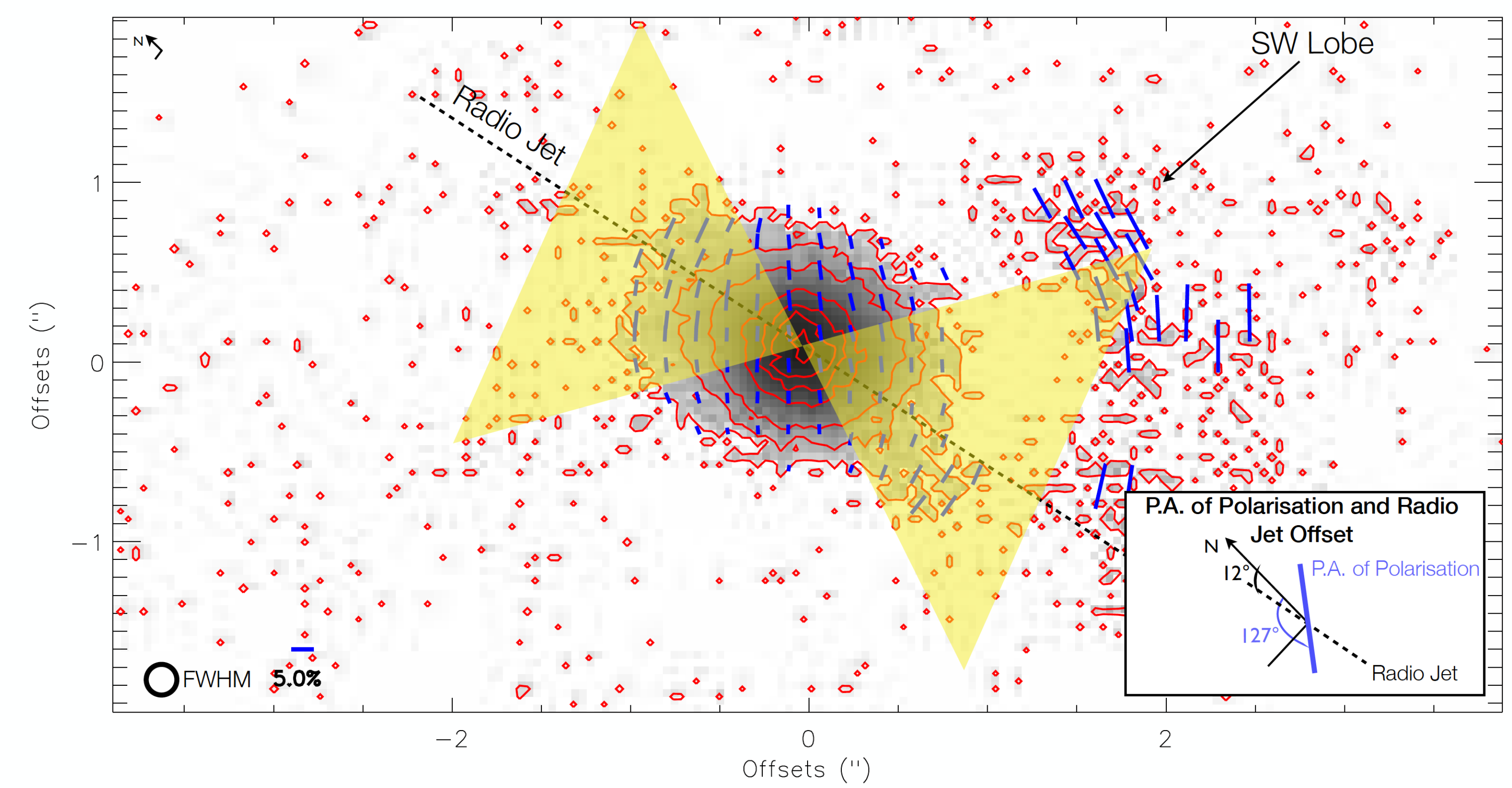}
\caption{2.2 $\mu$m imaging-polarimetric observations of NGC 1068 using MMT-Pol in conjunction with the AO secondary mirror on the 6.5-m MMT \citep{2015MNRAS.452.1902L}. The 8$''$ $\times$ 4$''$ polarized flux image (grayscale) at K$'$ with overlaid polarization vectors is shown. The polarized flux image has a pixel scale of 0.043$''$. The polarization vectors with $P/\sigma_{p} > 3$ were binned to a 4 $\times$ 4 pixels box (0.172$''$). A vector scale showing 5\% polarization and the FWHM of observations are shown in the lower left. The polarization angle of the core (blue) and the radio jet axes (dashed line) with respect to the North are shown in the inset.}
\label{fig5_2}
\end{figure}

Near-IR (NIR) polarimetric studies \citep{1995MNRAS.272..513Y,1997MNRAS.288..375P,1999MNRAS.303..209L,2001MNRAS.328L..32L,2001MNRAS.327..459L,2002ApJ...574...95S,2003ApJ...591..714W,2013MNRAS.431.2723L,2015MNRAS.452.1902L} of several AGN have shown that the NIR polarization is attributed to the extinction along the LOS by the passage of radiation of the active nuclei through hot dust (T $\sim$ 800-1500 K) in the inner edge of the torus. Furthermore, \citet{2013MNRAS.431.2723L,2015MNRAS.452.1902L} have shown that the magnetic field strength (few mG) and geometry (aligned with the torus axis onto the plane of the sky) within the torus of a few AGN can be determined through NIR (1$-$2.5 $\mu$m) polarimetry. \citet{2015MNRAS.452.1902L} performed imaging-polarimetric observations with MMT-Pol mounted on the Cassegrain focus, in conjunction with the adaptive secondary mirror and AO system, on the 6.5-m MMT. This configuration allowed the authors to obtain diffraction-limited (0.12$''$ at 2.2 $\mu$m) polarimetric observations with a sensitivity of 0.05$\pm$0.03\% in the degree of polarization (Fig. \ref{fig5_2}). These studies have shown the potential of high-angular resolution and sensitive polarimetric observations in the NIR. Due to the flux sensitivity of the current suite of IR polarimetric instruments in 8-m class telescopes, only a few objects are observable. This limits the statistical analysis required to obtain general and/or extraordinary magnetic properties. The high-angular resolution and the D$^{4}$ advantage of GMT will 1) minimize starlight dilution from the host galaxy, which is important to estimate the intrinsic polarization of the AGN, and 2) increase the number of objects, allowing us to measure magnetic properties within the torus of AGN.

\textbf{Dust properties and AGN-Host galaxy interaction:}  The recent work by Lopez-Rodriguez et al. (2015c, 2016) examine the nature of NGC 1068 using high-angular ($\sim$0.4$''$) resolution MIR (7.5-13 $\mu$m) imaging and spectropolarimetry with CanariCam\footnote{CanariCam is an imaging, spectroscopy, polarimetry and coronography MIR (7.5$-$25 $\mu$m) instrument on the 10.4-m GTC. See: \url{http://www.gtc.iac.es/instruments/canaricam/canaricam.php}} on the 10.4-m Gran Telescopio CANARIAS (GTC) showed the potential of MIR polarimetric techniques. This study was able to 1) distinguish between different dust grain compositions in and around the AGN, 2) investigate jet-host galaxy interaction, and 3) put physical constraints on the torus properties (e.g., dust grain sizes, dust temperatures, clump dust densities, etc.).  An IR imaging- and spectro-polarimetric capability in the GMT will be essential to study the dust properties in and around AGN. These observations will also help us to investigate the fraction of thermal and non-thermal polarization mechanisms in the central regions of AGN.

\section{Intranight polarimetric variability of blazars}
When AGN are viewed along the jet axis, they are referred to as blazars. Blazars are dominated by jet emission, with a minor contribution from the accretion disk, particularly during jet quiescence. The jet emission covers the entire electromagnetic spectrum and is strongly polarized due to particles spiralling around magnetic field lines, creating synchrotron radiation \citep{1978PhyS...17..265B}. The properties of this polarized light can be used to explore the jet structure and probe regions close to the jet formation site. Blazar jets cannot be spatially resolved. Therefore, detailed studies of the polarization can provide information about jet structure that would not be possible otherwise.

Blazars can show variability on a variety of timescales, from years down to minutes. The intranight optical variability (INOV) of blazars has been the focus of numerous campaigns over the past 30 years \citep[e.g.][]{miller1989,sagar2004,ojha2021} but is not a well-understood characteristic of their emission. Many models have been proposed to explain the different levels of variability observed over $\sim$ hour time scales, which predict time lags between different wavebands, color evolution, and polarization degree and angle changes. Differing mechanisms are proposed to produce the observed behavior, including geometric changes related to the jet angle and Doppler factors of emitting blobs, intrinsic changes relating to particle energy distributions, and extrinsic changes, including microlensing from objects outside the blazar system along our line of sight \citep{2015MNRAS.451L..21B}. However, the polarimetric data quality available for this type of analysis often lacks the sensitivity to make definitive statements about the granular variability.

Many bright blazar sources have been studied over long timescales for many decades, particularly with photometric observations. For bright blazar sources like BL~Lac, small telescopes can be used to probe fine intranight polarimetric variability (e.g., $\sim6\%$; \cite{2023MNRAS.522.3018B}). However, to probe the variability of a wider range of blazar sources at fainter magnitudes, it is necessary to utilize imaging polarimetry on large telescopes. \\






\subsection{Instrument Requirements}
\noindent \textbf{Science Program:} \newline
Revealing the role of the magnetic fields in the Galaxy-black hole connection.

\vspace{2ex}
\noindent \textbf{Linear and/or Circular Polarimetry:} \newline
Linear polarimetry

\vspace{2ex}
\noindent \textbf{Wavelength Coverage:} \newline
$1-12~\mu$m 

\vspace{2ex}
\noindent \textbf{Field-of-View:} \newline
$<2"\times2"$

\vspace{2ex}
\noindent \textbf{Observation Type (e.g.\ imaging polarimetry, spectropolarimetry, high-contrast):} \newline
Imaging-polarimetry, spectro-polarimetry. Ideally IFU-polarimetry.

\vspace{2ex}
\noindent \textbf{Target Brightness (magnitudes):} \newline
$14-12$ at K (2.2 $\mu$m) 

\vspace{2ex}
\noindent \textbf{Spectral Resolution:} \newline
$200-1000$

\vspace{2ex}
\noindent \textbf{Spatial Resolution:} \newline
$21$ mas at K (2.2 $\mu$m)

\vspace{2ex}
\noindent \textbf{Required S/N or S/N per Wavelength Bin:} \newline
$>400$ in Stokes I, which will provide an accuracy in the degree of polarization $<0.3$\%.

\vspace{2ex}
\noindent \textbf{Level of Polarization per Spatial and/or Spectral Element:} \newline
$>1$\%

\vspace{2ex}
\noindent \textbf{Acceptable Instrumental Polarization and Absolute Error (IP(\%)  $\pm$ error(\%)):} \newline
IP $<0.3$\%, absolute error $<0.3$\%.

\vspace{2ex}
\noindent \textbf{Acceptable Linear to Circular Crosstalk:} \newline
$<0.2$\% at FOV center.

\vspace{2ex}
\noindent \textbf{Additional Requirements:} \newline
N/A.

\bigskip
\section{Polarization of Lyman Alpha Nebulae}


\begin{figure}[ht!]
\begin{center}
\mbox{
  \includegraphics[height=0.25\textwidth]{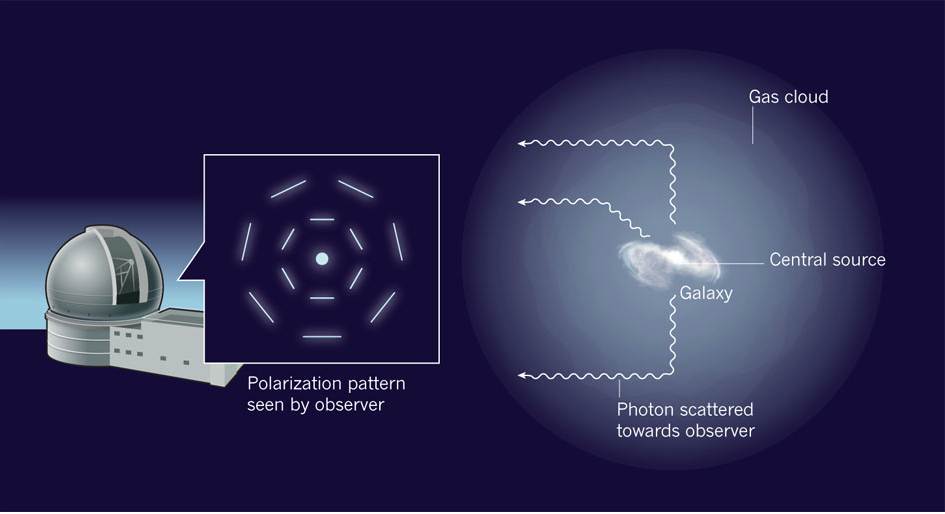}
  \includegraphics[height=0.25\textwidth]{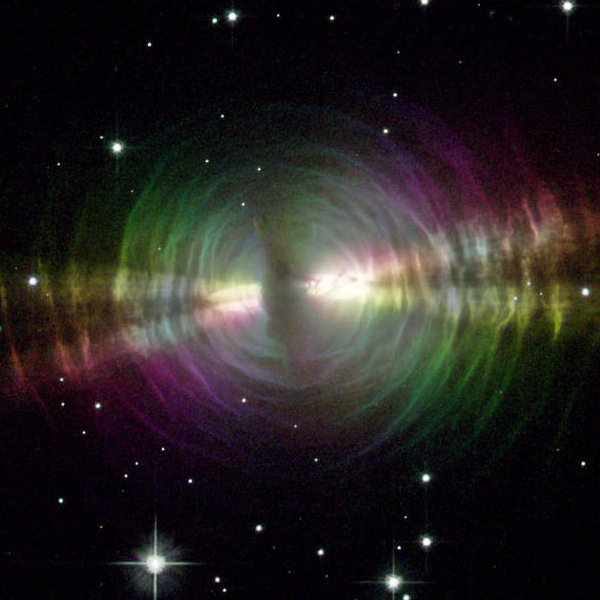}
  \includegraphics[height=0.25\textwidth]{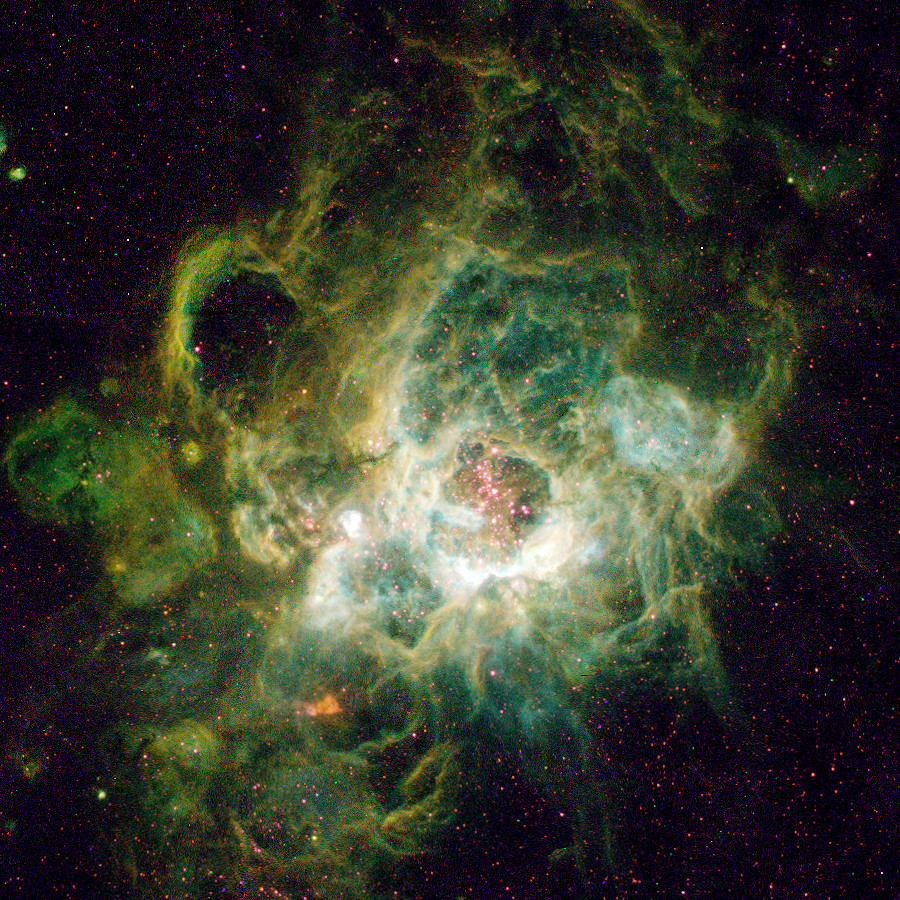}
}
\caption{
\small 
Using polarization maps to compare signatures of resonant scattering versus photoionization/recombination models for the origin of \lya nebulae.
{\bf (Left)} \lya photons produced by a central, embedded source---the galaxy in this picture---are resonantly scattered by the surrounding gas cloud. Photons observed furthest from the source have scattered perpendicular to their initial direction and are strongly polarized. Observers will see 1) a polarization pattern of concentric rings centered on the source (\citealt{Dijkstra-Loeb2008}, image credit: \citealt{Bower2011}) and 2) an increasing degree of polarization with radius. Thus, polarimetric mapping tells us where the observed \lya photons originate and why those photons produce a \lya nebula. 
{\bf (Center)} An analog in the local universe: resonant scattering is represented here by dust-scattered emission in the Egg Nebula. As outlined in the left panel, scattering produces coherent, concentric rings of polarization that strengthen with radius.
{\bf (Right)} A local HII region demonstrates how a nebula could be illuminated by photoionization/recombination without resulting polarization.
}
\label{fig:lya_polarimetry}
\end{center}
\end{figure}

\begin{figure}[h]
\begin{center}
\mbox{
   \includegraphics[width=\textwidth]{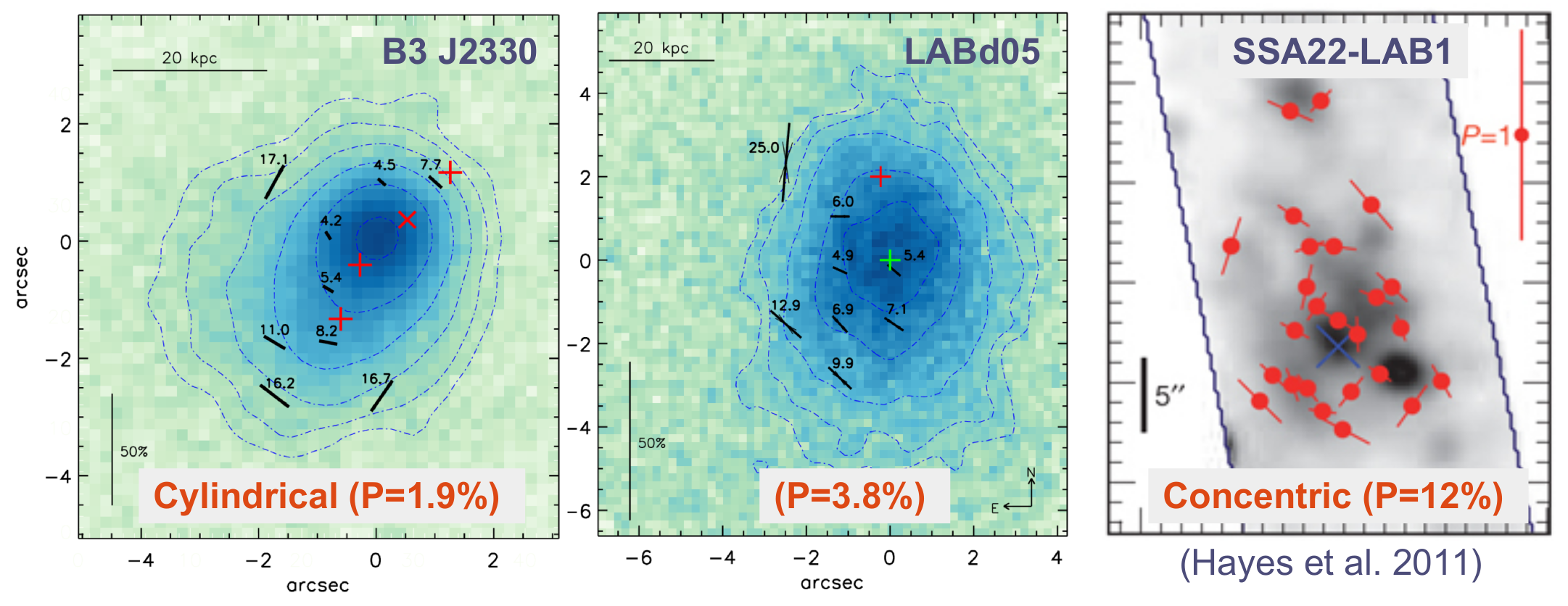}
}
\caption{
\small
{\bf Diverse Polarization Patterns in \lya Nebulae.}
{\bf (Left)} MMT/SPOL polarization measurements of the radio galaxy, B3 J2330+3927 \citep{You2017}.  Polarization detections at $>$2$\sigma$ are shown as black vectors. The length and direction of each vector represents the polarization fraction and direction, respectively. Significant polarizations are detected only along this blob's major axis (also the radio lobe axis; red crosses and x) and have angles perpendicular to that axis, with the polarization strength increasing with radius from the center.
{\bf (Middle)} \lya polarization map for LABd05, a radio-$quiet$ blob with an offset, obscured AGN \citep{Kim2020}. The green and red crosses represent the location of the \lya peak and AGN, respectively. The polarization pattern suggests that the peak center, not the AGN, is the source of the \lya photons, which are then resonantly scattered by the cloud.
{\bf (Right)} Concentric polarization pattern in SSA22-LAB1 \citep{Hayes2011}.
}
\label{fig:polmap}
\end{center}
\end{figure}

Hydrogen Ly$\alpha$ is the most prominent emission line and thus a powerful tool for studying the early universe at $z$ $>$ 2. Narrowband imaging surveys have revealed various \lya-emitting sources: compact \lya emitters \citep{Gawiser2007, Ouchi2008}, \lya blobs \citep[LABs;][]{Steidel2000, Matsuda2004, Yang2009, Yang2010}, and enormous \lya nebulae \citep[ELANe;][]{Hennawi2015, Cantalupo14, Cai2017}. \lya nebulae (LABs and ELANe) are typically extended over $\sim$100 kpc and have \lya luminosities of $\gtrsim$ $10^{44}$ \unitcgslum. They are believed to trace massive dark matter halos that will evolve into rich galaxy groups or even clusters today; as such, their embedded galaxies may be the progenitors of massive cluster galaxies, and their gas may be the precursor to the intra-group medium \citep[e.g.,][]{Badescu2017}.

While the extended \lya emission provides a unique chance to explore the circum-galactic medium (CGM) or inter-galactic medium (IGM) directly in emission, the origin of this \lya emission is still poorly constrained. The proposed power sources include photoionization by AGN \citep{Steidel2000, Arrigoni-Battaia2019}, cooling radiation from cold-mode accretion \citep{Trebitsch2016, Daddi2021}, shocks due to fast galactic outflows \citep{Cabot2016, Travascio2020}, and  scattering by the CGM of \lya photons produced by centrally embedded galaxies \citep{Hayes2011, Li2021}. Despite the range and  complexity of these proposed processes, 
why \lya nebulae glow has been narrowed to two classes of explanations: photoionization throughout the nebulae and the scattering of \lya photons produced at the center of the nebula. 

{\bf Polarimetry: A New Tool for Exploring the \lya-Luminous CGM.}
Polarimetric observations make it possible to discriminate between the proposed explanations for \lya nebulae. 
If embedded sources like AGN or star-forming galaxies photoionize the entire nebula, then \lya photons are produced by hydrogen recombinations throughout the cloud and in random directions. In this case, the nebula's \lya emission
will be only slightly polarized, if at all.
On the other hand, if a centrally embedded source produces \lya that is resonantly scattered by the cloud, sometimes into our sightline, the observed scattering angle defines the polarization, tangent to and increasing with the projected distance from the source (Figure \ref{fig:lya_polarimetry}). Photoionization and scattering might even co-exist in the same cloud if a \lya source photoionizes only local gas and escaping \lya is scattered into the sightline.

\citet{Hayes2011} observed a concentric polarization pattern around SSA22-LAB1, suggesting that scattering from a central source produces the extended \lya emission (Figure \ref{fig:polmap} right). 
Later, \citet{You2017} and \citet{Kim2020} conducted polarization mapping of a \lya nebula around a high-$z$ radio galaxy (HzRG) and another containing an offset, obscured AGN, showing that polarized \lya emission up to $\sim$10\% occurs for a range of embedded potential sources and that polarization map morphologies are diverse (Figure \ref{fig:polmap}). 
In particular, for B3 J2330+3927, the LAB around a HzRG  at $z$ = 3.087, \citet{You2017} found that the polarization vectors are aligned perpendicular to the major axis (also the direction of the radio jet).
In LABd05, a radio-$quiet$ blob with the offset, obscured AGN, \citet{Kim2020} found an asymmetric polarization pattern in which significant polarization was detected only toward the southeast. 
The polarized \lya emission in all three nebulae provides strong evidence that scattering by the neutral medium plays an important role in producing \lya halos.

\begin{figure}[ht!]
\includegraphics[width=\textwidth]{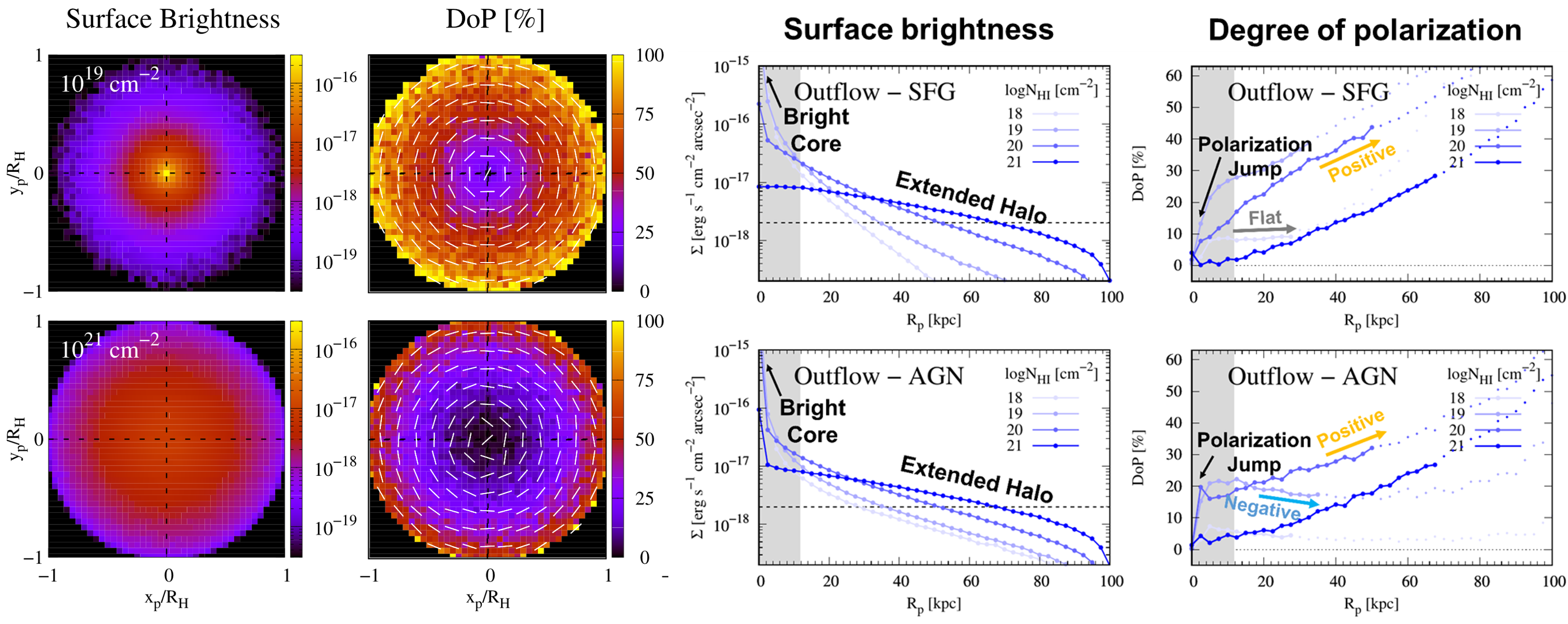}
\caption{Results from simulations of \lya radiative transfer from \cite{lya_chang23}. 
{\bf (Left)} Projected images of surface brightness and degree of polarization for H~I column densities $N_{\rm HI} = 10^{19}\, {\rm cm^{-2}}$ (top) and  $10^{21}\, {\rm cm^{-2}}$ (bottom).
{\bf (Right)} Radial profiles of surface brightness and degree of polarization in the range of $N_{\rm HI} = 10^{18-21}\,  {\rm cm^{-2}}$ for two types of \lya sources, a star-forming galaxy (SFG) with a $100\, \rm km\, s^{-1}$ \lya linewidth and an AGN with a $400\, \rm km\, s^{-1}$ linewidth. In both cases, the halo gas is modeled as expanding at $400\, \rm km\, s^{-1}$.
}
\label{fig:Lya_RT}
\end{figure}

{\bf GMT Polarimetry:  Connecting Galaxies to the Cosmic Web}.
Decoding the origin of the polarized \lya requires comparing sophisticated simulations of 
\lya radiative transfer (RT) with large and diverse samples of \lya nebulae.
\cite{lya_chang23} developed models for \lya RT in \lya nebulae, predicting 
surface brightness, kinematic, and polarization profiles  (Figure~\ref{fig:Lya_RT}).
Yet interpreting observations of scattered \lya through polarization patterns and other \lya properties is limited by the small number of samples.
With the advantages afforded by its large aperture and the low instrumental polarization resulting from its axisymmetric optics, the GMT will revolutionize \lya polarimetry of CGM. Currently, \lya polarimetry on 8m-class telescopes is limited to only the brightest \lya nebulae and requires significant telescope time (20-40h). With a set of narrowband or tunable filters, the GMT will enable an unprecedented and systematic imaging polarimetric survey of high-z \lya nebulae. The improved signal-to-noise and higher spatial resolution will increase sample sizes by an order of magnitude, push the polarization maps out to larger radii, and discern small-scale variations in polarization strength  and angle. As a result, the GMT will allow us to connect the CGM to the outerlying cosmic web and to the embedded galaxies and their AGN over $z=2$-6.

\bigskip

\subsection{Instrument Requirements}
\label{requirements_lyNeb}

\noindent \textbf{Science Program:} \newline
Polarization of \lya Nebulae at High Redshift

\vspace{2ex}
\noindent \textbf{Linear and/or Circular Polarimetry:} \newline
Linear polarimetry

\vspace{2ex}
\noindent \textbf{Wavelength Coverage:} \newline
3600\AA -- 9000\AA\ (redshifted \lya line for $z$ = 2--6)

\vspace{2ex}
\noindent \textbf{Field-of-View:} \newline
$\sim$1\arcmin\ (even larger FOV is preferred)

\vspace{2ex}
\noindent \textbf{Observation Type (e.g.\ imaging polarimetry, spectropolarimetry, high-contrast):} \newline
Imaging polarimetry, spectropolarimetry, and ideally, IFU-polarimetry

\vspace{2ex}
\noindent \textbf{Target Brightness (magnitudes):} \newline
\lya line surface brightness $\sim$ $10^{-18}$ \unitcgssb

\vspace{2ex}
\noindent \textbf{Spectral Resolution:} \newline
$R \sim 100$ (narrowband imaging)

\vspace{2ex}
\noindent \textbf{Spatial Resolution:} \newline
1\arcsec\ or better

\vspace{2ex}
\noindent \textbf{Required S/N or S/N per Wavelength Bin:} \newline
Imaging polarimetry: 5$\sigma$ per measurement aperture\newline
Spectropolarimetry: 5$\sigma$ per integrated line flux

\vspace{2ex}
\noindent \textbf{Level of Polarization per Spatial and/or Spectral Element:} \newline
10\% -- 20\%

\vspace{2ex}
\noindent \textbf{Acceptable Instrumental Polarization and Absolute Error (IP(\%)  $\pm$ error(\%)):} \newline
$<$ 0.1\%

\vspace{2ex}
\noindent \textbf{Acceptable Linear to Circular Crosstalk:} \newline
N/A

\vspace{2ex}
\noindent \textbf{Additional Requirements:} \newline
A suite of narrowband filters or Fabry-Perot system (i.e., tunable filter)

\bigskip
\section{Weak lensing and optical polarization of galaxies}
\label{Ch4_weaklensing}

Galaxies are teeming with dust, free molecules, and electrons, which are the main source of polarization in the optical bands. Optical light can become polarized in a normal galaxy mainly due to two mechanisms: scattering and dichroism. Scattering can be of different types (Mie, Rayleigh, Thompson) - corresponding to different scattering agents (dust, molecules, electrons, respectively), and can induce high levels of polarization with polarization angles perpendicular to the last direction of flight of the radiation producing a typical centrosymmetric pattern centered in the radiating source. Dichroic absorption can occur when dust grains are aligned due to the presence of a magnetic field and attenuate radiation preferentially in the plane of polarization parallel to their major axes. In this case, the observed electric vector position angle (EVPA) is parallel to the magnetic field. The dust grains re-emit the radiation in lower frequencies (dichroic emission), with the radiation being strongly polarized in the direction of their long axes, i.e., perpendicular to the magnetic field. Optical polarization is also observed in the case of Active Galactic Nuclei (AGN) hosting jets. In this case, free electrons are accelerated in the jet, producing synchrotron radiation, which is polarized with an EVPA perpendicular to the B-field direction. When the jet is oriented toward our line of sight, the synchrotron is the dominant mechanism for the observed optical polarization. In other cases of AGN, observed polarization is due to a combination of synchrotron, scattering, and dichroism (see above).

{\bf A tremendous lack of data}. AGNs have been targeted by polarization studies due to their very interesting nature. Nevertheless, the number of objects which have been observed in the optical is not significant. Specifically, \cite{2020PhRvD.102d3008F} collected 1278 AGN sources from the literature, which is considered to be the largest catalog of optical polarization of AGN. Meanwhile, the lack of optical polarimetric data of normal galaxies is deafening. Only a handful of nearby galaxies have been studied in optical polarization, offering precious information on their morphology and the mechanisms that produce polarization. Moreover, integrated optical polarization data are strikingly restricted. To date, there are only 70 galaxies with observed optical polarization \citep{2012ApJ...748...17J}, and this small sample is largely biased towards specific morphological types. Evidently, the area of integrated galactic optical polarimetry is largely a virgin territory.

 Obtaining a large set of optopolarimetric measurements for extragalactic sources will undoubtedly be invaluable for the astronomical society as a whole. The nature of the data will be entirely novel and will allow for exploitation in numerous areas of interest, but here, we focus on the potential impact on weak lensing.

{\bf Polarization in weak lensing}
Weak gravitational lensing is of utmost importance for the estimation of cosmological parameters. The projected image of background elliptical sources, such as inclined spiral galaxies, are subject to distortions by foreground clusterings of matter, both bright and dark, called gravitational lenses. The weakly-lensed image suffers from slight changes in the intrinsic size, orientation, and ellipticity. The distortion introduced by the lens in the orientation (namely the sheer) is indicative of its gravitational potential and can be used to directly measure the dark matter distribution within the lens. A major problem in determining these distortions is the ignorance of the original sources' characteristics. This is where polarization studies become relevant. {\it While the shape of a background galaxy generally changes, EVPA is known to remain unaltered by gravitational lensing} \citep[e.g.,][]{1992grle.book.....S, 2005MNRAS.359L..19S}. Moreover, one expects the integrated polarization of the galaxy to be correlated (albeit in a non-trivial manner) with the shape of the galaxy. Therefore, the observed EVPA should give a proxy of the true orientation of the galaxy.
\citet{1999MNRAS.303...87A} have proposed that by using polarization, one can independently constrain the effect of lensing. They claim that if one can measure the image properties and polarization angle of a source, the difference between EVPA and projected galaxy orientation can be used to determine the distortion induced by the lens, ergo its strength. The uncertainty of the estimated sheer will depend on the uncertainty of the measurements and the number of observed galaxies. \citet{1999MNRAS.303...87A} argue that some tens of sources should give a satisfactory estimation, but this has never been demonstrated observationally.

There is another challenge that studies of weak gravitational lensing face, where polarization can be of use. To date, the ignorance of the intrinsic shape of the galaxies before they get lensed is treated statistically. In any given line of sight, hundreds of galaxies are used to calculate their average projected ellipticity. If galaxies' shapes are randomly distributed, the average observed ellipticity can be assumed to be solely a product of weak gravitational lensing. However, there are indications that galaxies are intrinsically aligned, which poses a challenge for these calculations \citep{2015PhR...558....1T,2015SSRv..193..139K}. To address this problem, cosmologists have employed theoretical models to account for and correct for the predicted alignments. Since EVPA can directly provide the intrinsic shape of the source, the degree of alignment of background galaxies should be projected in the distribution of their EVPA. This will yield a frank measurement, which could be potentially incorporated in future studies.

{\bf Theoretical models}
The above ideas rely on the main assumption that there is a correlation between EVPA and the shape of the galaxy. \citet{1997ApJ...477L..25W} assumed a symmetrical disk galaxy model and polarisation arising both from scattering and dichroic absorption, with the magnetic fields following the spiral arms. This model suggests that scattering produces polarisation perpendicular to the disk while dichroic absorption is parallel. \cite{2017A&A...601A..92P} assumed only polarisation from scattering and studied the spatial distribution of polarisation in dusty disk galaxies, which yielded a similar result. It follows that scattering and dichroic absorption are two competing processes regarding the production of polarised light. The polarisation angles (electric vector position angles - EVPAs) produced by the two mechanisms are perpendicular to each other and would be partially mutually excluded if we were to consider the integrated light of the galaxy. Hence, the total polarisation of the galaxy would be either parallel or perpendicular to its position angle, depending on the dominant process.

{\bf Early Observational tests}
Although there is a solid theoretical background that supports the scenario that the polarisation of the light of galaxies can be preferentially either parallel or perpendicular to the major axis, the global picture seems to be more complex.

Observational works on galactic optical polarisation are limited in number. In optical polarisation maps of face-on galaxies, polarization is mainly dominated by dust scattering with perhaps several regions arising from dichroic absorption, implying the existence of an ordered magnetic field in the disk \citep[e.g.,][]{1996QJRAS..37..297S}. In the cases of more inclined galaxies, polarization seems to be either parallel or perpendicular to the major axis. In some instances the EVPA does not change along the disc \citep[e.g.,][]{1996QJRAS..37..297S,1990IAUS..140..245S,1996A&A...308..713F}, while in others there is a transition between parallel and perpendicular position angles depending on the distance from the nucleus \citep[e.g.,][]{1996QJRAS..37..297S,1990IAUS..140..245S}, indicating the dominance of different processes in different parts of the galactic disk. Furthermore, \citet{1990IAUS..140..245S} showed that the behavior of polarisation position angles and polarisation degree may be dependent on the galaxy type.

\citet{2012ApJ...748...17J} measured the polarisation of the integrated light of spiral galaxies in the B band and found that more than half of their sample does not follow the expected trend. 
In fact, near-infrared polarimetric observations of resolved galactic disks reveal that polarisation patterns can be complicated, mainly arising from scattering with potentially different parts of the galaxy having different polarization mechanisms \citep[e.g.,][]{2000AJ....120.2920J, 2014ApJ...786...41M}

Demonstrating that polarimetric information can enhance the accuracy of weak lensing studies would have a tremendous effect. However, we still don't fully understand how optical polarisation is produced in nearby galaxies, and the basic assumption that the EVPA is always parallel or perpendicular to the major axis of the galaxy is not sufficiently backed by observational data. Spectropolarimetry in the optical will be a game changer in this regard since it will unveil the wavelength and spatial variations of polarisation within the galaxy and will offer a unique opportunity to disentangle and model the contributions of scattering and dichroism. Therefore, not only will it offer information on the relationship between EVPA and galaxy orientation, but it will also provide insights into the physics of dust and magnetic fields in galaxies.


\newpage

\subsection{Instrument Requirements}
\label{requirements_weaklens}

\noindent \textbf{Science Program:} \newline
Polarization and weak lensing

\vspace{2ex}
\noindent \textbf{Linear and/or Circular Polarimetry:} \newline
Linear

\vspace{2ex}
\noindent \textbf{Wavelength Coverage:} \newline
3000 -- 9000 \AA

\vspace{2ex}
\noindent \textbf{Field-of-View:} \newline
$\sim 5\arcmin$

\vspace{2ex}
\noindent \textbf{Observation Type (e.g.\ imaging polarimetry, spectropolarimetry, high-contrast):} \newline
Spectropolarimetry

\vspace{2ex}
\noindent \textbf{Target Brightness (magnitudes):} \newline
$m \sim 19.0$

\vspace{2ex}
\noindent \textbf{Spectral Resolution:} \newline
low, $\sim 10 - 20$ \AA

\vspace{2ex}
\noindent \textbf{Spatial Resolution:} \newline
N/A

\vspace{2ex}
\noindent \textbf{Required S/N or S/N per Wavelength Bin:} \newline
~100

\vspace{2ex}
\noindent \textbf{Level of Polarization per Spatial and/or Spectral Element:} \newline
$1 - 2$\%

\vspace{2ex}
\noindent \textbf{Acceptable Instrumental Polarization and Absolute Error (IP(\%)  $\pm$ error(\%)):} \newline
< 0.1\%

\vspace{2ex}
\noindent \textbf{Acceptable Linear to Circular Crosstalk:} \newline
The smaller, the better. Ideally, it can be calibrated out.

\vspace{2ex}
\noindent \textbf{Additional Requirements:} \newline
N/A

\newpage


\chapter{Solar system objects}
\label{sec:solarsystem}
    
%
%
\drop{A}s a ratio of intensities, the single-scattered polarization, both linear and circular, does not depend on the amount of the scattering material but reflects its compositional and structural characteristics only. Multiple scattering is accompanied by depolarization and interference of the single-scattered light that, together with the intensity of the scattered light, allows for estimating the optical depth and structural characteristics of the medium. These capabilities have made polarization a powerful tool to study properties of all types of cosmic dust, including cometary and interplanetary dust, planetary aerosols, and regolith on the surfaces of the solar system bodies. Being sensitive to asymmetries in the scattering medium, polarization is also the main tool in studying magnetic and electric fields. More detailed information on the topics discussed below can be found in \citep{2015psps.book.....K}, chapters 17-26. 

\section{Regolithic surfaces}
\begin{figure}[ht]
\includegraphics[angle=0,trim=0cm 0cm 0cm 0cm,scale=.35]{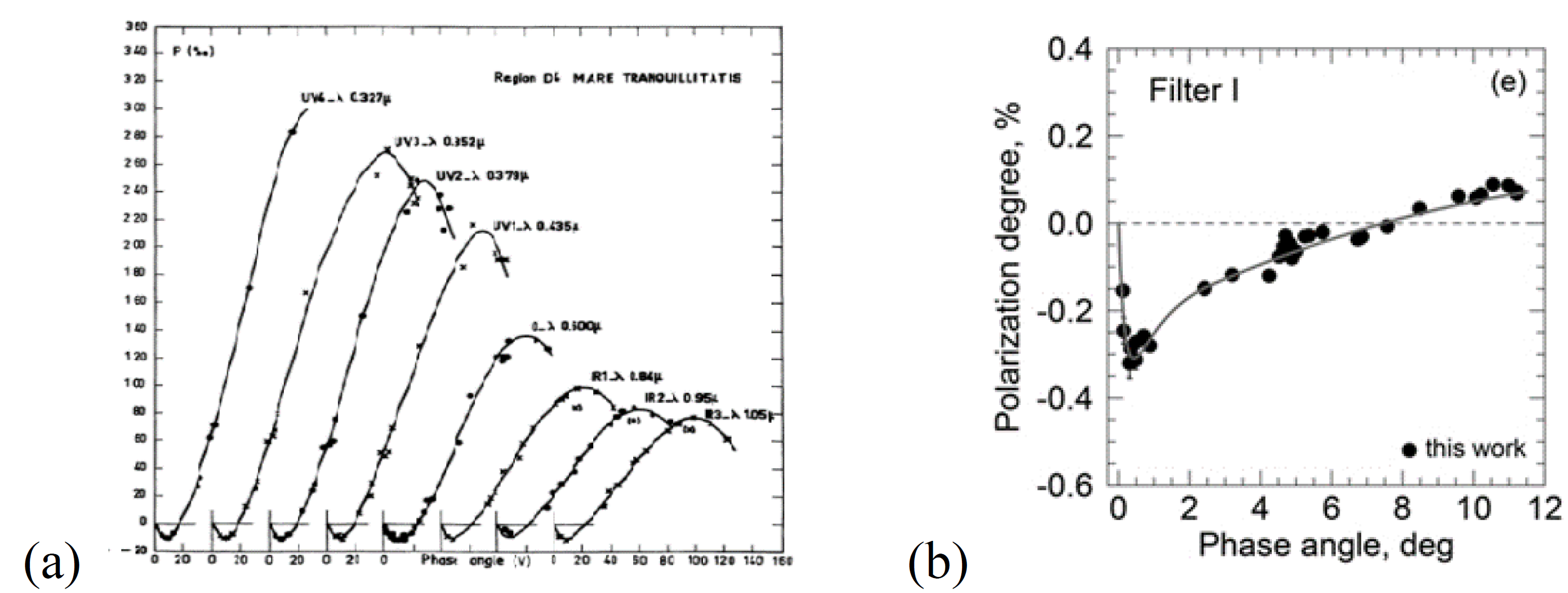}
\caption{Phase angle dependence of polarization: (a) for the Moon in different wavelengths \citet{1971A&A....10...29D}; (b) for Europa in Filter I \citet{2022PSJ.....3..134K}.}
\label{Fig5.1}
\end{figure}

The Moon was the first solar system body whose polarization was studied in detail for a variety of scattering angles and wavelengths. This allowed predicting that its surface is covered by a blanket of dust \citep{1929PhDT.........9L} now known as ``lunar regolith.'' The regularities in lunar polarization, e.g., wavelength and phase dependence of polarization (Figure~\ref{Fig5.1}a), appeared to be similar for other rocky bodies and were used for the interpretation of polarimetric observations of Mercury, Mars, and asteroids. A significant outcome of polarimetric studies is Umov’s law, according to which polarization strongly correlates with albedo \citep[e.g.,][]{1979aste.book..170D, mishchenko2010polarimetric}, and references therein) that allows using polarimetric data to determine albedos, and thus taxonomic types and diameters of asteroids \citep{2012JQSRT.113.2552C}. The application of polarimetry has been recently extended to the Near-Earth Asteroids that are of high interest due to their hazardous nature \citep{2021plde.confE..35D}. The Moon still remains among the objects of interest as it is a perfect testing object for new polarimeters and mineralogical mapping; more interest in the Moon has been ignited by the NASA lunar program Artemis, which requires high spatial resolution imaging of the lunar surface \citep{2022cosp...44.3187D}.  
Icy bodies (satellites of giant planets, comets, KBOs) are also covered by regolith which includes or even is dominated by small particles of ice. Polarimetry of icy bodies reveals the so-called polarimetric opposition effect (POE), a narrow feature that is observed at small phase angles and has negative values (Figure~\ref{Fig5.1}b). POE is usually explained by the coherent backscattering effect \citep[e.g.,][]{muinonen1994coherent}. Its width and depth and their change with wavelength allow estimation of the composition and size of the particles and porosity of the regolith \citep{muinonen2004coherent, muinonen2010coherent, 2009ApJ...705L.118M}. One of the striking results is the characterization of KBO using their POE and the discovery of their polarimetric duality \citep{2008A&A...491L..33B}. Note that the POE has also been observed for Saturn’s rings \citep{1993ApJ...411..351M} and high-albedo E-type asteroids \citep{2006JQSRT.100..325R}.

Detailed polarimetric studies of the surfaces of solar system bodies, e.g., high-resolution imaging in several wavelengths, enable a detailed characterization of the surface composition and structure and can be used for mineralogical studies or revealing differences between specific bodies or surface areas. 

\section{Planetary aerosols}
Unlike planetary regolith, which is dominated by densely packed particles, aerosols in planetary atmospheres are sparsely distributed and are more homogeneous and regularly shaped. This makes it easy to interpret polarimetric measurements of planetary atmospheres using radiative transfer techniques and simple light-scattering theories, e.g., the Mie theory. In particular, polarimetric data allowed \citet{1974JAtS...31.1137H} to determine that Venus is veiled by a cloud layer of micron-sized particles formed by sulfuric acid. Another application is related to studying Martian aerosols, especially during Martian dust storms. Similar techniques were applied to aerosols in the atmospheres of giant planets. They allowed not only the characterization of the aerosol particles but also the determination of the atmospheric structure (presence of layers, clouds, and haze) and estimation of the pressure at different levels.  

Unlike regolithic surfaces, planetary aerosols rapidly evolve and react to changing conditions. This is why polarimetric monitoring of planetary aerosols allows one not only to identify differences between aerosols in different parts of an atmosphere (e.g., spots and bands on Jupiter) but also to study planetary meteorology, including transient events (e.g., collisions with comets, such as the collision of SL9 with Jupiter)  as well as seasonal and secular (e.g., climatic) variations. It was shown that ``polarimetric data obtained over a wide range of scattering angles may provide definitive constraints on aerosol shape''\citep{2008MNRAS.384...64D}. This approach allowed \citet{1991Icar...90..330W} to prove, based on polarimetric data from Voyager, that aerosols in Titan’s atmosphere are aggregates of organic composition. In the specific case of the Jovian satellite Europa, imaging polarimetry may provide information of astrobiological value \citep{sparks2022probing}. 

\section{Cosmic dust}
\begin{figure}[ht]
\includegraphics[angle=0,trim=0cm 0cm 0cm 0cm,scale=.30]{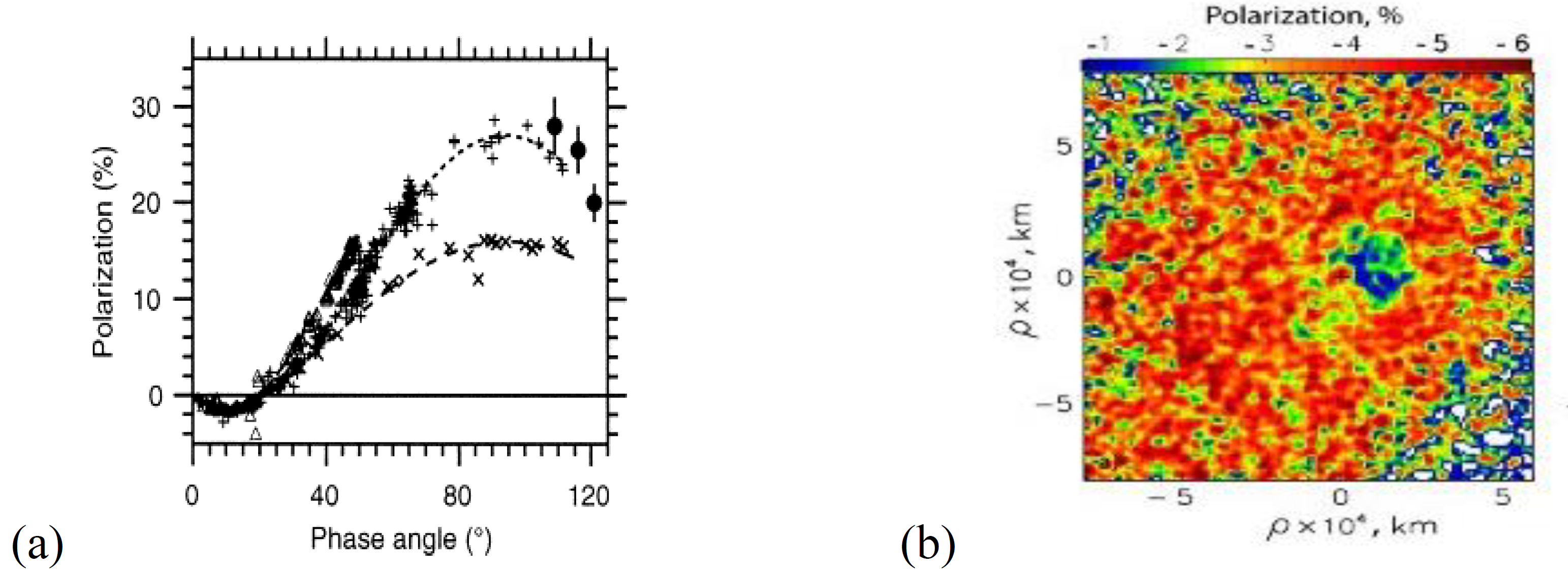}
\caption{polarization of comets: (a) cumulative dependence of polarization on phase angle (the angle supplementary to scattering angle) for a set of comets \citep{2003JQSRT..79..661H}; (b) polarization map of comet C/2011KP36 (Spacewatch), the cross indicates the cometary nucleus and the blue area is a polarimetric jet that was later attributed to an outburst of carbon dioxide ice \citep{2021A&A...651A..29I}.}
\label{Fig5.2}
\end{figure}

Early measurements of cometary polarization date as far back as 1819 and 1835 (Arago); polarization of the Zodiacal cloud was first detected by Dufay in 1925. These discoveries revealed that in both cases, one was dealing with some kind of dust, and the inferred properties of this dust can give us clues to the properties of planetesimals (comets) and interplanetary dust, which represents the solar-system debris disk. Later it was shown that all comets and interplanetary dust demonstrate angular dependences of polarization of a very similar shape (Figure~\ref{Fig5.2}a). \citet{1978A&A....65..265G} showed that such a curve cannot be obtained by modeling the cometary or zodiacal dust as spherical particles and that the best fit is provided by dark fluffy (porous) particles. Thus the polarimetry revealed the nature of the cometary and interplanetary dust particles (IDPs), which later was confirmed by direct studies of IDPs \citep[e.g.,][]{1987RvGeo..25.1527M} and dust in comet Wild 2 \citep{2008EM&P..102..447F}. Detailed analyses of cometary polarimetric data helped determine the size of constituents (monomers) in cometary aggregates and their composition \citep{2003A&A...407L...5K, 2006A&A...449.1243K}. Later it became possible to explain two types of comets, with lower and higher polarization maxima as seen in Figure~\ref{Fig5.2}a. It appeared that low-polarization comets have more compact dust particles than those showing the higher polarization maximum \citep{2007A&A...463.1189K}. A statistical analysis of cometary orbits has shown that more porous particles are typical for new and Halley-type comets, whereas Jupiter Family Comets have more compact dust ― a result easily expected from our understanding of the evolution of these two types of comets. 

Interestingly, curves very similar to those shown in Fig.\ref{Fig5.2}a are also typical of atmosphereless bodies (asteroids, Moon, Mercury) and Mars due to the ``dusty'' nature of their regolith. However, there is a significant dif\/ference between the spectral behavior of polarization of cometary dust and atmosphereless bodies: comets show an increase in polarization with wavelength, whereas the other objects show a decrease \citep{2004come.book..577K}. Later it was found that polarization decreases with wavelength for some comets \citep{2008JQSRT.109.1384K} and at longer (NIR) wavelengths \citep{2004AJ....127.2398K}.  \citet{kolokolova2011photopolarimetric} showed that this represents a difference in aggregate porosity, a negative polarimetric color being typical of more compact aggregates. Moreover, the wavelength at which the polarimetric color changes its sign can be used to estimate the aggregate porosity. Unfortunately, cometary spectropolarimetric data are almost completely absent despite the high importance of this type of data. 

Polarimetry of comets with a high spatial resolution (imaging polarimetry) opened up new horizons, allowing us to see a detailed polarimetric map of the coma and discover polarimetric jets, which are seen only in polarization (Fig.\ref{Fig5.2}b). Studying such jets is the most direct way to explore the relationship between areas on the nucleus from which the dust originates and the compositional/structural characteristics of the dust particles, thus revealing the structure of the nucleus itself. The polarimetric maps also provide information about the changes the dust undergoes and reveal its sublimation and fragmentation, helping to determine the composition of cometary refractories and volatiles. 

To summarize, the most informative polarimetric study of comets is imaging spectropolarimetry. This also refers to studying any medium that contains dust particles, e.g., the atmosphere of Titan and Saturn’s rings. This methodology can also be used to characterize interplanetary dust at different heliocentric distances, elongations, and latitudes thus reconstructing the solar-system debris disk. 

\section{Magnetic fields}
 Magnetic fields manifest themselves in polarization through the linear and circular polarization caused by the alignment of particles and polarization in spectral lines of excited atoms or ions. 
 
Since the majority of realistic dust particles can be characterized by some elongation, they can be aligned when affected by some mechanism of alignment, among which the magnetic field is the most widely spread and strongest one \citep[see,][]{2003JQSRT..79..881L}. Dust particles in the vicinity of planets, especially those with a significant magnetosphere, such as Jupiter and Saturn, can be aligned by the planetary magnetic field. The resulting polarization of the scattered light allows one not only to estimate the strength of the field but also to map it. This technique is widely used in studies of star-forming regions and molecular clouds \citep{2000prpl.conf..247W}.

Besides grain alignment, polarization can result from the ability of atoms to be aligned in their ground or metastable state by anisotropic radiation. As the aligned atoms precess in the magnetic field, the absorption lines become polarized with the polarization direction revealing the direction of the magnetic field \citep{2008ApJ...677.1401Y, 2012JQSRT.113.1409Y}. The polarization in spectral lines caused by such an alignment can be used to reveal the 3D orientation of the magnetic field. 

Another opportunity to study planetary magnetic fields is provided by the polarized light of their aurorae and airglow. Aurorae are produced by photons generated by ionized atoms regaining electrons or returning from an excited state to the ground one. Because polarization is strongest in the direction perpendicular to the magnetic field lines, polarimetric observations of aurorae, provide a way to determine the magnetic field configuration. Auroral polarization can also be used to:
\begin{itemize}[label=$-$]
     \item monitor the solar wind intensity, thus monitoring ``space weather;'' 
     \item reveal the energies of the precipitating particles and the altitude of emissions; 
     \item reveal the abundances of the line-producing species and their seasonal fluctuations.
   \end{itemize}  
Polarization of the auroral emission lines has been already demonstrated for Earth, Jupiter, Saturn, Uranus, and Neptune \citep{2012GeoRL..39.7105L, 2000AdSpR..26.1551B}, opening the way to study their magnetic fields. Even more exciting is the discovery of aurorae on Mars and Venus. These planets do not have an intrinsic magnetic field and their aurorae are associated with magnetic anomalies on the surface of Mars \citep[see,][]{2017JGRA..122.2260C}, and in the ionosphere of Venus \citep[see,][]{2021GeoRL..4895545X}. Studying their polarization opens new opportunities for geological and atmospheric studies of Mars and Venus.    
     
\section{Astrobiology}
 Unlike linear polarization, which becomes non-zero upon almost any scattering from particles and surfaces, non-zero circular polarization reveals a mirror asymmetry of the scattering medium \citep{hulst1981light}. One of the fundamental natural asymmetries is homochirality, the domination of molecules of a specific handiness (left-handed in amino acids and right-handed in sugars) in molecules of biological origin, whereas non-biological molecules demonstrate both handiness present in equal amounts. It was suggested that the homochirality could be a result of asymmetric photolysis \citep[e.g.,][]{2001AdSpR..27..313H,2022OLEB...52...93G}, and, thus, can exist already in prebiological molecules in the molecular clouds and protoplanetary disks. Asymmetry of the molecules of biological nature can indicate itself in the circular polarization of the light they scatter, making circular polarization a possible marker of life. 
 
Some time ago a significant interest was drawn to circular polarization observed in comets \citep{2007Icar..186..317R}, and immediately the hypothesis was suggested that the origin of this circular polarization is related to the presence of homochiral organic materials similar to those found in meteorites \citep[e.g.,][]{pizzarello2000non}. The origin of cometary circular polarization is not completely clear yet, although some recent studies attributed it to the dust particles aligned in the solar magnetic field \citep{kolokolova2016clues}.  

\begin{figure}[ht]
\includegraphics[angle=0,trim=0cm 0cm 0cm 0cm,scale=.4]{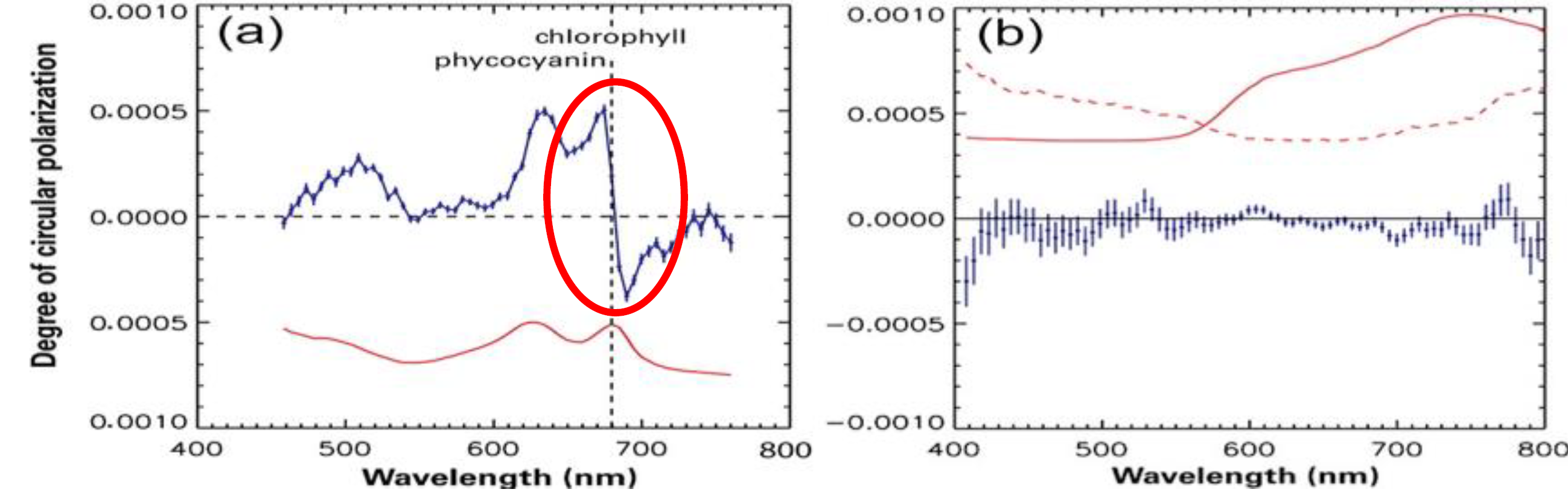}
\caption{Spectral behavior of circular polarization (blue) and absorption (red) for biological, maple leaf, (a) and non-biological, mineral, (b) samples \citep{2009PNAS..106.7816S}.}
\label{Fig5.3}
\end{figure}

Even more interesting would be to search for homochiral organics in other life-supporting environments, e.g., on Mars, Europa, and Titan. The rationale for this search has been supported by laboratory measurements of colonies of terrestrial microorganisms \citep{2009PNAS..106.7816S,sparks2009circular} which always produced circular polarization upon transmission or reflection of light, whereas various mineral samples always showed it to be zero (Figure~\ref{Fig5.3}). Moreover, for those organisms, a very special spectral trend has been measured:  circular polarization experiences a jump and even sign change within absorption bands (e.g., the 655 nm chlorophyll absorption band for cyanobacteria and vegetation). Such a jump is typical for the circular polarization within the absorption bands of homochiral molecules which makes it different from a stable circular polarization resulting from the alignment of particles. Thus, spectral studies of circular polarization can be a powerful tool for the search for life in the Solar system and other planetary systems and for the search for prebiological organics elsewhere.

\newpage
\section{Instrument Requirements}
\noindent \textbf{Science Program:} \newline
Solar System Objects

\vspace{2ex}
\noindent \textbf{Linear and/or Circular Polarimetry:} \newline
Linear and circular, the whole Stokes vector.

\vspace{2ex}
\noindent \textbf{Wavelength Coverage:} \newline
0.4 - 1 $\mu$m, although it will be very beneficiary to extend it to 5 $\mu$m (to cover Z-L spectral bands).

\vspace{2ex}
\noindent \textbf{Field-of-View:} \newline
A large field of view is interesting for comets, but comet tails become rapidly faint, so there is not a big case to have something larger than a few arcmin. A large field of view may be important to measure the background that needs to be subtracted, but of course, one has to make sure that the optics do not introduce instrumental polarisation at the edge of the field of view, as it happens for instance in the case of FORS @ VLT.

\vspace{2ex}
\noindent \textbf{Observation Type (e.g.\ imaging polarimetry, spectropolarimetry, high-contrast):} \newline
1) Imaging polarimetry
2) Low-resolution spectropolarimetry. An imaging spectropolarimeter would be most beneficial.

\vspace{2ex}
\noindent \textbf{Target Brightness (magnitudes):} \newline
In the range -12 (Moon) to 24 (KBO).

\vspace{2ex}
\noindent \textbf{Spectral Resolution:} \newline
100 may be enough for the solar system, but 1000-2000 is necessary for stars.
To indicate the reverse of circular polarization in absorption bands may require a spectral resolution of > 500.

\vspace{2ex}
\noindent \textbf{Spatial Resolution:} \newline
Spatial resolution may need to be high for bright objects (planets, Moon) but in most science cases there will not be enough photons for high spatial resolution.

\vspace{2ex}
\noindent \textbf{Required S/N or S/N per Wavelength Bin:} \newline
N/A

\vspace{2ex}
\noindent \textbf{Level of Polarization per Spatial and/or Spectral Element:} \newline
0.01\%.

\vspace{2ex}
\noindent \textbf{Acceptable Instrumental Polarization and Absolute Error (IP(\%)  $\pm$ error(\%)):} \newline
For linear polarization, an accuracy better than 0.05\% is necessary; for circular polarization, it should be better than 0.01\% 

\vspace{2ex}
\noindent \textbf{Acceptable Linear to Circular Crosstalk:} \newline
 If we are measuring linear polarization in a source that is not circularly polarized, cross-talk will never be a big issue. If we measure a small circular polarization in a source that we know is highly linearly polarized, then even a small x-talk will be problematic. To be safe, cross-talk should stay below a few percent, ideally less than 1\%. 

\vspace{2ex}
\noindent \textbf{Additional Requirements:} \newline
N/A

\chapter{Exoplanets and circumstellar disks}
\label{sec:disks}
  
%
%
\section{Exoplanets}

\subsection{Probing Exoplanet Magnetic Fields}

 Observations of an exoplanet's magnetic field would allow us to holistically understand the planet by providing constraints on difficult-to-study planetary properties. To start, knowledge of a planetary magnetic field can provide robust constraints on a planet's interior structure, including its composition and thermal processes \citep{G2015,Zarka2018haex,Zarka2015SKA}. Historically, initial insights into the interior structure of the Solar System's gas giants were derived from the understanding that they possess magnetic fields \citep{Hubbard1973}. Additionally, magnetic fields will also affect (either enhancing or suppressing) atmospheric escape \citep{G2015,Zarka2018haex,Egan2019}. Magnetic drag could also be an important factor for atmospheric dynamics and evolution \citep{Perna2010a,Beltz2023}. Most importantly, planetary magnetic fields may be one of the many ingredients for understanding planetary formation \citep{Lovelace2008,Batygin2018,Jia2023}. Finally, magnetic fields could potentially play a crucial role in understanding the potential habitability of Earth-like exoplanets by providing context for atmospheric escape mechanisms \citep{Lazio2019,Zarka2018haex}, providing protection against cosmic rays \citep{Gr2015}, and giving clues about the presence of tectonic plates \citep{Shahar2019}. 

Despite decades of efforts, the direct detection of magnetic fields on exoplanets has remained elusive \citep{G2015}. There are many methods proposed to constrain the magnetic fields of exoplanets. Radio observations are among the best methods to detect exoplanetary magnetic fields \citep{Zarka2007,Zarka2015SKA,Gr2015}. Many of the other methods are prone to false positives \citep{G2015,Turner2016a,Route2019}. Recently, the first possible detection of an exoplanet, $\tau$~Boo~b, in the radio was reported \citep{Turner2021}. However, despite their promise, radio observations are limited by radio frequency interference, variations in the Earth's ionosphere, and quenching of the emission mechanism \citep{Turner2017pre8,G2015,Griessmeier17PREVIII}; besides, the signals can only be detected out to $\sim$20 pc from the Sun with current instruments \citep{Turner2019}. Therefore, complementary methods to radio observations are needed. 

Recently, \citet{o2020} proposed a new method for detecting magnetic fields in exoplanets. The method is based on measuring polarization in the helium line at 1083 nm observed during transits of exoplanets. As stellar radiation passes through a magnetized exoplanet atmosphere, linear polarization is created via a mechanism called the Hanle effect and circular polarization via the Zeeman effect \citep{TrujilloBueno2005}. The Hanle and Zeeman effects are highly complementary probes of magnetic fields. The Hanle effect is primarily sensitive to the plane-of-the-sky component of the magnetic field that induces in-transit linear polarization of the helium line on the order of a few times $10^{-4}$, for a broad range of magnetic field strengths. The line-of-sight component of the magnetic field, on the other hand, induces circular polarization via the Zeeman effect, which is directly proportional to the magnetic field strength. Hence, in order to constrain the magnetic field strength and geometry, observations of all Stokes parameters are needed. One of the spectral features in which both Hanle and Zeeman effects operate and whose diagnostic power has been demonstrated over the years in spectropolarimetric observations of the Sun \citep{TrujilloBueno2005} is the neutral helium line triplet at the wavelength of 1083~nm. This method requires high-spectral resolution so that we can separate the radial velocity of the planetary spectral features from those of its host star and the Earth’s atmosphere. 

Strong absorption signals in the helium 1083~nm line have recently been detected in several dozen transiting exoplanets ranging from super-Earths to hot Jupiters (e.g. \citealt{Spake2018,OrellMiquel2023,Allart2023}). Most of this sample cannot be studied using radio observations. Also, the number of planets with detected helium is expected to rise with the onset of large-scale helium-dedicated surveys \citep{Jentink2023}. Existing spectropolarimeters like \textit{SPIRou} can currently study the magnetic fields of hot Jupiters assuming field strengths from Jupiter's ($\sim5$~G) to magnetic fields observed in brown dwarfs ($\sim$kG; \citealt{Kao2016}). These field strengths are expected to produce signals on the order of $10^{-4}$ or greater in each polarization. There are currently several ongoing observing campaigns to study the magnetic fields of exoplanets using this technique. Nevertheless, in order to constrain the magnetic fields of Neptune to Earth-like planets we need sensitivities on the order of $10^{-6}$. Therefore, future spectropolarimeters on 30-m telescopes (such as the \textit{GMT}) are needed to study the magnetic fields of a wide diversity of exoplanets.  

It might be possible to extend this method to other magnetically active lines (e.g., calcium, sodium, potassium, water, iron), some of which are very common in exoplanet atmospheres \citep{Roberge2018}. All these lines are used commonly on the Sun and other stars to study magnetic fields with planetary field strengths \citep{Landi2004,Ramos2008}. This modeling will be done in the coming years. For example, \citet{Kuzmychov2013} found that magnetic splitting of individual chromium hydride (CrH) lines shows an asymmetry on the order of 10$^{-4}$ due to the Paschen-Back effect at planetary field strengths. Recently, CrH was detected for the first time in an exoplanet atmosphere using high-resolution observations\citep{Flagg2023}; however, the CrH signal was only seen by cross-correlating 1000's of lines.

\subsection{Instrument Requirements}

\noindent \textbf{Science Program:} \newline
Studying the magnetic fields of exoplanets 

\vspace{2ex}
\noindent \textbf{Linear and Circular Polarimetry:} \newline
Linear and Circular (All Stokes parameters are needed)

\vspace{2ex}
\noindent \textbf{Wavelength Coverage:} \newline
$300-2000$~nm

\vspace{2ex}
\noindent \textbf{Field-of-View:} \newline
All sources are point sources

\vspace{2ex}
\noindent \textbf{Observation Type (e.g.\ imaging polarimetry, spectropolarimetry, high-contrast):} \newline
Spectropolarimetry

\vspace{2ex}
\noindent \textbf{Target Brightness (magnitudes):} \newline
$m_\mathrm{V} = 7 - 12$

\vspace{2ex}
\noindent \textbf{Spectral Resolution:} \newline
100,000 

\vspace{2ex}
\noindent \textbf{Spatial Resolution:} \newline
N/A

\vspace{2ex}
\noindent \textbf{Required S/N or S/N per Wavelength Bin:} \newline
2000 S/N per wavelength bin 

\vspace{2ex}
\noindent \textbf{Level of Polarization per Spatial and/or Spectral Element:} \newline
N/A

\vspace{2ex}
\noindent \textbf{Acceptable Instrumental Polarization and Absolute Error (IP(\%)  $\pm$ error(\%)):} \newline
IP $<$ 0.1$\%$\\
Absolute Error: 10$^{-6}$ or 0.0001$\%$

\vspace{2ex}
\noindent \textbf{Acceptable Linear to Circular Crosstalk:} \newline
$<$0.1$\%$

\vspace{2ex}
\noindent \textbf{Additional Requirements:} \newline
The measurement is differential between the out-of-transit and in-transit signal, so we can achieve our science goals without requiring extremely precise absolute polarimetric sensitivities. Also, all the in-transit data will be combined together.

\section{Circumstellar Disks}
Scattered light from circumstellar disks is expected to be polarized due to asymmetries in their structure or scattering/absorption by the dust grains. The polarized fraction as a function of the scattering angle is found to be sensitive to the detailed dust grain properties. Thus, when combined, total and polarized flux measurements help constrain the geometrical and scattering properties of the disks \citep[e.g.][]{2020AJ....160...79A} more than with either one alone. Additionally, Polarized Differential Imaging (PDI) can improve sensitivity to polarized sources relative to unpolarized star-light, improving the effective contrast ratio \citep{2015Perringpi}. 

\subsection{Protoplanetary Disks}
The magnetic fields play a significant role in the formation and evolution of the protoplanetary disks. High-contrast Linear and circular polarimetric observations will help constrain these magnetic fields in addition to the scattering properties of disks \cite{heese2020characterization,stolker2016scattered}. 

\subsection{Debris Disks}
Polarization observations of debris disks have been carried out using the current ground-based high-contrast imaging polarimeters and Advanced Camera for Surveys coronagraph in Hubble Space Telescope (HST) in optical and near-IR wavelength regions\citep{Esposito2020,crotts2021deep,hull2022polarization,chen2020multiband,engler2017hip,milli2017hr4796a,graham2007signature,hom2020first,maness2009hubble}. \citet{esposito2020debris} conducted a four-year survey of 104 stars and obtained polarization observations of 35 debris disks in the near-IR wavelengths using the Gemini Planet Imager \citep[GPI,][]{macintosh2014first} at the Gemini Telescope,. In addition to this, SPHERE/ZIMPOL, SPHERE/IRDIS \citep{beuzit2006sphere} and NaCo at the VLT, and Subaru Coronagraphic Extreme Adaptive Optics (SCExAO)/HiCIAO at the Subaru telescope have also been used to image debris disks in polarization in the near IR and optical wavelengths to constrain the disk properties. 
\begin{figure}[!ht]
    \centering
    \includegraphics[width=0.7\textwidth]
    {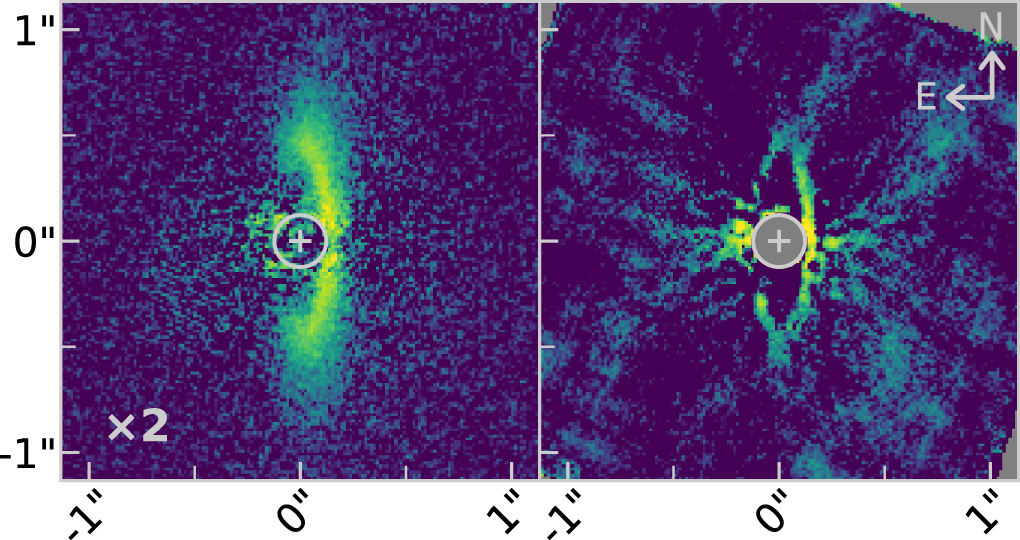}
    \includegraphics[width=0.7\textwidth]
    {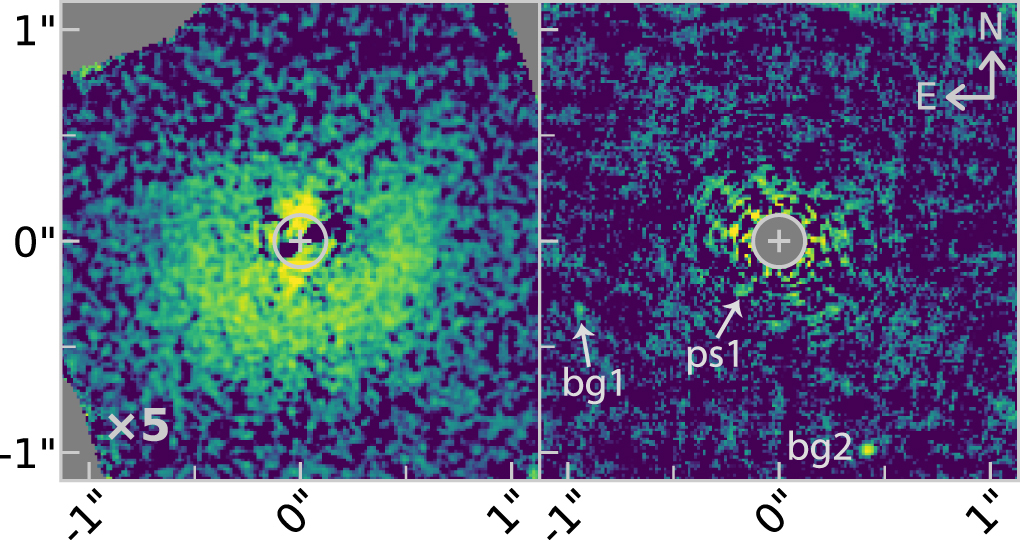}
    \caption{Polarized intensity, ${{ \mathcal Q }}_{\phi }$ (left) and total intensity (right) in H-band observed though GPIES for HD 117214 (top panel) and HD 156623  (bottom panel) debris disks are shown from \cite{Esposito2020}.}
    \label{fig:my_label}
\end{figure}
An extreme adaptive optics high-contrast imaging polarimeter for GMT, GMagAO-X, will be able to reach a raw contrast of $10^{-5}$ and post-processed contrast of $10^{-8}$ in the visible and near IR bands (V-H bands) and can facilitate polarimetric imaging of fainter debris disks in addition to improving the contrast ratio. (The detailed science cases for the high-contrast imaging polarimetry using GMagAO-X are under preparation \textit{Haffert, S et. al.,})
\section{Instrument Requirements}
\noindent \textbf{Science Program:} \newline
High-contrast imaging polarimetric observations of debris disks

\vspace{2ex}
\noindent \textbf{Linear and/or Circular Polarimetry:} \newline
Linear

\vspace{2ex}
\noindent \textbf{Wavelength Coverage:} \newline
$0.4-10~\mu$m

\vspace{2ex}
\noindent \textbf{Field-of-View:} \newline
10\arcsec

\vspace{2ex}
\noindent \textbf{Observation Type (e.g.\ imaging polarimetry, spectropolarimetry, high-contrast):} \newline
high-contrast

\vspace{2ex}
\noindent \textbf{Target Brightness (magnitudes):} \newline
$1-2$ mJy/$\rm arcsec^2$

\vspace{2ex}
\noindent \textbf{Spectral Resolution:} \newline
N/A

\vspace{2ex}
\noindent \textbf{Spatial Resolution:} \newline
N/A

\vspace{2ex}
\noindent \textbf{Required S/N or S/N per Wavelength Bin:} \newline
N/A.

\vspace{2ex}
\noindent \textbf{Level of Polarization per Spatial and/or Spectral Element:} \newline
$10-30$\%

\vspace{2ex}
\noindent \textbf{Acceptable Instrumental Polarization and Absolute Error (IP(\%)  $\pm$ error(\%)):} \newline
$\sim 3 - 4~\pm$ 0.1\% (static)

\vspace{2ex}
\noindent \textbf{Acceptable Linear to Circular Crosstalk:} \newline
<0.1\%

\vspace{2ex}
\noindent \textbf{Additional Requirements:} \newline
Extreme adaptive optics and coronagraphy



    
\begin{appendices}

\chapter{Polarimetric modeling of the GSMTs}
\label{sec:modeling}
\drop{A}s polarimetry is a photon-starved technique, with the next-generation of 30m class telescopes, i.e.\ the Extremely Large Telescope \citep[ELT,][]{cayrel2012elt}, the Thirty Meter Telescope \citep[TMT,][]{nelson2006tmt},  and the Giant Magellan Telescope \citep[GMT,][]{bernstein2014overview}, one can aim to obtain high accuracy polarimetric data for very faint targets. However, one significant challenge for accurate polarimetry is the change in polarization introduced by the telescope optics \citep{tinbergen2005astronomical}. Instrumental polarization (IP) is the polarization introduced by the telescope and instrument optics, and cross-talk (CT) is the conversion of input linear polarization to circular polarization or vice versa by the telescope and instrument optics. For a Cassegrain telescope, IP and CT  are found to be on the order of $0.1\%$ and increase to a few percent for Nasmyth telescopes \citep{keller2002instrumentation}. These effects depend on the coating and the angle of incidence (AOI) on the optical surface. For some of the existing and future polarimetric instruments, the polarization effects are modeled using polarization ray tracing algorithms developed by \cite{chipman1995mechanics} and \cite{sen1997instrumental} to understand the nature and variation of these effects and to design efficient calibration and mitigation strategies \citep{harrington2017polarization,van2020polarimetric,harrington2011deriving,sen1997instrumental,maharana2022walop}. 

In the context of the next generation, giant segmented mirror
telescopes (GSMTs), modeling of the polarization effects from telescope optics of the TMT and ELT have been done and presented. For the TMT with a monolithic primary mirror \cite{anche2018analysis} and \cite{atwood2014polarimetric} estimated the variation of instrumental polarization of 4.5\% to 0.6\% and crosstalk of 73\% to 11\% for wavelengths from 0.4~$\mu$m to 2.5~$\mu$m for an on-axis star. Similarly, for the ELT with a monolithic primary mirror, \cite{de2014instrumental} estimated an instrumental polarization and crosstalk of 6\% and 30\%, respectively, for zenith angle of 0\textdegree and wavelength of 0.55 $\mu$m.
The polarization effects from the telescope/instrument optics also give rise to polarization aberrations, such as polarization tilt, astigmatism, and piston, that affect high-contrast imaging instruments. \cite{anche2023polarization} has presented the polarization aberrations for all three GSMTs in the context of high-contrast imaging of exoplanets. 

When trying to achieve accurate polarimetry, the IP and CT effects from the telescope are usually calibrated using observations of polarized and unpolarized standard stars.  However, for GSMTs, at least 50\% of the observing time could be lost for the calibration and overheads \citep{de2014instrumental}. Thus, modeling the polarization effects from the telescope optics is critical to planning the instrument port and technical requirements of the polarimeter, calibration, and mitigation strategies. 
\vspace{5mm}

\begin{hibox1}
Comparing different technical requirements of the science cases shows that most of the polarimetric observations with GMT require linear Stokes measurement ($Q$ and $U$) using a moderate resolution imaging spectropolarimeter in the wavelength range of 0.35--1.0 $\mu$m with the field-of-view of 1--2 arcmin with the acceptable level of instrumental polarization of $\sim$ 0.05\% and crosstalk of $\sim$ 0.1\%. 
\end{hibox1}
 \section{Modelling the polarization effects from the Giant Magellan Telescope}
\label{sec2}

The polarization effects from the telescope and instrument optics are estimated using a polarization ray tracing algorithm described in detail in \cite{anche2018analysis,chipman1995mechanics,harrington2011deriving}. Here, we perform the ray tracing of 10000 rays through the telescope mirrors using Zemax Optics studio and estimate the polarization effects and telescope Mueller matrices in Python. We use the optical design of GMT with seven 8.365-m primary mirror segments, a secondary mirror of seven 1.05-m segments, and a 0.3-m Nasmyth fold mirror. The entire list of parameters of the optical design of GMT, along with the other GSMTs, is shown in Table \ref{tab:tele-param} from \cite{anche2023polarization}. GMT will have instruments mounted at both the direct Gregorian (DG) and folded port (FP) focus locations, for which we estimate the polarization effects using a bare Aluminum coating on all three mirrors. 
\begin{table*}[!ht]
\centering
\begin{tabular}{|llll|}
\hline
\multicolumn{1}{|l|}{Telescope   parameters}            & \multicolumn{1}{l|}{\textbf{ELT}}               & \multicolumn{1}{l|}{\textbf{TMT}}                   & \textbf{GMT}                         \\ \hline
\multicolumn{1}{|l|}{Field-of-View   (arcmin)}   & \multicolumn{1}{l|}{10}                & \multicolumn{1}{l|}{15}                    & 20                          \\ \hline
\multicolumn{1}{|l|}{Wavelength of   operation ($\rm \mu$m)}       & \multicolumn{1}{l|}{0.32-25}   & \multicolumn{1}{l|}{0.32-28 }       & 0.32-25              \\ \hline
\multicolumn{4}{|l|}{}                                                                                                                                                    \\ \hline
\multicolumn{1}{|l|}{Primary   mirror-M1}             & \multicolumn{1}{l|}{Elliptical concave}  & \multicolumn{1}{l|}{Hyperbolic   concave}  & Elliptical concave            \\ \hline
\multicolumn{1}{|l|}{Diameter (m)}                        & \multicolumn{1}{l|}{39}                & \multicolumn{1}{l|}{30}                    & 25                          \\ \hline
\multicolumn{1}{|l|}{Conic constant}                  & \multicolumn{1}{l|}{-0.996}            & \multicolumn{1}{l|}{-1.00097}              & -0.998                      \\ \hline
\multicolumn{1}{|l|}{Segment type}                    & \multicolumn{1}{l|}{Quasi Hexagonal}   & \multicolumn{1}{l|}{Quasi Hexagonal}       & Circular                    \\ \hline
\multicolumn{1}{|l|}{Segment size (m)}                    & \multicolumn{1}{l|}{1.45}              & \multicolumn{1}{l|}{1.44}                 & 8.365                      \\ \hline
\multicolumn{1}{|l|}{Segment gap}                     & \multicolumn{1}{l|}{4-mm}               & \multicolumn{1}{l|}{2.5-mm}                 & 0.345-m                      \\ \hline
\multicolumn{1}{|l|}{Number of   segments}            & \multicolumn{1}{l|}{798}               & \multicolumn{1}{l|}{492}                   & 7                           \\ \hline
\multicolumn{4}{|l|}{}                                                                                                                                                    \\ \hline
\multicolumn{1}{|l|}{Secondary   mirror-M2}           & \multicolumn{1}{l|}{Hyperbolic convex}            & \multicolumn{1}{l|}{Hyperbolic convex}                & Elliptical concave                     \\ \hline
\multicolumn{1}{|l|}{Diameter (m)}                        & \multicolumn{1}{l|}{4.1}               & \multicolumn{1}{l|}{3.1}                   & 1.05                        \\ \hline
\multicolumn{1}{|l|}{Conic constant}                  & \multicolumn{1}{l|}{-2.28962}          & \multicolumn{1}{l|}{-1.381}                & -0.716927274                \\ \hline
\multicolumn{1}{|l|}{Segments}                        & \multicolumn{1}{l|}{No}                & \multicolumn{1}{l|}{No}                    & Yes                         \\ \hline
\multicolumn{1}{|l|}{Number of   Segments}            & \multicolumn{1}{l|}{NA}                & \multicolumn{1}{l|}{NA}                    & 7                           \\ \hline
\multicolumn{1}{|l|}{Active/Adaptive control}                 & \multicolumn{1}{l|}{Yes}               & \multicolumn{1}{l|}{No}                    & Yes                         \\ \hline
\multicolumn{4}{|l|}{}                                                                                                                                                    \\ \hline
\multicolumn{1}{|l|}{Tertiary   mirror-M3}            & \multicolumn{1}{l|}{Circular Concave}           & \multicolumn{1}{l|}{Flat}       &  Flat               \\ \hline
\multicolumn{1}{|l|}{Diameter (m)}                        & \multicolumn{1}{l|}{4}                 & \multicolumn{1}{l|}{3.5$\times$2.5}               & 0.3                          \\ \hline
\multicolumn{1}{|l|}{Inclined angle(\textdegree)}                  & \multicolumn{1}{l|}{0}              & \multicolumn{1}{l|}{45}                    & 45                          \\ \hline
\multicolumn{4}{|l|}{}                                                                                                                                                    \\ \hline
\multicolumn{1}{|l|}{Quaternary   mirror-M4}          & \multicolumn{1}{l|}{Flat}              & \multicolumn{1}{l|}{NA}                    & NA                          \\ \hline
\multicolumn{1}{|l|}{Diameter}                        & \multicolumn{1}{l|}{2.38$\times$2.34}         & \multicolumn{1}{l|}{}                      &                             \\ \hline
\multicolumn{1}{|l|}{Inclined angle(\textdegree)}                  & \multicolumn{1}{l|}{7.25}             & \multicolumn{1}{l|}{}                      &                             \\ \hline
\multicolumn{4}{|l|}{}                                                                                                                                                    \\ \hline
\multicolumn{1}{|l|}{Fifth   Mirror-M5}               & \multicolumn{1}{l|}{Flat}              & \multicolumn{1}{l|}{NA}                    & NA                          \\ \hline
\multicolumn{1}{|l|}{Diameter}                        & \multicolumn{1}{l|}{2.6$\times$2.1}           & \multicolumn{1}{l|}{}                      &                             \\ \hline
\multicolumn{1}{|l|}{Inclined angle (\textdegree)}                  & \multicolumn{1}{l|}{37.25}             & \multicolumn{1}{l|}{}                      &                             \\ \hline
\multicolumn{4}{|l|}{}                                                                                                                                                    \\ \hline
\multicolumn{1}{|l|}{Proposed coating}& \multicolumn{1}{l|}{Protected Silver}& \multicolumn{1}{l|}{Protected Silver}&
\multicolumn{1}{|l|}{Bare Aluminum} \\ \hline
\multicolumn{4}{|l|}{}  \\ \hline
\multicolumn{1}{|l|}{Final focal ratio}                  & \multicolumn{1}{l|}{f/17.48}             & \multicolumn{1}{l|}{f/15}  & f/8.2 \\ \hline
\multicolumn{1}{|l|}{Instrument   mounting locations} & \multicolumn{1}{l|}{Nasmyth platforms} & \multicolumn{1}{l|}{Nasmyth platforms}     & Direct Gregorian            \\ \hline
\multicolumn{1}{|l|}{}                                & \multicolumn{1}{l|}{A and B}           & \multicolumn{1}{l|}{A and B}               & Folded port                 \\ \hline
\multicolumn{1}{|l|}{}                                & \multicolumn{1}{l|}{}                  & \multicolumn{1}{l|}{}                      & Auxiliary ports             \\ \hline
\end{tabular}
\caption{Telescope parameters for the three telescopes \citep{cayrel2012elt,nelson2006tmt,bernstein2014overview}. Table reproduced from \cite{anche2023polarization}}
\label{tab:tele-param}
\end{table*}

\begin{figure}[!h]
    \centering
    \includegraphics[width=1\textwidth]{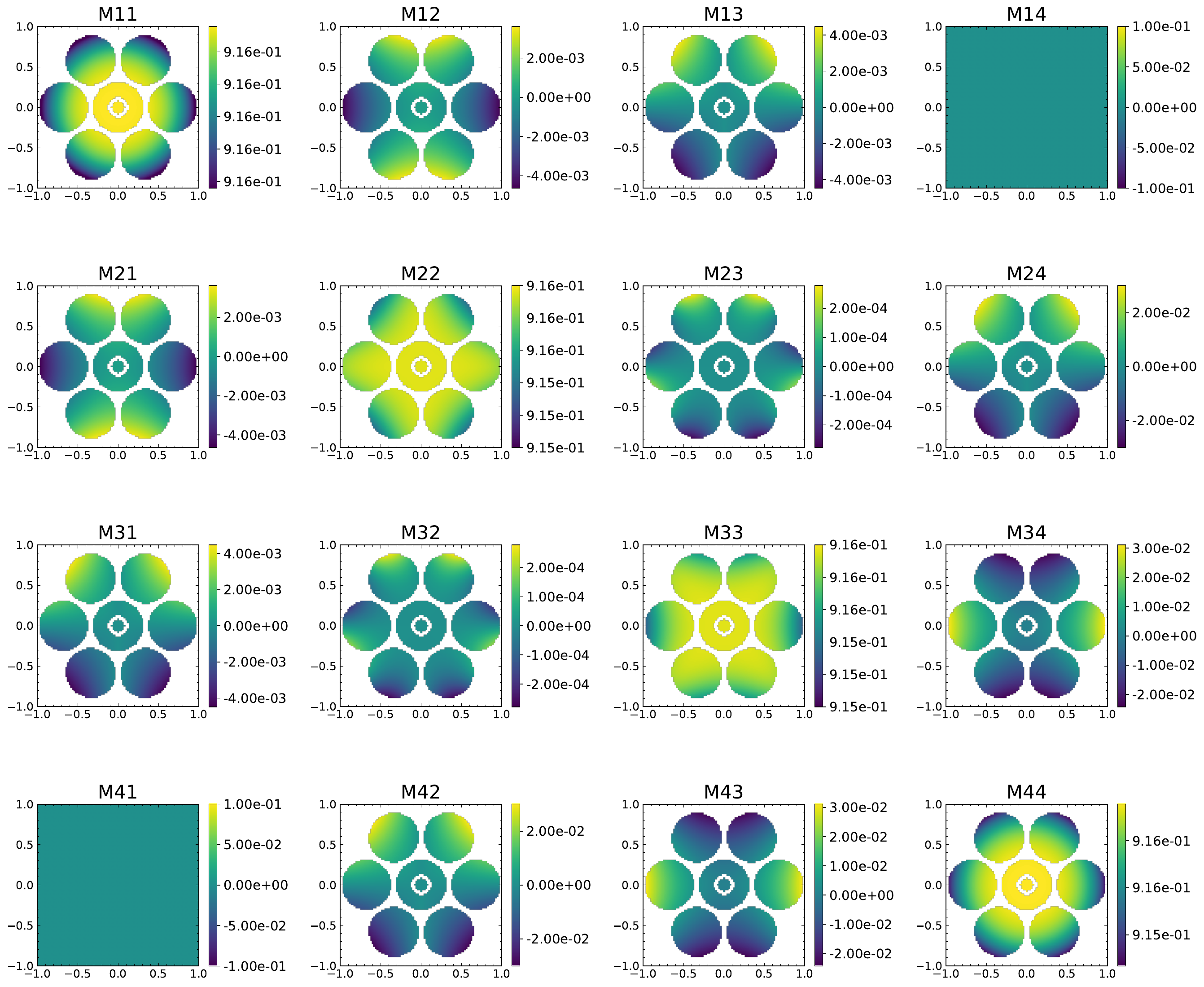}
    \caption{The Mueller matrix of the primary mirror of GMT for the on-axis rays in the $V$ band. The $x$ and $y$ axis in each panel correspond to the normalized pupil coordinates ($px$ and $py$). $M11$, $M22$, $M33$, and $M44$ show the azimuthal symmetry with values increasing from the center to the edge of the mirror. The azimuth anti-symmetry is seen in all the elements except the diagonal elements, and $M14$ and $M41$ are found to be zero. For $M21$ and $M31$, which contribute to the instrumental polarization, one quadrant of the mirror exhibits positive values while the other exhibits negative values thereby compensating for one another and resulting in negligible instrumental polarization. Similar behavior is also seen for the crosstalk terms ($M43$, $M34$, and $M42$).}  
\label{fig: Mueller-pri}
\end{figure}

The evolution of the polarization through any optical system can be represented using the 4$\times$4 Mueller matrices with the input and output polarization defined as Stokes parameters \cite{clarke2009stellar}. Figure \ref{fig: Mueller-pri} shows the Mueller matrix of the primary mirror of GMT on the normalized pupil coordinates. $M21$ and $M31$ indicate the instrumental polarization, $M42$, $M43$, and $M34$ indicate the linear to circular crosstalk. The diagonal elements ($M11$, $M22$, $M33$, \& $M44$) correspond to $I\rightarrow I$, $Q\rightarrow Q$, $U\rightarrow U$, and $V\rightarrow V$ respectively. The diagonal elements follow the variation of Fresnel reflection coefficients $r_p$ and $r_s$ on the mirror surface. The elements corresponding to the instrumental polarization and crosstalk show positive and negative values in consecutive quadrants, which result in negligible values when averaged for all the on-axis rays at the primary focus as shown for the TMT primary mirror in \cite{anche2018analysis}. A similar Mueller matrix is obtained after the Gregorian secondary of the GMT. Most science cases require a FOV of 1--2$\arcmin$ and observations spanning the optical and near IR wavelength regions. Therefore, we estimate the Mueller matrices for different field positions and wavelengths of interest. 
 \subsection{Instrumental Polarization (IP) and Crosstalk (CT) at the direct Gregorian focus over a 5~arcmin FOV}
\begin{figure}[!h]
    \centering
    \includegraphics[width=1\textwidth]{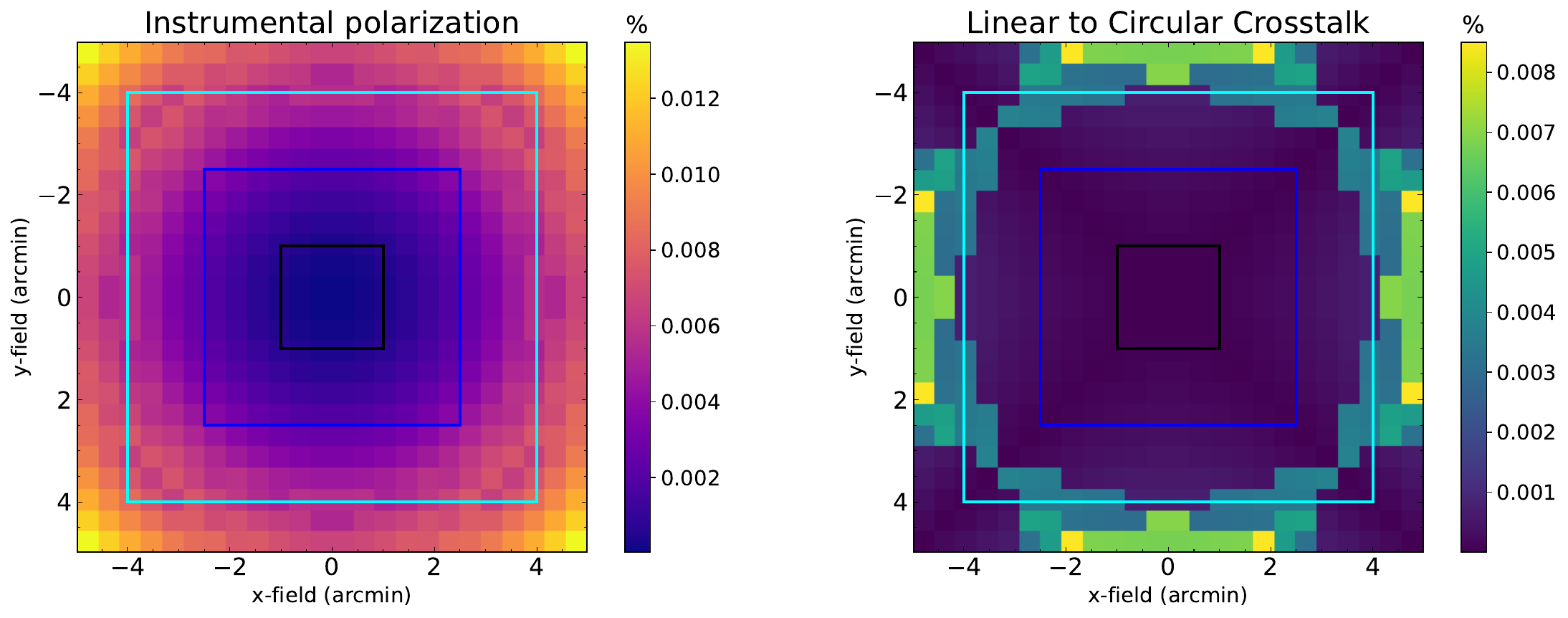}
    \caption{Instrumental polarization and crosstalk estimated at the Gregorian focus of GMT for a 5$\arcmin$ field-of-view at 0.54~$\mu$m. The three squares correspond to a 1$\arcmin$, 2.5$\arcmin$, and 4$\arcmin$ FOV. }
    \label{fig:polgreg}
\end{figure}

\begin{figure}[!h]
    \centering
    \includegraphics[width=1\textwidth]{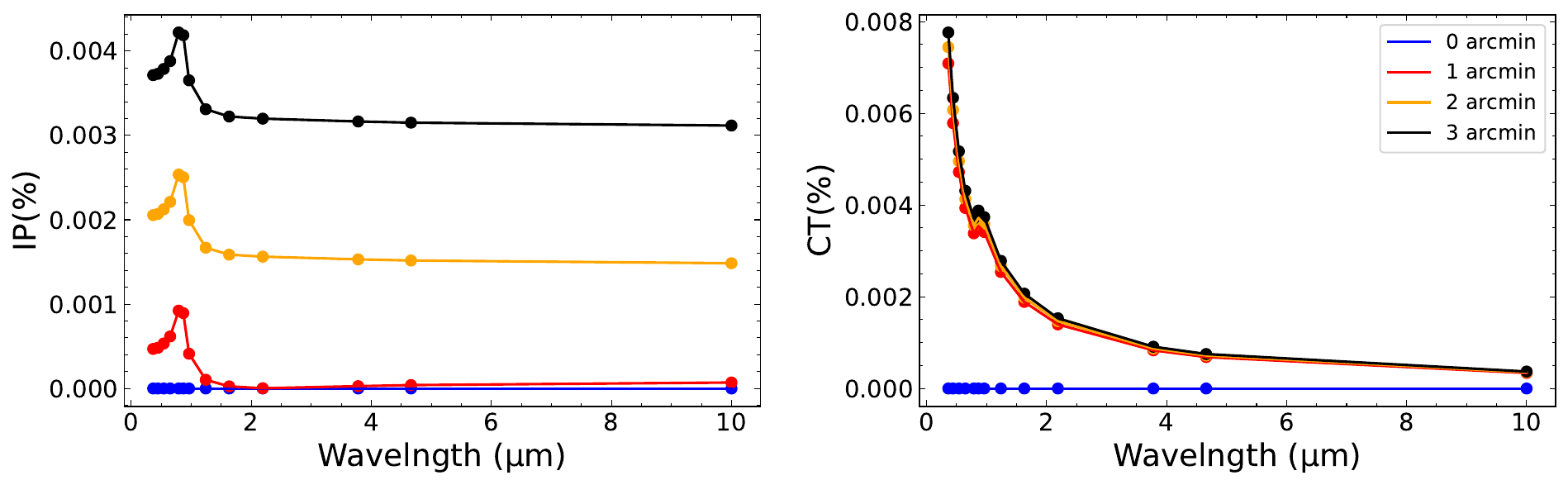}
    \caption{Variation of the instrumental polarization and crosstalk estimated at the direct Gregorian focus of GMT for wavelengths 0.35~$\mu$m to 10~$\mu$m.}
    \label{fig:polgregfield}
\end{figure}

Using the polarization ray tracing algorithm, we estimate the IP and CT for a field-of-view of 5$\arcmin$ at a wavelength of 0.54~$\mu$m. The variation of IP and CT over the 5$\arcmin$ FOV is shown in Figure \ref{fig:polgreg}. The field is divided into a 24 $\times$ 24 array of field positions, and the IP and CT increase with field angle as the asymmetry increases. For FOV $\sim$ 4$\arcmin$, the IP and CT values are estimated to be $<$ 0.01\%, which meets the requirements specified in the science cases. Figure \ref{fig:polgregfield} shows the variation of IP and CT over the wavelength of 0.35--10~$\mu$m for different field angles. The IP values increase in the optical and peak in the near-IR wavelength region and then decrease from the near-IR and mid-IR wavelengths. This behavior is directly related to the variation of the refractive index of aluminum. The CT values decrease with the wavelength and show a small bump near 0.85~$\mu$m (at the position where IP shows the maximum value). 
\section{Comparison of IP and CT at the Nasmyth focus for all three GSMTs }
\begin{figure}[!h]
\centering
    \includegraphics[width=0.8\textwidth]{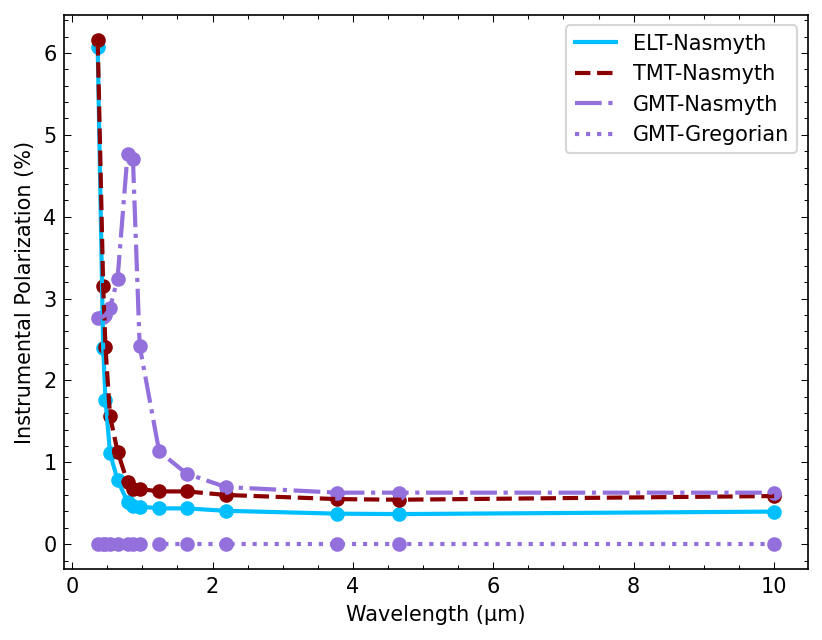}
    \caption{Variation of instrumental polarization with wavelength for all the three GSMTs. We use Gemini coating (Ag+$\rm Si_3N_4$) on all the mirrors of ELT and TMT, and bare aluminum for GMT.}
    \label{fig: IP-NasmythfocusGSMTs}
\end{figure}

We estimate the Mueller matrices for all three GSMTs at the Nasmyth focus using the polarization ray tracing algorithm. For ELT, we use the Silver+5.5~nm $\rm Si_3N_4$  on all five mirrors; for TMT, we use Silver+8.5~nm $\rm Si_3N_4$ on all three mirrors, and finally, for GMT we use bare aluminum on all the mirrors.  Figure \ref{fig: IP-NasmythfocusGSMTs} shows the variation of IP in the wavelength range of 0.36--10~$\mu$m. TMT has the highest IP of 6.2\% at 0.364~$\mu$m, followed by ELT and GMT. The IP of ELT and TMT is seen to reduce with the wavelength, whereas, for GMT, the IP increases with wavelength up to 0.8~$\mu$m and then decreases. Between 0.78--0.86~$\mu$m, GMT has the highest IP due to the aluminum absorption at those wavelengths. Beyond 0.86~$\mu$m, the IP for GMT reduces with wavelength; however, it remains higher than ELT and TMT. 

\begin{figure}[!h]
\centering
    \includegraphics[width=0.8\textwidth]{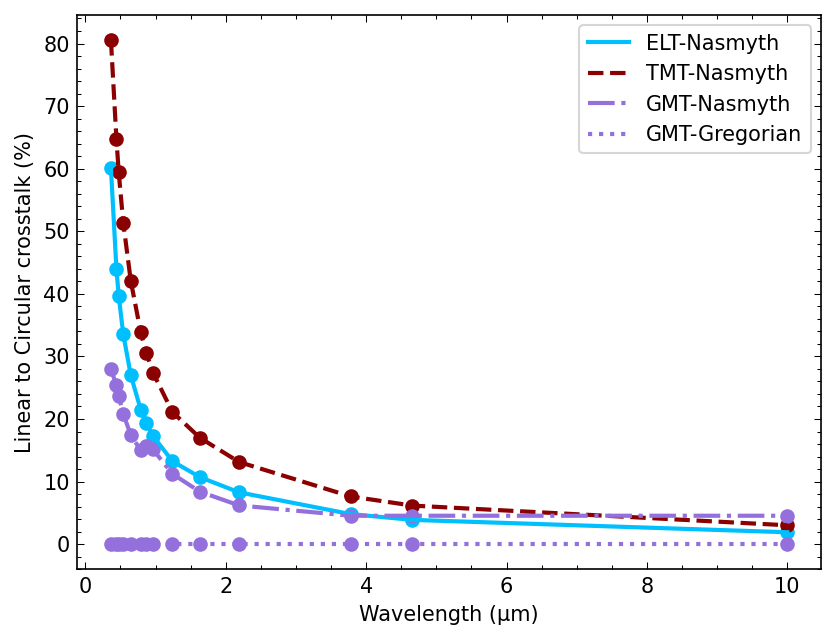}
    \caption{Variation of crosstalk with wavelength for all the three GSMTs. We use Gemini coating (Ag+$\rm Si_3N_4$) on all the mirrors of ELT and TMT, and bare aluminum for GMT.}
    \label{fig: CT-NasmythfocusGSMTs}
\end{figure}

Figure \ref{fig: CT-NasmythfocusGSMTs} shows the variation of CT with wavelength for all the three GSMTs. Similar to the case of IP, TMT shows the highest crosstalk $\sim$ 80\%, followed by ELT with $\sim$ 60\% and GMT with $\sim$ 30\% at 0.36~$\mu$m.  As TMT and ELT mirrors have a protective Silver coating, they give rise to higher CT compared to the bare Aluminum coating of GMT. Although CT reduces with wavelength, it is still estimated to be $\sim$ 10\% in the near-IR and mid-IR wavelength regions for all the GSMTs. Thus, overall, GMT will have similar polarization effects at the folded port instrument location as the other two GSMTs. The advantage of GMT is the availability of the direct Gregorian focus where the IP and crosstalk are very small and within the requirements for the science cases, as shown in Figures \ref{fig: IP-NasmythfocusGSMTs} and Figure \ref{fig: CT-NasmythfocusGSMTs}. 
\vspace{5mm}
\begin{hibox1}
Considering the complexities of polarization effects, mitigation, and the development of calibration strategies, it is favorable to design a polarimeter for the GMT at the direct Gregorian focus because the IP and CT $<$ 0.01\%. 
\end{hibox1}

\chapter{Instrument Options and Technical Requirements}
\label{sec:instrument_options}
\section{Possible routes for the polarimetry with GMT} 
Before the design of a second-generation polarimeter for GMT, a trade study is required to investigate the possibility of incorporating a polarimetric capability into one of the first-generation instruments of GMT, which are listed below:    
\begin{itemize}
    \item Giant Magellan Telescope Consortium Large Earth Finder (G-CLEF) \cite{szentgyorgyi2016gmt}. A high resolution, high precision radial velocity multi-object spectrograph operating in the range 0.35--0.95~$\mu$m.
    \item Giant Magellan Telescope Multi-object Astronomical and Cosmological Spectrograph (GMACS). A multi-object medium resolution wide field spectrograph operating in the range of 0.32--1.0~$\mu$m \cite{depoy2014update}
    \item Giant Magellan Telescope Integral-Field Spectrograph (GMTIFS) \cite{sharp2016gmtifs}. A diffraction-limited spectrograph operating in the range 0.9--2.5~$\mu$m
    \item Giant Magellan Telescope Near-IR Spectrograph (GMTNIRS) \cite{jaffe2016gmtnirs}. A single object echelle spectrograph operating in the 1.1--5.4~$\mu$m. 
    \item Commissioning Camera (ComCam) \cite{crane2020conceptual}. An alignment and image quality assessment instrument with 6$\arcmin$ FOV operating in the range of 0.36--0.95~$\mu$m.
\end{itemize}
Among these, GMTIFS and GMTNIRS operate only in the near/mid-IR wavelength region, and both are fed by the tertiary mirror that introduces significant instrument polarization and crosstalk, as shown in the previous section. GCLEF, the first instrument on GMT, is a fiber-fed instrument and is already in the stage of development and testing. ComCam and GMACS are both direct Gregorian instruments and their designs meet many of the technical requirements outlined in the science cases.  These two instruments are the best candidates for implementing a polarimetric capability. 

\section{ComCam as a polarimeter}

\begin{figure}[ht]
    \centering
    \includegraphics[width=1\textwidth]{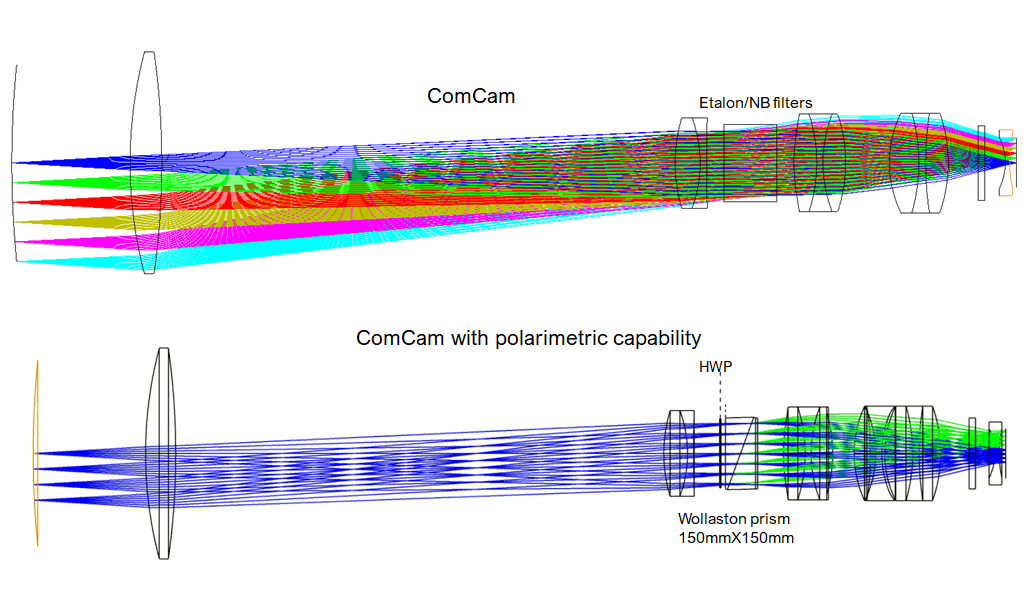}
    \caption{The top panel shows the current ComCam design \citep{crane2020conceptual}, and the bottom panel shows the ComCam design with the polarimetric capability. The first two collimator lenses form the pupil at the location of the etalon/narrow band (NB) filter. To convert the instrument to an imaging polarimeter, a half-wave plate (HWP, for linear polarimetry) and  Wollaston prism can be added at the location of etalon.}
    \label{fig:comcam-pol}
\end{figure}

Figure \ref{fig:comcam-pol} shows the current ComCam optical design from \cite{crane2020conceptual}. The collimator system consists of a field lens and the cemented doublet to reimage the telescope pupil plane where the narrowband filters or a Fabry-Perot etalon can be inserted into the beam. The cemented triplets and the field flattener form the camera, and a broadband filter can be moved into the beam before the field flattener.  To incorporate an imaging polarimetric capability, we propose the addition of a rotating half- (for linear polarimetry) or quarter- (for circular polarimetry) wave plate with a Wollaston prism at the location of the Fabry-Perot etalon and analyze the performance. As the Wollaston prism splits the beam into ordinary and extraordinary beams, a field mask/aperture stop must be placed before the field lens or Wollaston prism to avoid overlapping the beams. Additionally, a polarization grating (described below) with a rotating quarter-wave plate (QWP) can be inserted at the location of the Wollaston prism to incorporate the spectropolarimetric capability. 

\subsection{Requirements}
The ComCam imaging system requirements are given in terms of the full width at half maximum (FWHM) and the root mean square (RMS) radius ($r_{\text{RMS}}$). The relationships are defined as follows (this is an approximate estimation based on Gaussian functions):

\begin{equation}
\boldsymbol{\sigma} = r_{\text{RMS}}
\end{equation}

\begin{equation}
\text{FWHM} = 2.355\boldsymbol{\sigma} = 2.355 \, r_{\text{RMS}}
\end{equation}

Thus,

\begin{equation}
r_{\text{RMS}} = 0.425 \, \text{FWHM}
\end{equation}

The ComCam imaging requirements at different wavelengths are:

\begin{itemize}
    \item At 440~nm: 
    \begin{equation*}
    \text{FWHM} < 0.14\arcsec = 35.7 \, \mu\text{m} \quad \Rightarrow \quad r_{\text{RMS}} = 15.2 \, \mu\text{m}
    \end{equation*}
    \item At 550~nm: 
    \begin{equation*}
    \text{FWHM} < 0.12\arcsec = 30.6 \, \mu\text{m} \quad \Rightarrow \quad r_{\text{RMS}} = 13.0 \, \mu\text{m}
    \end{equation*}
    \item At 790~nm: 
    \begin{equation*}
    \text{FWHM} < 0.10\arcsec = 25.5 \, \mu\text{m} \quad \Rightarrow \quad r_{\text{RMS}} = 10.8 \, \mu\text{m}
    \end{equation*}
\end{itemize}

The ComCam requirements are more stringent than the expected seeing conditions because the instrument will be used to assess the ground layer adaptive optics (GLAO) performance. For GMT-Pol, the image quality requirements are not as stringent since it will typically operate in seeing-limited mode.

The median seeing at the GMT site on Cerro Las Campanas ranges from 0.64$\arcsec$ in summer to 0.73$\arcsec$ in winter \citep{thomas-osip2008}. Assuming the delivered image quality (\(S_d\)) of an instrument or the measured seeing is the combination of the site seeing (\(S_a\)) and the instrument image quality (\(S_i\)) added in quadrature:

\begin{equation}
S_d = \sqrt{S_a^2 + S_i^2}
\end{equation}

Where:

\[
\begin{aligned}
S_d &= \text{delivered seeing} \\
S_a &= \text{atmospheric seeing} \\
S_i &= \text{instrument seeing (image quality)}
\end{aligned}
\]

If we require the delivered seeing (\(S_d\)) to be no more than 10\% worse than the median atmospheric seeing (\(S_a\)):

\begin{equation}
S_d = 1.1 \times S_a
\end{equation}

Then:

\begin{equation}
S_i < 0.46 \times S_a
\end{equation}

For the best median atmospheric seeing of 0.64$\arcsec$:

\begin{equation}
S_i < 0.29"
\end{equation}

Therefore, for GMT-Pol the image quality requirement is:

\begin{itemize}
    \item At all wavelengths: 
    \begin{equation*}
    \text{FWHM} < 0.29\arcsec = 74.0 \, \mu\text{m} \quad \Rightarrow \quad r_{\text{RMS}} = 31.5 \, \mu\text{m}
    \end{equation*}
\end{itemize}

\subsection{Field-of-View}
At all wavelengths (\(\lambda\)) the ComCam field-of-view (FOV) is 6.0$\arcmin$ in diameter. For imaging polarimetry with GMT-Pol, two images, the ordinary and extraordinary, must fit on the same detector. Therefore:

\begin{equation}
\text{FOV}_\text{img} < 3.0\arcmin \times 6.0\arcmin
\end{equation}

\subsection{Polarization grating}

A new form of polarization optics called polarization gratings \citep[PG;][]{2010PASP..122.1471P} promise to enable spectropolarimetric observations across a wide wavelength range (0.4$-$40~$\mu$m).  Optical and near-IR PGs are already commonly used in applications such as liquid crystal displays, nonmechanical optical beam steering, and telecommunication. As \cite{2010PASP..122.1471P} described, MIR PGs are in development at a lower cost than previously available polarimetric components. A PG is a thin-film beamsplitter (less than 300~$\mu$m thick) that uses a periodic variation in the retardance of the material. This device is functionally analogous to a Wollaston prism, which simultaneously acts as a spectroscopic and polarimetric disperser for circularly polarized light. Using a quarter wave retarder (quarter wave plate, QWP, or analog) to convert linearly to circularly polarized light, PGs can be used as linear polarimetric analyzers \citep{2010PASP..122.1471P}. PGs offer high throughput, high levels of birefringence, can be produced within a few weeks, and are relatively inexpensive to produce (a few hundred dollars). Additionally, PGs can be produced in a very large format; they are currently produced up to 150~mm in diameter in North Carolina State University's (NCSU) academic research setting, and larger sizes of up to 300~mm are under development. A single PG can cover GMT's 300~mm collimated beam, reducing instrumental cost and components compared to a 3$\times$3 mosaic Wollaston-prism system. PGs have been tested to obtain polarization observations of the twilight sky in the wavelength range of 500--900~nm by \cite{2016SPIE.9908E..2IM}. As a proof of concept, a near-IR ($1-5$ $\mu$m) PG is currently in use with the MMT-Pol polarimeter\footnote{MMT-Pol is a 1--5~$\mu$m imaging- and spectropolarimeter mounted at the Cassegrain focus of the 6.5-m MMT. Further information can be found at \url{https://www.mmto.org/instrument-suite/mmtpol/}}.
 \begin{figure}[!h]
\includegraphics[angle=0,trim=-5cm 0cm 0cm 0cm,scale=.50]{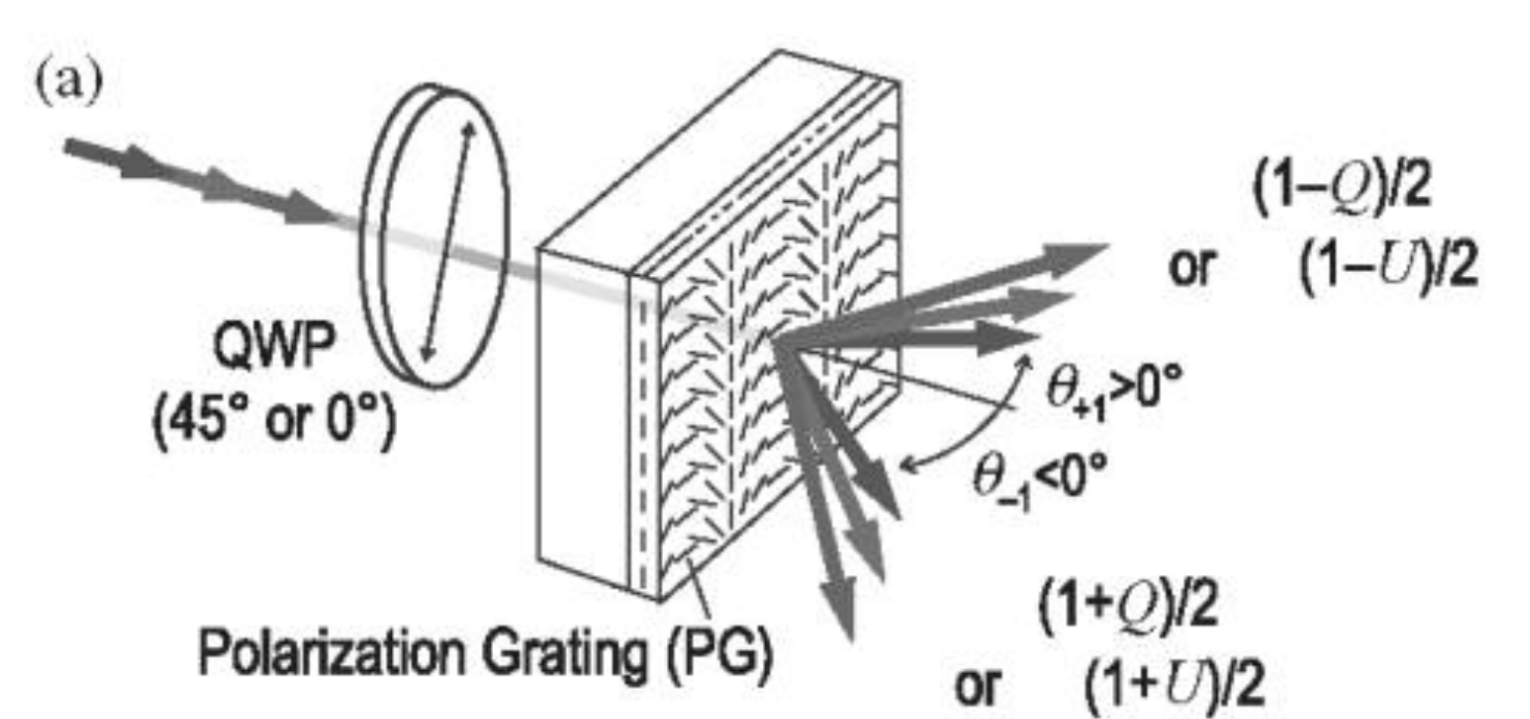}
\caption{Basic PG behavior and optical elements.
A QWP sensitizes the diffracted intensities to linearly polarized light (i.e., Stokes parameters $Q$ and $U$). 
Incident linearly polarized radiation passes through the QWP, converting it to circularly polarized light, which is then split by the PG equally into the two first orders. The periodic pattern of lines on the PG indicates the local orientation of the extraordinary axis of the birefringent liquid crystal polymer \citep{2010PASP..122.1471P}.}
\label{fig131_1}
\end{figure}
\section{Polarimeter for GMT: GMT-Pol}   
Depending on the outcome of the trade study, another possible route for the polarimetry with GMT will be the design of GMT-Pol, a stand-alone imaging spectropolarimeter as a second-generation instrument at the direct Gregorian focus port. GMT-Pol is proposed to operate from 380--1000~nm with a 5$\arcmin$ FOV and a moderate spectral resolution. Depending on the science target, either a Wollaston prism or a polarization grating can be used as the analyzer for imaging or spectro-polarimetry, respectively.

\vspace{5mm}

\end{appendices}

\bibliography{bibliography1,ISMbibliography,exoplanets}{}
\bibliographystyle{aasjournal}

\end{document}